\documentclass[mnsc,nonblindrev]{informs3}

\OneAndAHalfSpacedXI 

\usepackage{mathtools}
\usepackage{comment}
\usepackage{booktabs}
\usepackage{natbib}
 \bibpunct[, ]{(}{)}{,}{a}{}{,}%
 \def\bibfont{\small}%
\TheoremsNumberedThrough     
\ECRepeatTheorems

\EquationsNumberedThrough    

\begin{document}


\RUNAUTHOR{Cong, Xie, and Zhang}

\RUNTITLE{Knowledge Accumulation, Privacy, and Growth in a Data Economy}

\TITLE{Knowledge Accumulation, Privacy, and \\Growth in a Data Economy\thanks{We thank Jasmine Cheng, Kay Giesecke, Shiyang Huang, Laura Veldkamp, Ji Yan, Buyuan Yang, Liyan Yang, and the anonymous associate editor and referees for their feedback. We also thank conference and seminar participants at the Midwest Macroeconomics Meeting, Australasian Finance and Banking Conference, Delhi School of Economics and the Econometric Society Winter School, Machine Lawyering's Conference for helpful comments and discussions. Cong and Xie are grateful for the financial support from the Ewing Marion Kauffman Foundation and from the National Natural Science Foundation of China (Grant No. 71973076), respectively. The contents of this publication are solely the responsibility of the authors.}
}

\ARTICLEAUTHORS{
	\AUTHOR{Lin William Cong}
	\AFF{SC Johnson College of Business, Cornell University, Ithaca, NY 14853, \EMAIL{will.cong@cornell.edu}} 
	\AUTHOR{Danxia Xie}
	\AFF{Institute of Economics, Tsinghua University, Beijing, China 100084, \EMAIL{xiedanxia@tsinghua.edu.cn}}
	\AUTHOR{Longtian Zhang}
	\AFF{School of International Trade and Economics, Central University of Finance and Economics, Beijing, China 102206, \EMAIL{zhanglongtian@cufe.edu.cn}}
} 

\ABSTRACT{%
We build an endogenous growth model with consumer-generated data as a new key factor for knowledge accumulation. Consumers balance between providing data for profit and potential privacy infringement. Intermediate good producers use data to innovate and contribute to the final good production, which fuels economic growth. Data are dynamically nonrival with flexible ownership while their production is endogenous and policy-dependent. Although a decentralized economy can grow at the same rate (but are at different levels) as the social optimum on the Balanced Growth Path, the R\&D sector underemploys labor and overuses data---an inefficiency mitigated by subsidizing innovators instead of direct data regulation. As a data economy emerges and matures, consumers' data provision endogenously declines after a transitional acceleration, allaying long-run privacy concerns but portending initial growth traps that call for interventions.
}%


 \HISTORY{Initially written in October 2019 and first submitted on April 16, 2020.}

\maketitle

%

\newpage
\section{Introduction}
\label{intro}
Data not only help produce new products and services, but also are used in research and development and knowledge creation, which in turn improves production efficiency.\footnote{Data have become a key input to many technology and financial firms \citep[e.g.,][]{Veldkamp2019,Ibarra2018,cong2019alternative}; \citet{Manyika2011} estimate that big data research is believed to save over 100 billion Euros for Europe, and reduce medical care cost of the United States by 8\% or 300 billion dollars annually.} Endogenous generation, dynamic nonrivalry, and flexible ownership of data distinguish them from labor and capital, with implications for labor market allocations and policies. Meanwhile, the proliferation of big data applications often comes at the expense of consumer privacy and is associated with discrimination and misuse.\footnote{Cambridge Analytica and Facebook's data scandal presents a salient example \citep{Confessore2018}. The United States have the highest ratio of data records stolen relative to population \citep[over 6 billion stolen records, exceeding the population by 19 times,][]{DataBreach2020}. Data from \textit{China Judgements Online} and \textit{China Academy of Information and Communications Technology} also reveal that civil and criminal cases of data leakage have been increasing in lock-steps with the growth of China's digital economy.} Despite the introduction of data privacy laws ranging from the General Data Protection Regulation (GDPR) to the California Consumer Privacy Act (CCPA) to Japan's Act on the Protection of Personal Information, we know little about how data (mis)usage, digital infrastructure, and privacy regulation affect knowledge accumulation and the growth of a dynamic economy where data are the new key driver.

To fill in this gap, we build on \citet{Romer1990} to develop an endogenous growth model of the data economy. Our key innovation lies in that consumer data add to R\&D and knowledge accumulation. At the same time, data are by-products of economic activities with potential privacy issues, which differ from other input factors such as labor or capital because growth can endogenously feed back to data generation. Consumers in our baseline model choose the quantity of data to sell to intermediate firms, cognizant of potential information leakage and misuses. Innovative intermediate firms utilize the raw data for research that contributes to the final good production. Specifically, data are transformed into intermediate goods (information products included)---a feature absent in other models. Data can generate spillovers through knowledge accumulation, which is further enhanced when they are traded over time and used by multiple parties with little reproduction costs (both static and dynamic nonrivalry). These effects are moderated in the model by the disutility from potential data privacy violation.

We show that a decentralized economy grows at the same rate as the social optimum on the Balanced Growth Path (BGP). But social welfare and consumer surplus are strictly lower due to underemployment and data overuse in the R\&D sector. Monopolistic markups in the production of intermediate goods lead to a crowding out of labor in R\&D \citep{Jones1995}, producers then compensate the underemployment of labor in R\&D by more aggressive utilization of data. The crowding-out of labor crowds in data usage in R\&D to a socially excessive level. R\&D labor employment and data usage can thus deviate significantly from those in the social planner's solution, especially in the initial phase of BGP (albeit less severe in the long run).

Different from recent studies such as \citet{Jones2020}, data are overused even when they are nonrival and owned by consumers who directly factor in the disutility from data leakage or abuse. Direct regulation on data usage comes at the cost of economic growth and constitutes a wealth transfer from future generations to current generations. Rather than taxing data overuse, our model reveals that subsidizing R\&D wage or intermediate producers are more effective at mitigating the social inefficiency. Moreover, because data expand the innovation possibility frontier which exhibits diminishing returns to scale, historical data usage reduces the benefit of future data usage, potentially leading to a declining data provision per capita in the long run. 

As the economy transitions into a steady growth, data provision by consumers can undergo accelerated growth before declining. Importantly, a low initial growth may limit data generation even in the planner's solution, which further delays the transition to high growth stages in the long run---a form of growth trap. Interventions such as foreign aid for digital infrastructure development can help escape the trap, but only to the extent that the data generating constraint is still binding, i.e., when economic activities translate into limited volumes of data.

Our paper primarily contributes to the emerging literature on information and the data economy. Relative to earlier studies on the social value, sales, and property rights of information \citep[e.g.,][]{Hirshleifer1971,Admati1990,Murphy1996}, recent studies focus on connecting digital information with privacy issues \citep[e.g.,][]{Akcura2005,Casadesus2015} or the nonrivalry of data and competition \citep[e.g.,][]{Ichihashi2020,Easley2019}. We differ by being the first to connect data usage to knowledge accumulation and endogenous growth with privacy considerations. Furthermore, our model complements studies microfounding data privacy concerns \citep[e.g.][]{ichihashi2020online,Liu2020}, and is broadly consistent with empirical patterns on the correlation of data economy size with regulation, privacy issues, and declining labor income share in the United States and around the world \citep{Karabarbounis2014,Tang2019,Abis2020,Barkai2019,Liao2020}. 
 
As such, our theory adds to the large literature on economic growth, especially recent studies embedding data into growth models by allowing them to directly enter production. For example, \citet{Jones2020} highlights the underutilization of data due to its nonrivalry and the importance of giving data property rights to consumers; \citet{Farboodi2020} emphasizes that data have bounded returns in long-run growth. We complement \citet{Jones2020} by allowing data to facilitate knowledge accumulation in a semi-endogenous growth model \citep{Jones1995,Jones2016} and accounting for the consumers' data privacy concerns in the spirit of \citet{Stokey1998} and \citet{Acemoglu2012}.\footnote{Unlike the growth literature focusing on firms' property rights, we allow consumers to own and sell data.} Through a mechanism different from \citet{Farboodi2020}, we also find that data play a limited role in the long run. 


\section{The Model}
\label{mod}

We incorporate data production, data-based innovation, knowledge accumulation, and data privacy concerns into a macroeconomic model to characterize economic growth. Our dynamic ``data economy'' consists of representative agents who are both consumers and workers, innovative intermediate producers, and a final good producer. Time is continuous and infinite. 

\subsection{Representative Consumers}
\label{hh}

A population of homogeneous representative consumers (or households) grow at a constant rate $n$ and is $L(t)$ at time $t$. Besides consumption, they each choose in each period to supply one unit of labor inelastically in either the R\&D sector (intermediate good production) or final good production. 

Each consumer produces data as by-products of consumption \citep[e.g.,][]{Veldkamp2005} and can sell the data to intermediate good producers.\footnote{``Selling'' can be broadly interpreted: For example, not having to pay for the use of Gmail can be viewed as compensation to consumers for allowing Google to use their data in ways delineated in the usual terms and conditions. In Section \ref{ownershipSection}, we allow firms to own data as in \citet{Jones2020} and derive additional insights in addition to showing robustness of the main results.} However, the data involve personal information and risk leakage and misuse, leading to a disutility that they consider when selling data. We follow the literature \citep[e.g.,][]{Jones2020} to allow data to fully depreciate in every period in the baseline model, but relax this assumption in Online Appendix \ref{app:accumulation} and further discuss it in Section \ref{sec:nonrivalry}.

Each consumer's utility maximization problem is then:
\begin{equation}
	\max_{c(t),\varphi(t)}\int_0^{\infty}e^{-(\rho-n) t}\left[\frac{c(t)^{1-\gamma}-1}{1-\gamma} - \varphi(t)^\sigma \right] \mathrm{d}t,
	\label{prefer}
\end{equation}
subject to
\begin{equation}
	\dot a(t) = (r(t)-n)a(t) + w(t) + p_\varphi(t) \varphi(t) - c(t), \forall t \in [0,\infty),
	\label{budget}
\end{equation}
and
\begin{equation}
	\frac{\dot \varphi(t)}{\varphi(t)} \leq \frac{\dot c(t)}{c(t)}+s.\label{dataBound}
\end{equation}
Here, $c(t)$ is the per capita consumption level at time $t$ and $\varphi(t)$ is the quantity of data a consumer provides to potential intermediate producers for R\&D. $\rho$ is the consumers' discount rate, the reciprocal of $\gamma$ (also indistinguishable from the coefficient of risk aversion) is the elasticity of intertemporal substitution (EIS) of consumption, and $\sigma$ parameterizes the disutility of data misuse or privacy violation, $\varphi(t)^{\sigma}$, which also depends on the quantity of data provided. $s$ represents the tightness of this constraint and depends on the digital infrastructure, legal development, and privacy regulation policy of the country. We normalize $s$ to zero in the baseline and discuss comparative statics with respect to $s$ later to understand the impact of policy intervention. 

In the budget constraint (\ref{budget}), $a(t)$ is the asset held by a consumer at time $t$ and $r(t)$ is its interest rate. $w(t)$ and $p_\varphi(t)$ are the time-$t$ wage for labor and price of data, respectively. The relevant variables for transitional dynamics are the growth rates of data contribution and consumption. To see this, we can derive the system's evolution in the form of Euler equations from the Hamilton:
\begin{equation}\label{motC}
	\frac{\dot c(t)}{c(t)} = \frac{r(t)-\rho}{\gamma}
\end{equation}
and
\begin{equation}\label{motphi}
	\frac{\dot p_\varphi(t)}{p_\varphi(t)} -(\sigma-1) \frac{\dot \varphi(t)}{\varphi(t)}= r(t) -\rho.
\end{equation}
Constraint (\ref{dataBound}) therefore requires the growth rate of data provision to be bounded by the corresponding growth rate of consumption, which is natural and ensures analytical tractability. Moreover, (\ref{dataBound}) directly implies that $\varphi(t)\leq \chi c(t)$ for some constant $\chi>0$, which other recent studies feature. In other words, data are by-products of economic activities and cannot exceed a fixed proportion of consumption activities.  

\subsection{The Final Good Producer}
\label{final}

A representative final good producer operates in a competitive environment with a production function,
\begin{equation}
	Y(t)= L_E(t)^\beta \int_0^{N(t)} x(v,t)^{1-\beta} \,\mathrm{d}v ,
	\label{profunc}
\end{equation}
where $L_E(t)$ is the labor employed and $N(t)$ is the number of varieties of intermediate goods used in the final good production at time $t$. $x(v,t)$ is the total amount of intermediate good of variety $v$ which can only be used in the final good production for one period. We can view price $p_{x}(v,t)$ as the rental fee of patents. Finally, $\beta$ is the elasticity coefficient of labor in the final good production. The final good producer's profit maximization over labor employed and the amount of each intermediate good yields the first order conditions:
\begin{equation}
	x(v,t) = \left[ \frac{1-\beta}{p_x(v,t)} \right]^{\frac{1}{\beta}} L_E(t),
	\label{FOCx}
\end{equation}
and
\begin{equation}
	w(t) = \beta L_E(t)^{\beta-1} \int_0^{N(t)}x(v,t)^{1-\beta}\,\mathrm{d} v.
	\label{FOCLE}
\end{equation}

\subsection{Intermediate Producers}
\label{inter}

An unlimited pool of potential intermediate producers decide whether and how much to conduct research, the success of which gives them monopoly over the intermediate product developed. An intermediate producer therefore enters the market by conducting R\&D and evaluating the prospective profit from success less the costs of labor and data as inputs for R\&D. We solve the intermediate producers' problem through backward induction. 

\paragraph{\textbf{Production phase.}}
\label{prointer} Upon R\&D success, each entrant produces a distinct variety of intermediate goods in a monopolistic market. The present value of an intermediate good of variety $v$ is
\begin{equation}\label{paval}
	V(v,t) = \int_t^\infty \exp{\left(- \int_t^s r(\tau) \,\mathrm{d}\tau \right)} \pi(v,s) \,\mathrm{d}s,
\end{equation}
where the profit from the intermediate good of variety $v$ in a single period of time $t$ is
\begin{equation}
	\pi(v,t) = p_{x}(v,t)x(v,t) - \psi x(v,t).
	\label{piexp}
\end{equation}
Here, $\psi$ is the marginal cost of this production process, which is set as constant in this economy.

Substituting (\ref{FOCx}) into (\ref{piexp}) and taking derivative with respect to $p_{x}(v,t)$ yields the optimal price of each variety of intermediate good:
\begin{equation}
	p_{x}(v,t)=\frac{\psi}{1-\beta},
	\label{px}
\end{equation}
which is the same among different varieties and different periods. Next, substitute (\ref{px}) into (\ref{FOCx}):
\begin{equation}
	x(v,t)= \left[\frac{(1-\beta)^2}{\psi} \right]^{\frac{1}{\beta}}  L_E(t) \equiv x(t),
	\label{x}
\end{equation}
which implies that production quantity is independent of the variety.

\smallskip

Substituting (\ref{px}) and (\ref{x}) into (\ref{profunc}) and (\ref{piexp}), respectively, we get
\begin{equation}\label{piL}
    \pi(v,t) =  \frac{\psi^{1-\frac{1}{\beta}}\beta}{(1-\beta)^{1-\frac{2}{\beta}}} L_E(t) \equiv \pi(t)
\end{equation}
and
\begin{equation}
	Y(t) = \left[\frac{(1-\beta)^2}{\psi}\right]^{\frac{1}{\beta}-1} N(t)L_E(t).
	\label{Y}
\end{equation}
Also, we derive the wage rate as:
\begin{equation}\label{wage}
	w(t) = \beta \left[\frac{(1-\beta)^2}{\psi}\right]^{\frac{1}{\beta}-1} N(t).
\end{equation}

\paragraph{\textbf{Entry and R\&D phase.}}
\label{entinter} Potential intermediate producers enter by conducting R\&D using both labor (researchers, including those working on computing, and AI to allow us to use data more efficiently), $L_R(t)$, and data purchased from consumers, $\varphi(t)L(t)$.\footnote{In some sense, data acquired by firms are their intangible capital, akin to customer capital discussed in recent studies \citep[e.g.,][]{Dou2019}.} We assume that intermediate producers pay a ``data processing cost'' before they use data for R\&D:
\begin{equation}
    \theta \varphi(t)^{\phi}, \quad \textit{where} \quad \theta\geq 0 \quad \textit{and}\quad \phi>1. \label{processingCost}
\end{equation}
For simplicity, we set $\theta=0$ in the baseline model, and solve in Section \ref{ownershipSection} the general case in which $\theta>0$ and $\phi>1$ ensure that the cost function is increasing and convex. We need $\theta>0$ when we allow firm ownership of data because otherwise the solution is trivial: Firms certainly use up all the data if there is not any data processing cost.

We also specify the evolution of the aggregate innovation possibility frontier (number of varieties) as:
\begin{equation}
	\dot N(t) =\eta N(t)^\zeta [\varphi(t)L(t)]^\xi L_R(t)^{1-\xi}=\eta N(t)^\zeta \varphi(t)^\xi l_R(t)^{1-\xi} L(t),
	\label{frontierori}
\end{equation}
where $\eta>0$ is an efficiency term of innovation, $\xi\in(0,1)$ represents the relative contribution of data and labor in innovating process, $0<\zeta<1$ captures the spillover effect of knowledge, $\varphi(t)L(t)$ corresponds to data provided by all the consumers in the period, $L_{R}(t)$ is labor employed in R\&D sector, and $l_R(t)=L_R(t)/L(t)$ denotes the fraction of labor employed in the R\&D sector.\footnote{We follow the convention in the literature to maintain constant return to scale of inputs; $\zeta>0$ corresponds to the positive external returns due to the cumulative nature of innovation \citep[e.g.,][]{Chang1995,Cong2020IPO}, whereas $\zeta<1$ follows the innovation of \citet{Jones1995} over \citet{Romer1990} to reflect that it becomes harder to come up with new varieties of intermediate goods as the variety expands.} Labor market clearing condition requires $L_{E}(t)+L_{R}(t)\leq L(t)$. 

The fact that data enter the R\&D of intermediate goods distinguishes our paper from studies such as \citet{Jones2020} that have data only entering directly into final good productions (similar to the $x(v,t)$ in our model). While it holds for data-intensive industries (e.g., self-driving cars) that use data directly as inputs, other industries instead use non-data intermediate goods as inputs. Data, in addition to R\&D labor, can be useful for creating those intermediate goods, which our model captures. As \citet{Romer1990} aptly puts, ``an intermediate-goods sector uses the designs from the research sector together with forgone output to produce the large number of producer durables that are available for use in final-goods production at any time.'' Our model speaks to non-data-intensive industries as well.

An intermediate producer decides how much labor $L_R(t)=l_R(t)L(t)$ and data $\varphi(t)L(t)$ to employ and purchase to maximize the expected net profit:
\begin{equation}\nonumber
	\max_{L_R(t),\varphi(t)} \eta N(t)^\zeta \varphi(t)^\xi l_R(t)^{1-\xi} L(t) V(t) - w(t) l_R(t) L(t) - p_\varphi(t) \varphi(t)L(t).
\end{equation}
The first order conditions yield two free-entry conditions: 
\begin{equation}\label{freeentryinfo}
	\eta \xi N(t)^\zeta \varphi(t)^{\xi-1} l_R(t)^{1-\xi} V(t) = p_\varphi(t),
\end{equation}
and
\begin{equation}\label{freeentrylab}
	\eta (1-\xi) N(t)^\zeta \varphi(t)^\xi l_R(t)^{-\xi} V(t) = w(t).
\end{equation}
Intermediate producers enter until the marginal benefits of adding data or labor equal to the marginal costs. 

\subsection{Equilibrium Definition}

An equilibrium in our model is an allocation in which all intermediate producers choose $\{p_x(v,t),\varphi(t),L_R(t)\}_{v\in [0,N(t)],t=0}^\infty$ to maximize the discounted value of profits, the evolution of $\{N(t)\}_{t=0}^\infty$ is determined by free entry, the evolution of $\{r(t),w(t),p_\varphi(t)\}_{t=0}^\infty$, is consistent with market clearing, the evolution of $\{L_E(t),x(v,t)\}_{v\in [0,N(t)],t=0}^\infty$ is consistent with the final good producer's profit maximization, and the evolution of $\{c(t),\varphi(t),L_E(t),L_R(t)\}_{t=0}^\infty$ is consistent with consumers' utility maximization.

\subsection{Distinguishing Features of Data}

Although the functional forms of how data enter the production function of intermediate goods resemble that of R\&D-specific labor, the role of data in economic growth is fundamentally different from labor or capital. It is worth clarifying the distinguishing features of data before we proceed to solve the model.

First, whereas population dynamics are exogenous in many growth models (such as the possible supply of labor), data are endogenized by consumption which is itself endogenous and depends on data usage in knowledge accumulation. Privacy regulations such as the GDPR and CCPA can affect the endogenous production and usage of data while population growth is more organic and hard to regulate with immediate effects. Also, note that neither capital nor labor usage causes disutility from privacy concerns. 

Another key distinguishing feature of data our paper highlights is dynamic nonrivalry. In \citet{Jones2020}, all firms need data for production and the nonrivalry is static and cross-sectional. In our model, only potential intermediate producer needs to use raw data in every period, although the entrants and incumbents are benefiting from the same data without incurring high reproduction costs. Because data are traded between intermediate producers entering in different periods, the focus is more on data nonrivalry over time. 

This is implicit in our setup: Even though data are fully depreciated every period, they contribute to knowledge accumulation over time in terms of varieties of intermediate goods, which differs from the literature that only allows labor to enter the evolution. Intuitively, a firm may innovate incrementally by observing other firms' previous data-based innovations. As such, data create knowledge spillovers to future periods by creating new varieties---a form of dynamic nonrivalry. 

One more distinguishing feature of data from R\&D-specific labor is that data can be owned by firms but labor cannot. Traditional firms use long-term labor contracts while recent on-demand labor or freelance services as seen in Uber, TaskRabbit, Scripted, and Amazon's Mechanical Turks require spot compensations \citep{serfs2018}. Neither entails firm ownership.

We further elaborate on data's endogenous production and usage as by-products of economic activities, nonrivalry under knowledge accumulation and creative destruction, and flexible ownership in Sections \ref{regulatoryPolicy}, \ref{sec:nonrivalry}, and \ref{ownershipSection}, respectively, as we characterize the equilibrium. 

\section{Data Economy on the Balanced Growth Path}
\label{ana}

We first solve the model along the Balanced Growth Path (BGP), which requires that all variables are growing at the same constant rate---a steady state of transformed variables that the literature focuses on. Constant growth then implies $r(t)=r^*$ (Eq.(\ref{motC})). We then identify inefficient data overuse and underprovision of R\&D labor, explore policy remedies, and discuss the implications of data nonrivalry and ownership.

\subsection{Growth Rate and Labor Share in a Decentralized Economy}

\begin{proposition}
	The economic growth of the decentralized economy on the balanced growth path does not exhibit scale effect, and the BGP growth rates can be expressed as follows
	\begin{equation}\label{ggrowfinal}
		g_c^* = g_y^* = g_N^* = g^* = \left[\frac{\sigma }{(1-\zeta)\sigma - \xi (1-\gamma)}\right] n.
	\end{equation}
	The constraint on data provision per capita never binds under BGP, and its growth rate is:
	\begin{equation}\label{gphigrow}
		g_\varphi^* = \frac{1-\zeta}{\xi} g^* - \frac{1}{\xi} n = \left[\frac{1-\gamma}{(1-\zeta)\sigma - \xi (1-\gamma)}\right] n < 0.
	\end{equation}
	\label{dcgrow}
\end{proposition}

Online Appendix \ref{proofdcgrow} contains the proof. Here, $g_\varphi^*$ is negative: As the economy grows on a BGP, the data each person contributes steadily decrease in the long run, although the aggregate data use can still grow.

Because we recognize data as an input into the innovation possibility frontier, the BGP growth rates are related to not only the population growth ($n$), but also consumers' EIS ($1/\gamma$), relative contribution of data to innovation ($\xi$), and knowledge accumulation in the form of varieties of intermediate goods. 

Specifically, note that all growth rates on BGP are ultimately driven by the exogenous population growth rate $n$; growth rates become zeros when $n$ becomes zero.\footnote{Like \citet{Jones1995}, our model does not suffer from the scale effect in, for example, \citet{Romer1990}, whose endogenous growth rates depend on population level.} In Online Appendix \ref{proofcond}, we derive the parameter ranges for a BGP equilibrium to exist and to be unique. We restrict our discussions within these parameter ranges throughout the paper.

Next, when $\gamma$ converges to 1, consumers' utility function converges to a logarithmic form as in \citet{Jones1995}, the BGP growth rate becomes, according to (\ref{ggrowfinal})
\begin{equation}\label{ggrowconv}
	g^*_{\gamma \rightarrow 1} = \frac{n}{1-\zeta}.
\end{equation}
Here, only knowledge accumulation and population growth rate influence the BGP growth rate. Yet, in contrast with $ g^*_{Jones} = n(1-\xi)/(1-\zeta)$ in \citet{Jones1995}, our BGP growth rate is higher under the same set of parameters because data add positively to the innovation possibility frontier.

In general, when $\gamma>1$ \citep[as empirical studies in the macro and behavioral literature consistently estimate, e.g.,][]{Lucas1969,Coen1969,Jorgensen2002}, we find that BGP growth rate increases with $\xi$, the extent that data influence innovation possibility frontier. Counter-intuitively, BGP growth rate also increases in $\sigma$, the severity of privacy concerns, and converges to (\ref{ggrowconv}). This general equilibrium effect stems from the fact that consumers in equilibrium require a higher growth rate to compensate for the disutility from greater privacy concerns or information leakage.\footnote{Similar phenomenon can also be found in \citet{Stokey1998}: Higher growth rates are needed in equilibrium when there is greater pollution because people dislike the harm caused by pollution and require compensation.} Finally, the relationship between $\gamma$ and growth rate is negative: Consumers prefer lower growth rate and thus less production when they are less willing to substitute current consumption with future consumption.

Besides growth rates, we derive the following result regarding labor shares in Online Appendix \ref{proofDClsh}:

\begin{proposition}\label{DClsh}
	In this decentralized economy, the share of labor employed by the R\&D sector $s_{D}$ is constant in BGP, which is determined by
	\begin{equation}\nonumber
		s_{D} = \frac{1}{1+\Theta_{D}}, \quad \textit{where} \quad \Theta_{D} = \frac{g^*\gamma + \rho-n}{g^*(1-\xi)(1-\beta)}.
	\end{equation}
\end{proposition}

Here, the subscript ``$D$'' stands for decentralized. A larger growth rate $g^*$ encourages firms to employ more labor in R\&D. When the population stops growing ($g^*=0$), labor in the R\&D sector becomes zero because without growth, using labor in R\&D only leads to distuility of consumers.

\subsection{Growth Rate and Labor Share under the Planner's Solution}
\label{sp}

We now derive the BGP growth rates and the shares of labor allocated in the two sectors under socially optimal allocations, which constitute a benchmark for comparison with those in the decentralized economy. The equilibrium in a decentralized economy is not socially optimal because of monopolistic competition and knowledge spillover (the First Welfare Theorem fails here).

A social planner maximizes the utility of representative consumer/household, (\ref{prefer}), subject to the resource constraint. The resource constraint requires that the aggregate consumption $ C(t)=c(t)L(t)$ equals the aggregate net output, which we denote by $\widetilde Y(t)$. In other words,
\begin{equation}\nonumber
	C(t) = \widetilde Y(t) \equiv L_E(t)^\beta \int_0^{N(t)} x(v,t)^{1-\beta}\,\mathrm{d}v - \int_0^{N(t)} \psi x(v,t)\,\mathrm{d}v.
\end{equation}

We first characterize the static allocation given $N(t)$ in every period. The social planner chooses the optimal level of intermediate goods input $[x(v,t)]_{v \in [0,N(t)]}$ at each time $t$ given the time paths of $C(t)$, $\varphi(t)$, and $N(t)$, which is equivalent to maximizing $\widetilde Y(t)$ with respect to $x(v,t)$. Thus, the optimal net output is:
\begin{equation}
	\widetilde Y_S(t) = \left(\frac{\psi}{1-\beta}\right)^{1-\frac{1}{\beta}} \beta N(t) L_E(t),
	\label{YtS}
\end{equation}
where the subscript ``$S$'' indicates the social planner's problem. Relative to the decentralized economy (\ref{Y}), $\widetilde Y_S$ is larger with a markup of $(1-\beta)^{1-\frac{1}{\beta}}/(2-\beta)$ given the same level of labor and technology. This difference comes from the monopoly power in a decentralized economy, which lowers the provision of intermediate goods and thus, the net output of the final good. 

Given (\ref{YtS}), the social planner solves (dropping the subscript ``$S$'' to simplify notation):
\begin{equation}
	\max_{c(t),\varphi(t)}\int_0^{\infty}e^{-(\rho-n) t}\left[\frac{c(t)^{1-\gamma}-1}{1-\gamma} - \varphi(t)^\sigma \right] \mathrm{d}t,
	\label{prefer2}
\end{equation}
subject to
\begin{equation}
	\dot N(t) =\eta N(t)^\zeta \varphi(t)^\xi l_R(t)^{1-\xi} L(t),
	\label{rcwan}
\end{equation}
\begin{equation}\label{rcsim}
	c(t) = \left(\frac{\psi}{1-\beta}\right)^{1-\frac{1}{\beta}} \beta N(t) l_E(t),
\end{equation}
\begin{equation}\label{rclabor}
	l_R(t) + l_E(t) =1.
\end{equation}
(\ref{rcwan}) is the innovation possibility frontier, (\ref{rcsim}) is the simplified resource constraint, and (\ref{rclabor}) requires labor market clearing. In Online Appendix \ref{proofgrowsp}, we derive the following proposition:
\begin{proposition}\label{growsp}
	With $\gamma>1$, BGP growth rates in the social planner's problem are the same as those in the decentralized economy.
\end{proposition}

Despite the lower net aggregate output relative to the planner's solution, the decentralized data economy grows at the same rate. The growth rates are determined by the final goods production and the usage of data. The former is the same under the decentralized and the planner's solutions. The latter exhibits a gap due to the markup created by the data price mechanism in the decentralized economy (as opposed to the planner's directly setting data usage), which causes the decentralized equilibrium to be less socially efficient. But, the gap is a constant and does not manifest in growth rates. 

Reminiscent of \citet{Jones1995}, growth rates alone cannot fully characterize the performance of an economy.  We thus also examine the labor share in R\&D and derive in Online Appendix \ref{proofSPlsh}:

\begin{proposition}\label{SPlsh}
	In the social planner's problem, the share of labor allocated in the R\&D sector $s_{S}$ is constant in BGP, which is determined by
	\begin{equation}
		s_{S} = \frac{1}{1+\Theta_{S}}, \quad \textit{where} \quad \Theta_{S} = \frac{(\sigma-\xi)n+\xi \rho }{\xi(1-\xi)g^*_S} - \frac{ (\sigma-\xi)(1-\zeta)}{\xi (1-\xi)}.
		\label{lRSP}
	\end{equation}
\end{proposition}

Without monopolistic production, the result differs from that in the decentralized economy. To ensure $s_{S}\in[0,1]$, i.e., $\Theta_{S}\geq 0$, we also note that the BGP growth rate under the planner's solution cannot be too high:
\begin{equation}\label{rangesp}
	0< g^*_S \leq \frac{1}{1-\zeta} \left( n + \frac{\xi}{\sigma-\xi} \rho \right).
\end{equation}
This upper limit consists of three components: knowledge accumulation, population growth, and a data-related term. Intuitively, a social planner does not want the growth of the economy to be too fast since higher growth rates require more data usage, which can create excessive data leakage or privacy violations in expectation. When $\gamma>1$, $g_{S}^{*}$ is given by (\ref{ggrowfinal}) and (\ref{rangesp}) always holds. But when $\gamma<1$, it is possible that the upper limit binds and the planner's solution features slower growth. 

\subsection{Misallocation and Data Overuse in a Decentralized Economy}
\label{compdcsp}

\begin{figure}[h]
	\begin{minipage}[t]{0.5\linewidth}
		\centering
		\includegraphics[width=\textwidth]{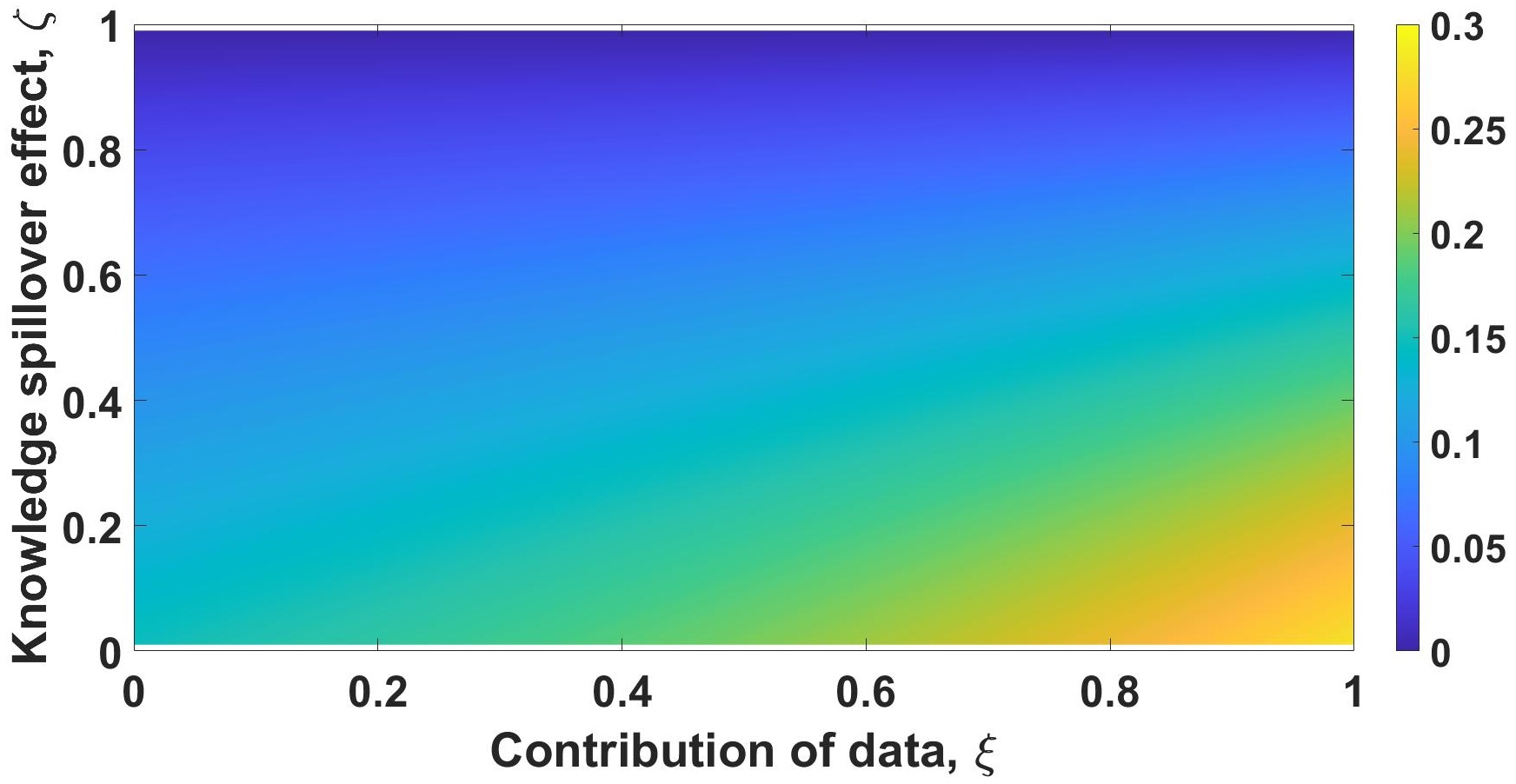}
		\par (a) $\sigma=1.5$
	\end{minipage}
	\begin{minipage}[t]{0.5\linewidth}
		\centering
		\includegraphics[width=\textwidth]{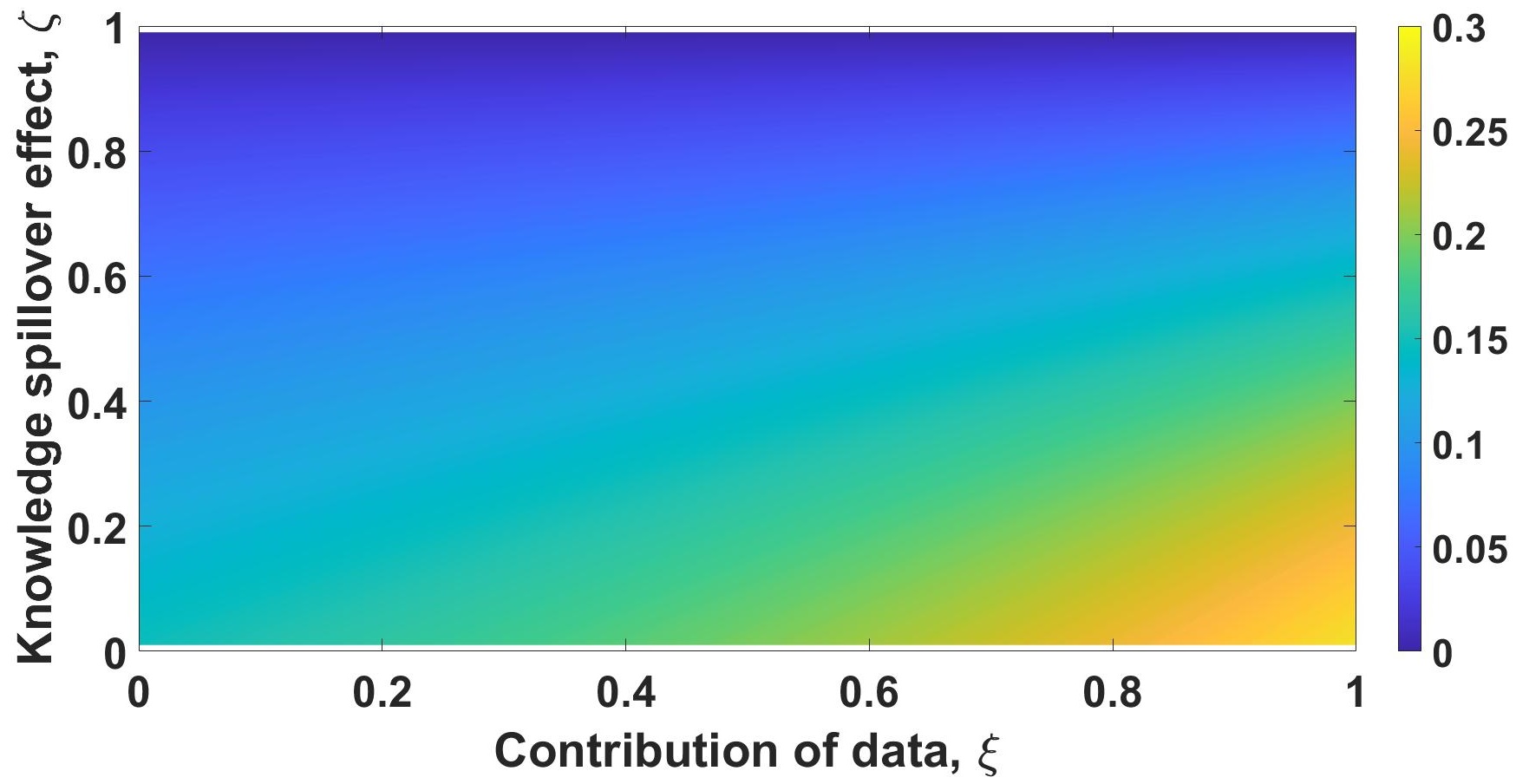}
		\par (b) $\sigma=2.5$
	\end{minipage}
	\begin{minipage}[t]{0.5\linewidth}
		\centering
		\includegraphics[width=\textwidth]{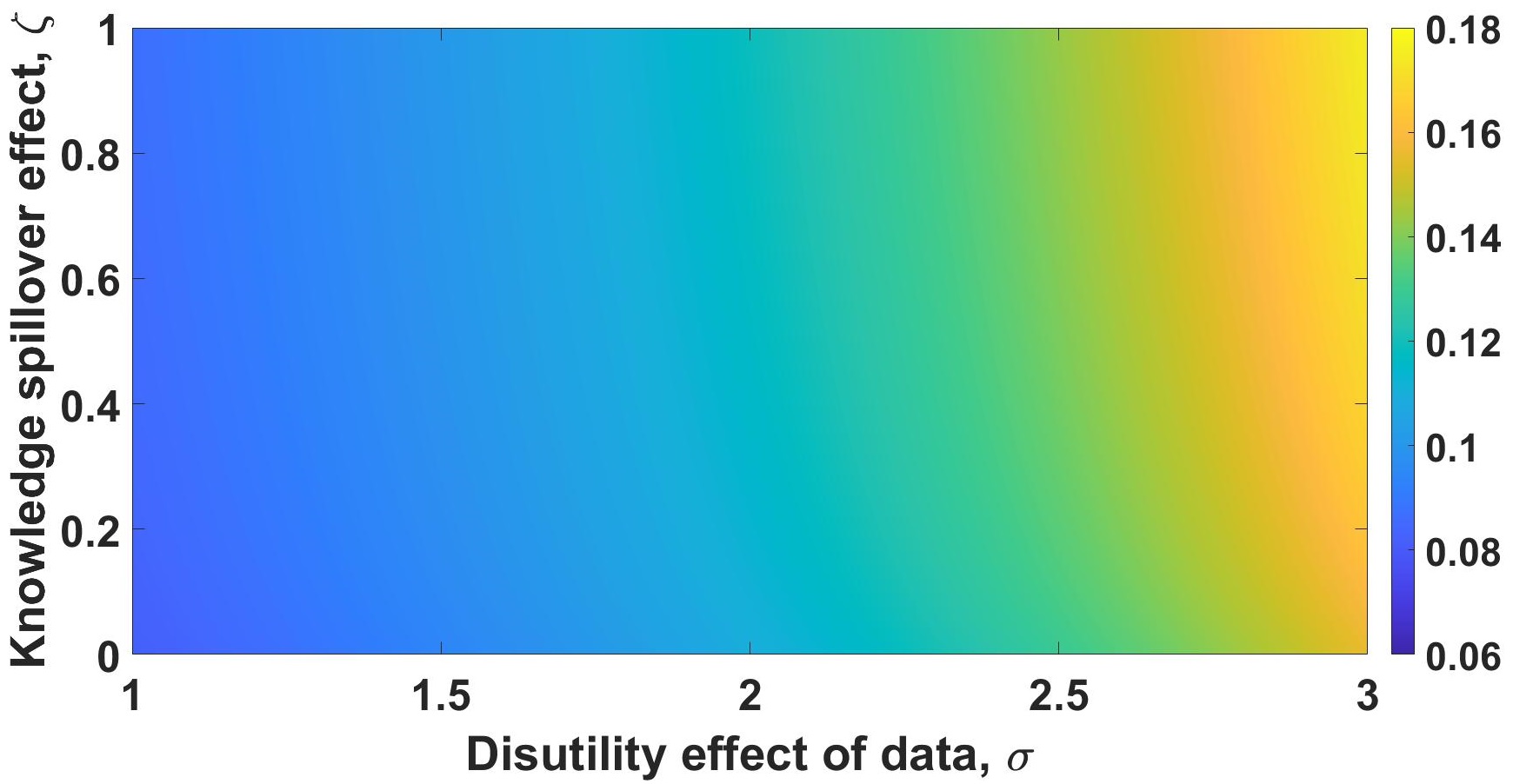}
		\par (c) $\xi=0.5$
	\end{minipage}
	\begin{minipage}[t]{0.5\linewidth}
		\centering
		\includegraphics[width=\textwidth]{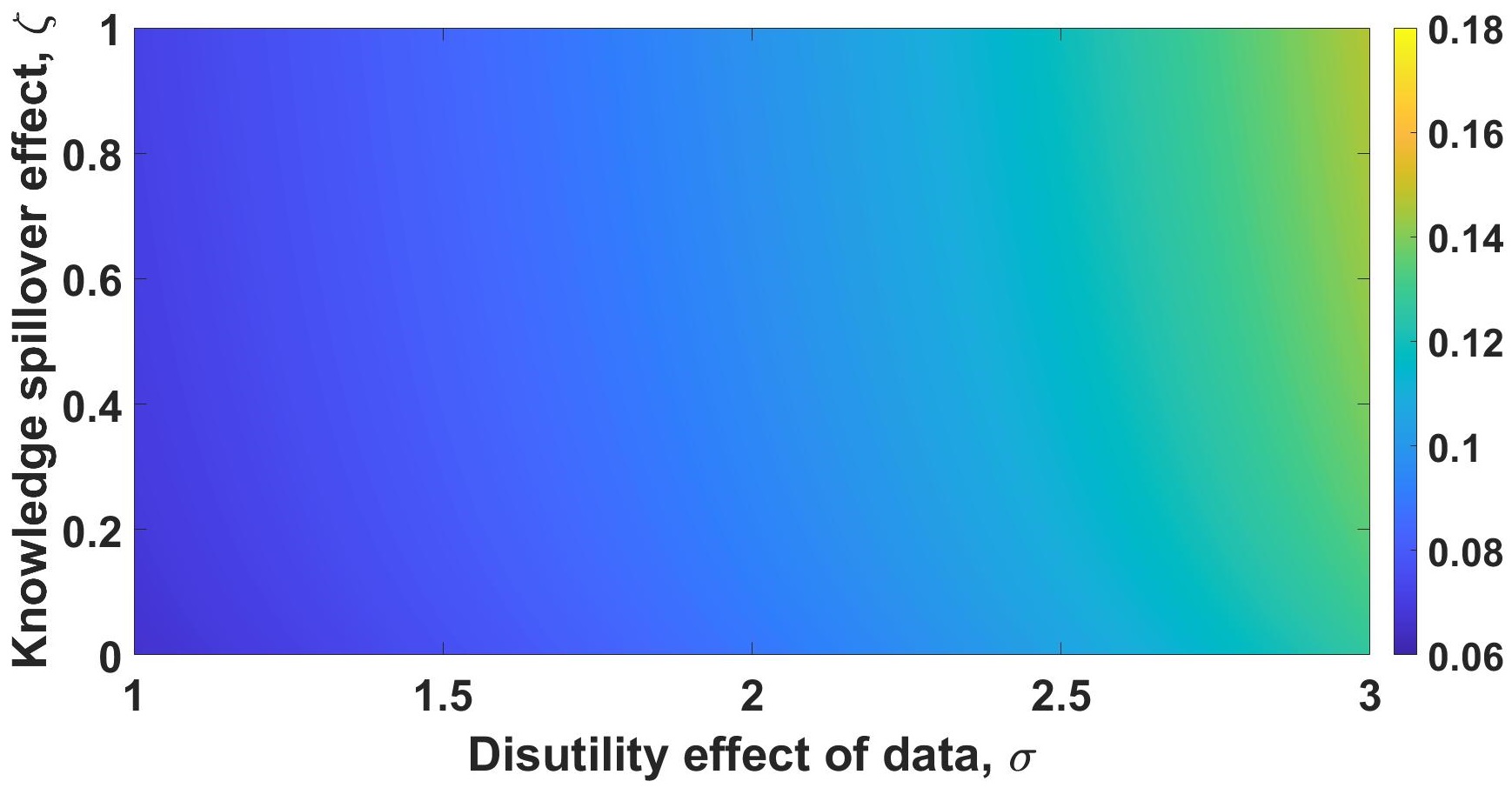}
		\par (d) $\xi=0.6$
	\end{minipage}
	\begin{center}
		\caption{Difference in labor employed in the R\&D sector between the two cases}
		\label{diffSPDC}
	\end{center}
	\noindent\small \textit{Notes:} The figure shows the difference of labor allocated in R\&D sector between decentralized economy and the social planner's problem. Light color represents larger differences (greater misallocation) and dark represents smaller ones. Other parameters are set as $n=0.02$, $\beta=2/3$, $\gamma=2.5$, $\rho=0.03$, which are standard values used in existing literature. $\xi\in [0,1]$, $\sigma>1$ ensures the convexity of disutility term in consumer's utility function and $\zeta \in [0,1]$ ensures the existence of BGP.
\end{figure}

Endogenous labor allocations between production and R\&D sectors influence other variables in equilibrium. In particular, with $n$, $\beta$, $\gamma$ and $\rho$ taking on standard values from the existing literature, the labor allocation in the R\&D sector in the social planner's problem is always larger than that in the decentralized economy. Figure \ref{diffSPDC} plots this difference as we vary the influence of data in innovation possibility frontier ($\xi$), knowledge accumulation through innovation possibility frontier ($\zeta$), and the severity of privacy concerns in the consumers' utility function ($\sigma$). The larger $\sigma$ is, the more the failure to internalize privacy concerns creates misallocation; the overuse of data and underallocation of labor in R\&D are especially severe when data are important for innovation (large $\xi$) and knowledge accumulation is slow (small $\zeta$).

Our model thus reveals a new source of inefficiency in the decentralized economy, even though the growth rates are the same as those in the planner's solution: undersupply of labor and overuse of data in the R\&D sector. This finding contrasts and complements recent studies such as \citet{Jones2020}, which predicts an underutilization of data due to its nonrival nature. In our setting, data are overused even when owned by consumers who directly factor in the disutility from data leakage and/or abuse. 

Similar to \citet{Jones1995}, the underallocation of labor in the R\&D sector comes from monopolistic markups in the production of intermediate goods. The final good producer employs more labor to compensate for the lower production and usage of intermediate goods, which in turn crowds out labor employed in R\&D. The crowding out of labor in R\&D is exacerbated because intermediate producers are less reliant on labor, once data enter the evolution of the innovation possibility frontier \citep[unlike][in which labor is the only input for the innovation possibility frontier]{Jones1995}. To maintain the same growth rate as in the planner's solution, intermediate producers have to compensate the underemployment of labor in R\&D by more aggressive utilization of data. In other words, this crowding-out effect of labor crowds in data usage in the R\&D sector to a socially excessive level. Given the parameter set we have, data usage is four to five times higher in the decentralized economy than in the social planner's problem.

\subsection{Data Generation   and Regulatory Policies} 
\label{regulatoryPolicy}

Different from R\&D-specific labor, the production of data is endogenized by consumption, which itself is also endogenous. More data supplied add to innovations on variety, which in turn spurs consumption (e.g., in BGP), which then further relaxes the data generating constraint. Such a feedback is absent when it comes to labor or capital in that consumption is typically decoupled from the exogenous evolution of population size or the growth in potential capital to be allocated. 
 
As such, privacy regulations reducing $s$ in (\ref{dataBound}) may affect the production and thus the usage of data.\footnote{Both the GDPR and the CCPA give individuals certain rights over the collection and usage of their personal information, but they differ in multiple aspects \citep[e.g.][]{CCPA2020}. For example, California's being a much larger economy implies more severe penalties than that of the GDPR, which maps to a tighter constraint (\ref{dataBound}).} However, reducing $s$ would reduce growth rate and would not lead to welfare improvement without sacrificing growth, if it improves welfare at all. A full analytical characterization is not tractable, not to mention that such an intervention entails intergenerational transfers in practice which can be controversial. We therefore restrict our discussion in this section to policy interventions that preserve growth rates in the planner's BGP solution.

We find that levying a tax on the usage of data alters the transitional dynamics but is ineffective in bringing the equilibrium allocations in decentralized economy closer to the social planner's solution because as discussed in Online Appendix \ref{proofdatatax}, it does not solve the underemployment of labor in R\&D but only slows down the economy before it eventually returns to the original BGP path. However, subsidizing labor wage rate in the R\&D sector or subsidizing intermediate producers in terms of profit proves to be effective. Appendices \ref{proofsubsidy} and \ref{prooftax} derive these optimal subsidies.

The intuition is that labor wage is pinned down by both the R\&D and production sectors, and a subsidy directly affecting the wage level alters the labor share. Specifically, because labor allocations are derived from equalizing wages in the R\&D sector and production sector, a government can apply taxes or subsidies to adjust the prices of factors to alleviate the overuse of data in intermediate producers' R\&D. For example, a subsidy of rate $0<\tau(t)<1$ to the R\&D sector for employing labor would modify (\ref{freeentrylab}) into
\begin{equation}\nonumber
\eta (1-\xi)N(t)^\zeta \varphi(t)^\xi l_R(t)^{-\xi} V(t) = \tau(t)w(t),
\end{equation}
while (\ref{wage}) remains unchanged. Then, one can derive the desirable $\tau(t)$ by equating the labor share to that in the social planner's problem.\footnote{In Online Appendix \ref{proofsubsidy}, we show that this subsidy rate should be a constant. Because only the changing rate of variables may influence growth rates (as seen in Online Appendix \ref{proofdcgrow}), applying a fixed subsidy rate does not introduce further distortions into the model.} 
To avoid overuses of data and potential privacy violations, intermediate producers should be incentivized to employ more labor for innovation with subsidies on wages in the R\&D sector, which would lower data usage.

In contrast, data price is pinned down by intermediate producers and consumers. The consumers' joint decision on consumption and data provision implies that they care about the growth rate of data provision, not the level, as seen in (5). In a sense, a direct tax on data purchase is decoupled from the data provision and because of that, would not alter the equilibrium labor share.\footnote{A non-constant tax rate would not work either because it alters the BGP growth rate in the decentralized economy which is originally the same as that in the social planner's solution.} Consequently, the underemployment in the R\&D and data overuse persist. 

Our findings have important policy implications because the current debate has centered on privacy regulations, which decisively affect the data economy's growth. We demonstrate how R\&D labor relates to the increasing use of data and how labor market policies can effectively reduce excessive data overuse to improve the social welfare. 

\subsection{Historical Data and Dynamic Data Nonrivalry}
\label{sec:nonrivalry}

One unique and important property of data we highlight thus far is dynamic nonrivalry. Beyond its implicit manifestation through the evolution of intermediate good varieties in the baseline specification, we now extend our discussion of dynamic nonrivalry by explicitly modeling the trading of historical data and the associated creative destruction. Specifically, instead of one potential intermediate producer using data for research (and entry) with full depreciation, we now assume:
\begin{itemize}
	\item Data generated at time $t$ can also be used and traded in the following $M>0$ periods. Historical data depreciate at a rate of $\delta$.
	\item Potential intermediate producers can purchase not only data from current consumers, but also historical data from existing intermediate producers entering in the past $M$ periods. The new and historical data bundles are perfect substitutes. 
	\item The owners of historical data determine the proportion of data sold to newcomers, fully recognizing potential creative destruction.
	\item If historical data are reused by new intermediate producers, disutility of contributing consumers is also cumulative according to the usage.
\end{itemize}

In a sense, data become a stock variable whereas labor is a flow variable in the intermediate good production function. Online Appendix \ref{app:nonrivalry} contains detailed derivation and discussion, which are summarized here:
\begin{proposition}\label{nonrivalry}
    Dynamic data nonrivalry under creative destruction as specified does not change the BGP growth rates or labor share in the decentralized economy and social planner's problem. However, incumbent intermediate producers are always willing to sell more than the social optimal level of historical data to entrants in the decentralized economy, as long as the negative effect of creative destruction does not come to an extreme level.
\end{proposition}

The implicit dynamic nonrivalry in our baseline model still exists: Data are transformed into new varieties of intermediate goods, and the existing level of varieties can affect the R\&D process in future. But, the proposition demonstrates that the key findings are robust to explicitly modeling data nonrivalry and the direct use of historical data. Therefore, the differences between our findings and \citet{Jones2020}'s are driven by the dynamic nature of data usage we model, not by the simple consideration of nonrivalry. Moreover, the proposition has implications for regulating data resale among intermediate producers, an interesting topic for future studies.

\subsection{Data Ownership}
\label{ownershipSection}

So far, we have allowed consumers to own data and the firms (intermediate producers) to acquire data by paying the consumers. We now consider cases with positive data processing cost $\theta>0$ under both consumer ownership and firm ownership (detail derivation in Online Appendix \ref{appendixOwnership}). $\theta>0$ would not affect much of the equilibrium outcomes when consumers own data as long as  the knowledge spillover effect is moderate, but consumers' EIS are reasonably moderate and the data processing cost is sufficiently convex (which holds true due to e.g., curse of dimensionality). The main new insight from the analysis is the positive BGP growth rate of data provision when firms own data. In other words, data usage under BGP could be increasing when firms own data, which is different from the decreasing trend when consumers own data. 

Intuitively, because firms no longer need to pay consumers for data usage, they no longer bear the disutility of potential privacy violation. Although they still pay a mechanical ``data processing cost'', it does not push the BGP growth rate of data provision to negative as we have seen in the case in which consumers own data. That said, under standard parameter values, e.g., $\xi=0.5$, $\zeta=0.85$, and $\phi>1$, the BGP growth rate of other variables, $g^*$, is still larger than $g^*_\varphi$. This is consistent with our baseline model: Data usage becomes trivial in BGP compared with consumption as time passes.

Moreover, Sections \ref{sec:nonrivalry} and \ref{ownershipSection} demonstrate that our finding on data overuse is robust to nonrivalry and ownership variation. This result differs from and complements \citet{Jones2020} which emphasizes the underutilization of data due to its cross-sectional nonrivalry and the importance of consumer data ownership. Because data add to innovation over time and have persist benefits in our model (instead of only entering contemporary production as in \citet{Jones2020}), firms are more willing to utilize and purchase data if consumers own data.

\section{Transitional Dynamics: A Numerical Analysis}
\label{numana}

The transitional dynamics of the data economy onto a BGP also reveal interesting patterns, as we demonstrate numerically with the baseline model setup. Even without regulations restricting the use of data, a decline in data usage could occur as the economy grows, which mitigates the concern of data privacy. At the same time, for data economies with minimal initial growth, the process of data generation dictates that they may be trapped in low-growth regimes for a long time without any intervention to improve digital infrastructure or relax privacy regulation. 

\subsection{Methodology and Calibration}

Similar to \citet{Jones2016}, we focus on the planner's solution in this numerical exercise mainly for tractability. This would be the situation if a benevolent government implements policies, balancing privacy protection and growth, such as the subsidy scheme we mentioned earlier. Instead of a formal calibration designed to replicate any country's data, this analysis is best viewed as an illustration of the basic transitional dynamics that are possible in our theoretical
framework.

We derive in Online Appendix \ref{diffsys} a system of differential equations that describe the dynamics of the planner's economy. They consist of a constraint on data generation (\ref{dataBound}) and three state-like variables whose interpretations and steady state values are shown in Table \ref{trandymean}. Other variables can be derived from them. We solve the system of differential equations using ``reverse shooting" \citep{Judd1998}\footnote{We start from the steady state and run the system backward according to the three differential equations. We first find the values of the three state-like variables that minimize the distance between their growth rates and zero, since they all converge to nonzero constants in the steady state. For other parameters, we choose the standard values in the existing literature or plausible values if they are not discussed in extant studies ($\xi$, $\sigma$, and $\zeta$). We set $\xi=0.5$ to indicate that data and labor contribute equally in creating new varieties; we discuss various choices of $\sigma$ in Online Appendix \ref{app:addres}; we set $\zeta$ to be a value less than one following \citet{Jones1995}.} Table \ref{trandypara} provides a summary of parametrization.

\begin{table}[h]
	\centering
	\caption{State-like variables for studying transitional dynamics}
	\resizebox{\textwidth}{!}{
	\begin{tabular}{cll}
		\toprule
		Variable & Meaning & Steady state value \\
		\midrule
		$g_N(t)$ & Growth rate of variety of intermediate goods & (\ref{ggrowfinal}) \\
		$g_\mu(t)$ & Growth rate of shadow price corresponding to technology change & $g_\mu^*=\frac{\sigma(1-\zeta)-\xi}{\xi}g_N^*-\frac{\sigma}{\xi}n$ \\
		$l_E(t)$ & Ratio of labor employed in production sector & (\ref{lRSP}) \\
		\bottomrule
	\end{tabular}}
	\label{trandymean}
\end{table}

\begin{table}[h]
    \centering
	\caption{Parameters for studying transitional dynamics}
	\resizebox{\textwidth}{!}{
	\begin{tabular}{clcc}
		\toprule
		Variable & Meaning & Value & Source \\
		\midrule
		$\beta$ & Contribution of labor in final good production function & $2/3$ & Standard \\
		$\gamma$ & Reciprocal of Elasticity of Intertemporal Substitution & 2.5 & Standard \\
		$\rho$ & Subjective discount factor & 0.03 & Standard \\
		$\xi$ & Contribution of data in innovation possibility frontier & 0.5 & Discretionary \\
		$\zeta$ & Knowledge accumulation through innovation possibility frontier & 0.85 & Discretionary \\
		$\sigma$ & Severity of consumers' privacy concern & 1.5 & Discretionary \\
		$n$ & Population growth rate & 0.02 & Standard \\
		$\eta$ & Efficiency term in innovation possibility frontier & 1 & Standard \\
		\bottomrule
	\end{tabular}}
	\label{trandypara}
\end{table}

\subsection{Results and Discussions}

\begin{figure}[h]
	\begin{minipage}[h]{0.5\linewidth}
		\centering
		\includegraphics[width=\textwidth]{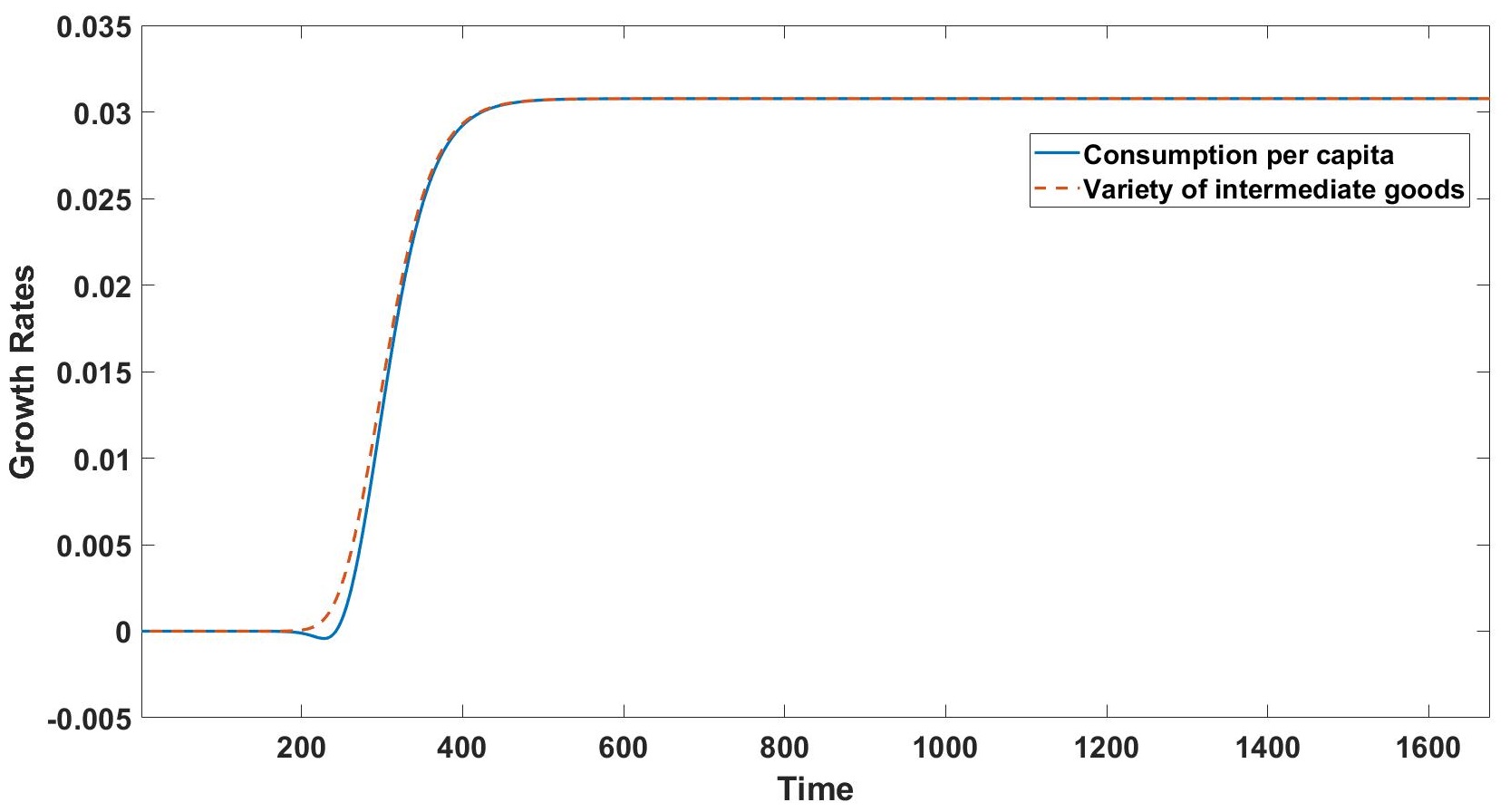}
		\par (a) Growth rates of consumption and intermediate goods varieties (without constraint)
	\end{minipage}
	\begin{minipage}[h]{0.5\linewidth}
		\centering
		\includegraphics[width=\textwidth]{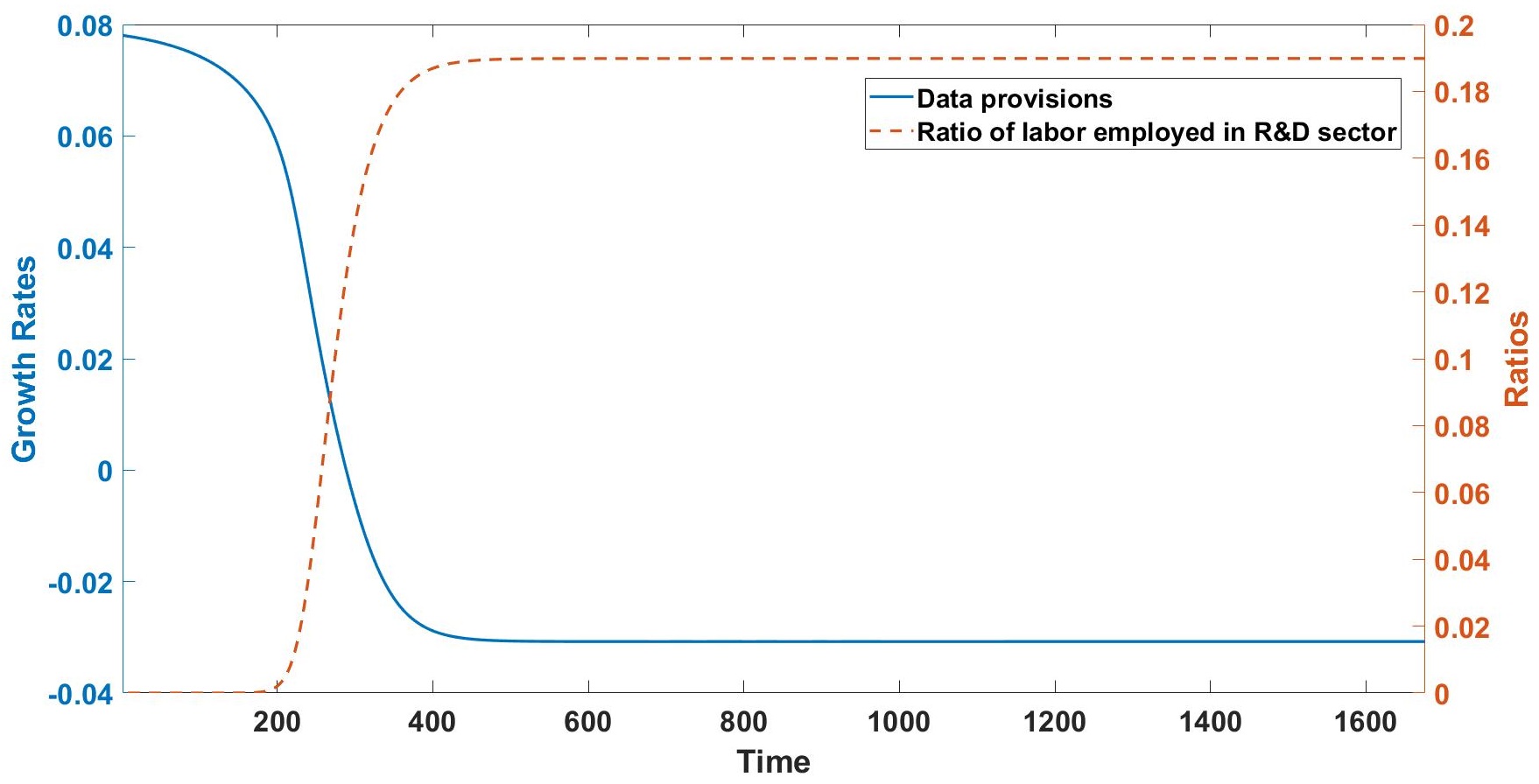}
		\par (b) Labor allocation and growth rate of data provision (without constraint)
	\end{minipage}
	\begin{minipage}[h]{0.5\linewidth}
		\centering
		\includegraphics[width=\textwidth]{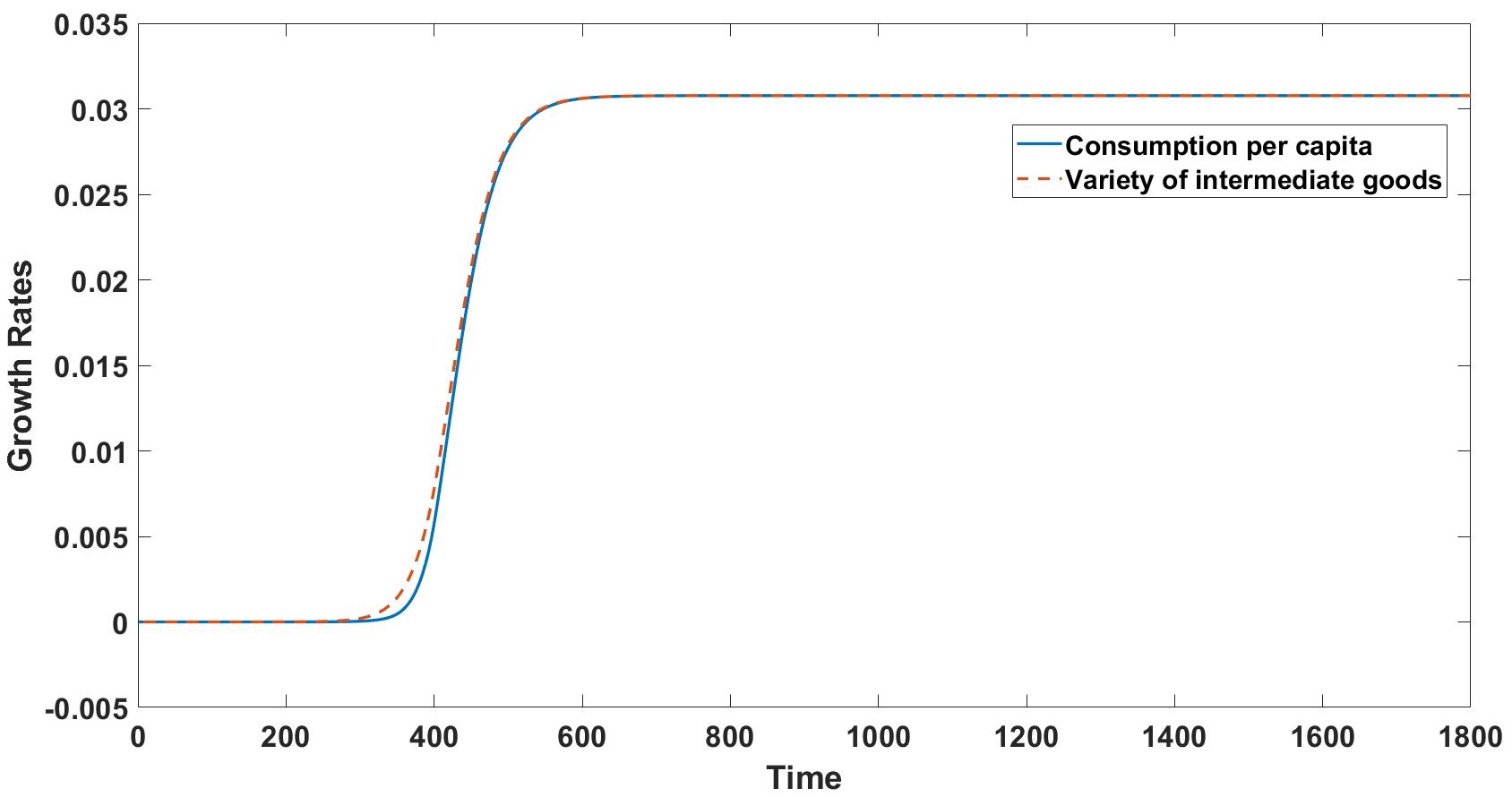}
		\par (c) Growth rates of consumption and intermediate goods varieties (with constraint)
	\end{minipage}
	\begin{minipage}[h]{0.5\linewidth}
		\centering
		\includegraphics[width=\textwidth]{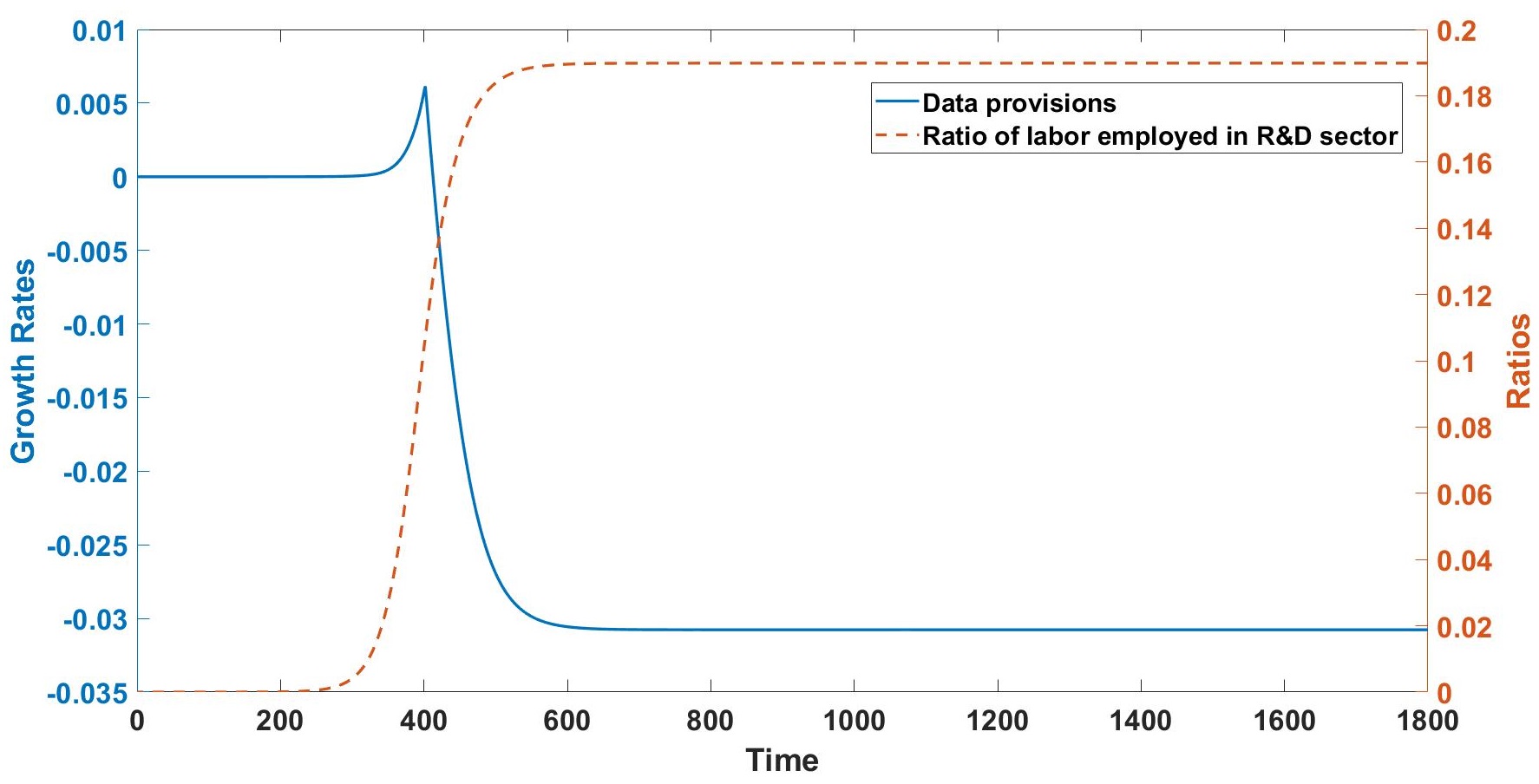}
		\par (d) Labor allocation and growth rate of data provision (with constraint)
	\end{minipage}
	\begin{center}
		\caption{Key variables along transitional path when $\sigma=1.5$}
		\label{trans}
	\end{center}
	\noindent \small \textit{Notes:} These figures show the transitional dynamics of the social planner's problem when $\sigma=1.5$, with and without the constraint of data provision. The economies undergo long but relatively steady states before finally reaching BGP at the end points.
\end{figure}

We can simulate the paths of the three state-like variables before reaching equilibrium for different values of $\sigma$, which stands for the extent of disutility caused by privacy leakage. We can then derive the paths of other variables like $c(t)$ and $\varphi(t)$. Figure \ref{trans} provides an illustration of the transitional dynamics for $\sigma=1.5$ with and without the data provision constraint here. Other cases exhibit similar patterns, which we discuss further in Online Appendix \ref{app:addres}.

Several robust patterns emerge. No matter where an economy starts before BGP, the growth rate of consumption and variety both move to the steady state levels, and a transition can happen relatively quickly once the economy reaches nontrivial growth (sufficiently away from zero), as (a) and (c) show. Moreover, as shown in (b) and (d), the growth rate of data provision decreases from positive to negative in BGP. For an economy starting from low growth, the provision of data increases rapidly for the accumulation of varieties of intermediate goods, which contributes to the final good production and growth in consumption. Along this transitional path, labor moves from production sector to R\&D sector, which reflects how labor is used to compensate for the decreasing provision of data. 

Notice that in Panel (a), different from the growth rate of variety, the growth rate of consumption undergoes some periods of negative growth before increasing to the positive growth rate without constraints on data provision. The temporary pain is common in the growth literature \citep[e.g.][]{Brock2010} and indicates that at the beginning of the adjustment to high BGP growth, labor moves out from the production sector, which causes the decrease of output and consumption. But data are unique because economic activities constrain their supply, as is clear from (d) in which data provision is binding from time 0 to 400. The temporary pain is absent because moving labor away from production is not as costly as before, due to declines in data provision (which lead to slower growth rates). 

The transitional dynamics is also broadly consistent with the empirical patterns in \citet{Abis2020} that the labor income share in knowledge work is decreasing. Formally, labor income share is decreasing in $l_R(t)$ in our model:
\begin{equation}\nonumber
    \text{Labor Income Share} = \frac{w(t) L(t)}{p_\varphi(t) \varphi(t) L(t)} = \frac{\eta (1-\xi) N(t)^{\zeta} \varphi(t)^\xi l_R(t)^{1-\xi} V(t) L(t)}{\eta \xi N(t)^{\zeta} \varphi(t)^\xi l_R(t)^{1-\xi} V(t) L(t)} \frac{1}{l_R(t)}= \frac{1-\xi}{\xi} \frac{1}{l_R(t)}.
\end{equation}
With similar extent of labor allocation in intermediate and final goods in production sectors under the decentralized economy, our numerical simulations reveal this decreasing trend of labor income share as $l_R(t)$ increases in transitional states.

\subsection{Implications for Growth Trap and Privacy Concerns}

Comparing the cases with and without the constraint of data provision, we find that constraint (\ref{dataBound}) binds for an economy with low initial growth rate. The per capita data contribution is also much lower than the case without the constraint. Importantly, for an economy starting with a growth rate close to zero, it takes almost 200 additional time periods (years) for it to reach BGP. Even after reaching BGP growth, the economy's output could be significantly lower due to the delay of accelerated growth. The intuition is straightforward: Initial low growth limits data generation, which negatively feeds back to growth because data constitute an input factor for innovations in intermediate good variety. 

An intervention to boost the initial growth rate is crucial for escaping such a growth trap. Meanwhile, we should notice that the social planner problem is at a national level. Thus, intervention to relax the data provision constraint (\ref{dataBound}) here is more at an international level, e.g., actions taken by the World Bank or IMF or fellow countries in European Union to improve digital infrastructure and data collection/storage efficiency or share expertise, which relax (\ref{dataBound}). Also, if the central planner is originally too budget/cash constrained to improve data infrastructure, foreign subsidies might help relax the binding constraint temporarily.

Global waves of technological innovations such as the Internet and AI take place when countries are having different population growth and different environments for technology transfer and spillover. This can lead to vastly different transitional dynamics. The tightness of data constraint $s$ does not affect the BGP growth rates, but affects transitional dynamics. For example, the United States and Europe are in the same stage of development when data economy emerges, but the difference in privacy protection (different $s$) makes them tread different paths, reaching BGP at different times.  

\begin{figure}[h!]
	\begin{center}
	\includegraphics[width=\textwidth]{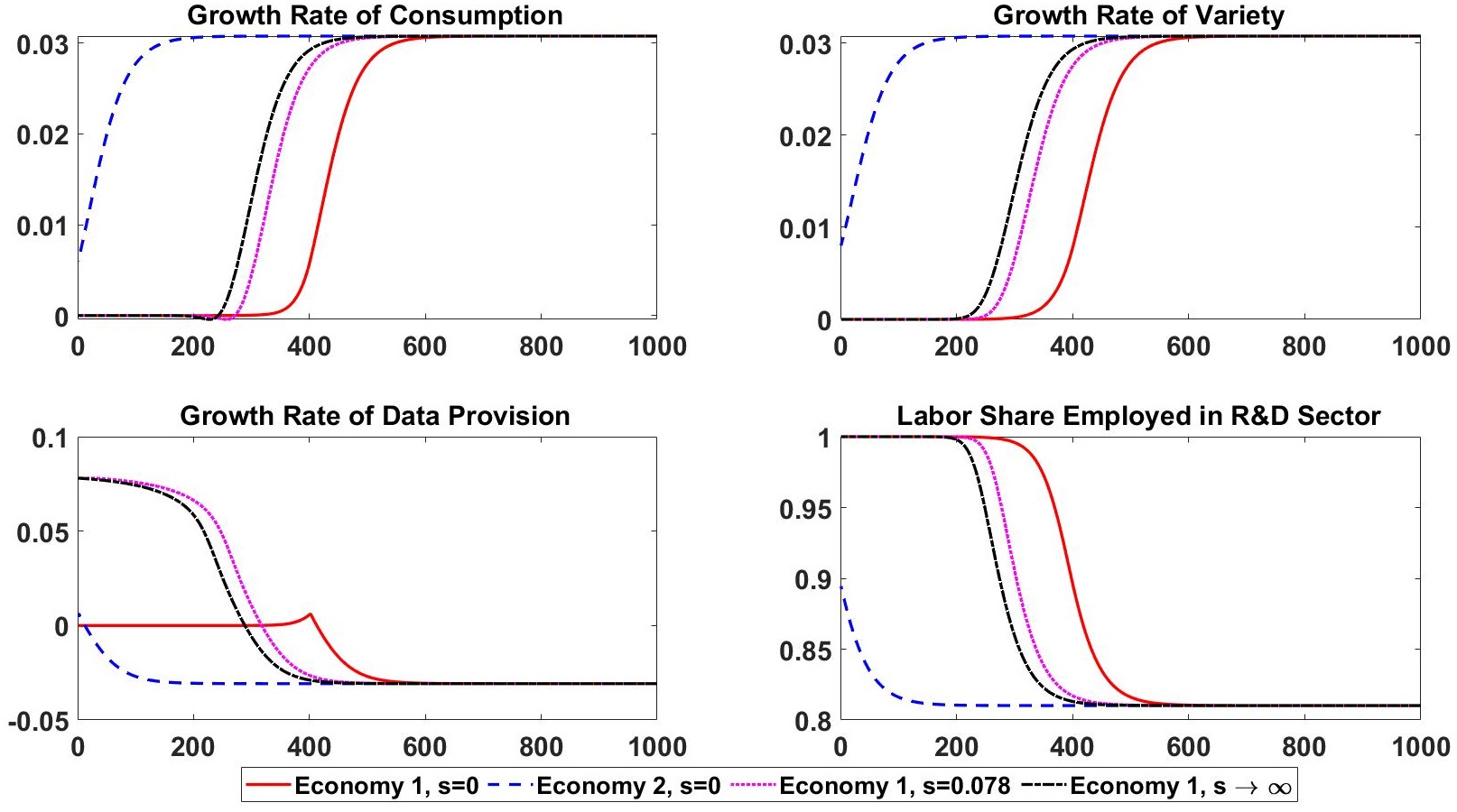}
	\caption{Different pre-BGP Paths Given Different Starting Points and Different Privacy Policies}
	\label{privacytrans}
	\end{center}
	\noindent\small \textit{Notes: }These panels show the transitional dynamics of two economies with the same parameters but different starting points. The full line and the dash line show the paths of the two economies with $s=0$, the dotted line shows the path of Economy 1 when its privacy policy becomes loose ($s=0.078$), and the dash dotted line shows the path of Economy 1 when privacy regulation is completely relaxed ($s\to \infty$).
\end{figure}

First, as the economy just starts to grow, data can significantly facilitate the transition to BGP, as illustrated in Figure \ref{privacytrans}: Economies 1 and 2 have the same parameters but start at different stages in their transitions to BGP, with Economy 1 trapped in near zero growth for a prolonged period of time. An intervention to enhance data facilities and digital infrastructure or to relax privacy regulation (increasing $s$) can help escape the growth trap earlier. However, even when such interventions fully relax (\ref{dataBound}), i.e., $s\to \infty$, they compensate the lagging behind Economy 2 only partially.

In the long run, because the knowledge accumulation parameter $\zeta$ is set to be smaller than one, it becomes less effective to provide more data for creating new varieties of intermediate goods as the existing varieties accumulate. Thus, the growth rate of data provision decreases after some periods of increase. As data become less productive in the innovation process, the economy substitutes the use of data with more labor to focus on better exploitation of data. Finally, as the economy matures (transition to BGP), the benefit of using data is diminishing but individual's privacy concerns remain, which reduce each consumer's endogenous data contribution. In that regard, privacy issues in the long run may not be as severe as current debates indicate. Instead, regulatory policies could focus on the overuse of data in the R\&D sector, as discussed in Section \ref{regulatoryPolicy}.

\section{Conclusion}
\label{conclsion}

We develop an endogenous growth model against the backdrop of the rise of big data and digital economy. Although a decentralized economy on the balanced growth path grows at the same rate as in the socially optimal allocations, data are inefficiently overused and R\&D is understaffed. Consumers suffer since they are inadequately compensated for potential information leakage and privacy violation. When consumers own data, data privacy concerns become allayed in the long run because the use of data eventually declines. However, less developed economies with low growth at the dawn of the data economy may face a new form of poverty trap that potentially warrants interventions. 

For the first time, we treat data as an input factor besides labor in the process of creating new varieties of intermediate goods, which subsequently fuel the production of final good and long-run growth. We highlight data's endogenous generation as by-products of economic activities, nonrivalry in a dynamic environment, and flexible ownership. For tractability and focus, we necessarily leave out certain aspects of the data economy such as final goods differentiation. Therefore, our findings should be taken as first-order benchmark results rather than foregone conclusions. 

%
%
%






\bibliographystyle{informs2014.bst}
\bibliography{Bigdata.bib}

\ECSwitch


\ECHead{Proofs and Extended Discussions}

\section{Proofs of Lemmas and Propositions}

\subsection{The Growth Rate of Labor Allocations in BGP}
\label{prooflem1}
\begin{lemma}
	In BGP, the ratios of labor employed in production sector and R\&D sector both converge to constants, thus their growth rates are both zero.
\end{lemma}
\proof{Proof of Lemma.}
	From the labor market clearing condition, we have $l_E(t)+l_R(t)=1$, and $0 \le l_E(t),l_R(t) \le 1$. Since we need the growth rates of all endogenous variables to be constant in BGP, if $l_E(t)$ or $l_R(t)$ grows in BGP, it becomes larger than 1 eventually, which contradicts the definitions. Meanwhile, if one of them grows at a negative rate in BGP, from the market clearing condition, the other one should grow positively, which also contradicts with the definition. Thus, we can conclude that, the only possible situation is that the two variables both remain constant in BGP, i.e.,
	\begin{equation}\nonumber
		\frac{\dot l_E(t)}{l_E(t)} = \frac{\dot l_R(t)}{l_R(t)} = 0.
	\end{equation}
	Q.E.D.
\endproof

\bigskip

This is a very general conclusion, which will be used for many times in the following proofs. Thus, we use it directly as given from now on. Notice that we can only use this conclusion in BGP, while in other non-BGP states, the growth rates of these two ratios may not be constant.

\subsection{Proof of Proposition 1: BGP Growth Rates in Decentralized Model}
\label{proofdcgrow}

\proof{Proof of Proposition 1.}
	Combine the free-entry condition of labor (\ref{freeentrylab}) and the wage determining equation from the final good producer side (\ref{wage}) together, and change it into the form of growth rate, we have
	\begin{equation}
		\zeta\frac{\dot N(t)}{N(t)} + \xi\frac{\dot \varphi(t)}{\varphi(t)} + \frac{\dot V(t)}{V(t)}= \frac{\dot N(t)}{N(t)}.
		\label{ECgr11}
	\end{equation}
	Recall that in BGP, $r(t)=r^*$, then the patent value shown in (\ref{paval}) is reduced to
	\begin{align}
	    V(t) =& \int_t^\infty e^{-r^*(s-t)} \pi(s) \mathrm{d}s \nonumber \\
	    =& \int_t^\infty e^{-r^*(s-t)} e^{n(s-t)} \pi(t) \mathrm{d}s \nonumber \\
	    =& \frac{\pi(t)}{r^*-n}, \nonumber
	\end{align}
	where the second equation comes from (\ref{piL}). Thus, the growth rate of $V(t)$ is
	\begin{equation}\label{ECVgrow}
		\frac{\dot V(t)}{V(t)} = \frac{\dot \pi(t)}{\pi(t)} = n.
	\end{equation}
	Plug (\ref{ECVgrow}) into (\ref{ECgr11}), we have
	\begin{equation}\label{ECgrphi}
		(\zeta-1)\frac{\dot N(t)}{N(t)} + \xi \frac{\dot \varphi(t)}{\varphi(t)} +  n = 0.
	\end{equation}
	
	Meanwhile, the per capita output growth rate of the economy $g(t)$ can be represented by the growth rate of final good production or variety (from (\ref{Y})), that is 
	\begin{equation}\nonumber
		g(t) = \frac{\dot y(t)}{y(t)} = \frac{\dot Y(t)}{Y(t)}-n=\frac{\dot N(t)}{N(t)}.
	\end{equation}
	Also, from the innovation possibility frontier (\ref{frontierori}), we have
	\begin{equation}\label{ECYgrow}
		g(t) = \frac{\dot N(t)}{N(t)} =\eta N(t)^{\zeta-1} \varphi(t)^\xi l_R(t)^{1-\xi} L(t).
	\end{equation}
	
	Next, we want to pin down the growth rate of $N(t)$ and $\varphi(t)$. Consider the free-entry condition of data (\ref{freeentryinfo}) and rewrite it into the form of growth rate, and plug in (\ref{ECVgrow}), we have
	\begin{equation}\nonumber
		\frac{\dot p_\varphi(t)}{p_\varphi(t)} = \zeta \frac{\dot N(t)}{N(t)} + (\xi-1)\frac{\dot \varphi(t)}{\varphi(t)} + \frac{\dot V(t)}{V(t)}
		= \zeta\frac{\dot N(t)}{N(t)} + (\xi-1)\frac{\dot \varphi(t)}{\varphi(t)} + n.
	\end{equation}
	Then, from the Euler equation of consumption (\ref{motC}) and the Euler equation of data (\ref{motphi}), we have
	\begin{align}
		&\frac{1}{\gamma}\left(\frac{\dot p_\varphi(t)}{p_\varphi(t)} - (\sigma-1) \frac{\dot \varphi(t)}{\varphi(t)} \right) \nonumber\\
		= &\frac{\zeta}{\gamma}\frac{\dot N(t)}{N(t)} + \frac{\xi-\sigma}{\gamma}\frac{\dot \varphi(t)}{\varphi(t)}  + \frac{1}{\gamma}n
		= \frac{\dot c(t)}{c(t)} = \frac{\dot y(t)}{y(t)} =\frac{\dot N(t)}{N(t)}, \nonumber
	\end{align}
	where the second equation comes from the fact that in BGP, the growth rate of consumption is the same as that of output. Rewrite the above equation, we get the following expression
	\begin{equation}
		\left(\zeta-\gamma\right)\frac{\dot N(t)}{N(t)} +(\xi-\sigma)\frac{\dot \varphi(t)}{\varphi(t)} +  n=0.
		\label{ECgrphi2}
	\end{equation}
	Then, we can derive the growth rate of $\varphi(t)$ and $N(t)$ from (\ref{ECgrphi}) and (\ref{ECgrphi2}), that is
	\begin{equation}\nonumber
		\frac{\dot N(t)}{N(t)} = \left[\frac{\sigma }{(1-\zeta)\sigma - \xi (1-\gamma)}\right] n
	\end{equation}
	and
	\begin{equation}\nonumber
		\frac{\dot \varphi(t)}{\varphi(t)} = -\frac{1}{\xi} n + \frac{1-\zeta}{\xi}\frac{\dot N(t)}{N(t)} = \left[\frac{ 1-\gamma }{(1-\zeta)\sigma-\xi (1-\gamma)}\right] n.
	\end{equation}
	Now, from (\ref{ECYgrow}), we have
	\begin{equation}\nonumber
		g^* = \left[\frac{\sigma }{(1-\zeta)\sigma - \xi (1-\gamma)}\right] n.
	\end{equation}
	Q.E.D.
\endproof

\subsection{Proof of Proposition 2: Labor Allocations in BGP in Decentralized Model}
\label{proofDClsh}

\proof{Proof of Proposition 2.}
	From (\ref{ECYgrow}) we know that in BGP, $N(t)^{\zeta-1} \varphi(t)^\xi L(t)$ is constant among all the periods $t$, thus
	\begin{equation}\nonumber
	    g^* = \eta N(t)^{\zeta-1} \varphi(t)^\xi (l_R^*)^{1-\xi} L(t), \forall t.
	\end{equation}
	Combine (\ref{freeentrylab}) and (\ref{wage}), then plug in (\ref{paval}) and (\ref{piL}), we have
	\begin{equation}\nonumber
		(1-\xi)(1-\beta) g^* ((l_R^*)^{-1}-1) = r^* - n
	\end{equation}
	Meanwhile, from (\ref{motC}) we can also derive $g^*$, then combine it with the above equation, we have
	\begin{align}
		g^* = \frac{1}{\gamma} (r^* - \rho) =& \frac{1}{\gamma} \left[ (1-\xi)(1-\beta) g^* ((l_R^*)^{-1}-1) + n - \rho \right] \nonumber\\
		\Rightarrow \quad l_R^* =& \frac{(1-\xi)(1-\beta)}{\gamma + (1-\xi)(1-\beta)+ \frac{\rho-n}{g^*}}. \nonumber
	\end{align}
	Let
	\begin{equation}\nonumber
		\Theta_{D} = \frac{g^*\gamma +\rho-n}{g^*(1-\xi)(1-\beta)},
	\end{equation}
	where, $g^*$ is shown in (20) in the main text. Then, the ratio of labor employed in R\&D sector is
	\begin{equation}\nonumber
		s_{D} = \frac{1}{1+\Theta_{D}}.
	\end{equation}
	Q.E.D.
\endproof

\subsection{Proof of Proposition 3: BGP Growth Rates in Social Planner's Problem}
\label{proofgrowsp}

\proof{Proof of Proposition 3.}
    We should first solve the static problem before we get to the dynamic problem. Taking other variables as given, the net output $\widetilde Y(t)$ can be maximized by choosing the optimal intermediate good usage $x_S(v,t)$ as
    \begin{equation}\label{xvt}
        x_S(v,t) = \left( \frac{\psi}{1-\beta}\right)^{-\frac{1}{\beta}} L_E(t).
    \end{equation}
    Then, substitute this into the definition of net output, we get the optimal net output which is shown in (\ref{YtS}). After that, the social planner's problem is set from (\ref{prefer2}) to (\ref{rclabor}). To solve this problem, we first set up the current-value Hamiltonian equation as
	\begin{equation}\nonumber
		\mathcal{H} = \frac{c(t)^{1-\gamma}-1}{1-\gamma} - \varphi(t)^\sigma + \lambda(t) \left[ \left(\frac{\psi}{1-\beta}\right)^{1-\frac{1}{\beta}} \beta N(t) l_E(t) - c(t) \right] +\mu(t) \eta N(t)^\zeta \varphi(t)^\xi l_R(t)^{1-\xi} L(t),
	\end{equation}
	where $\lambda(t)$ and $\mu(t)$ are the shadow prices corresponding to the constraints (\ref{rcsim}) and (\ref{rcwan}), respectively. Now, write the markup as
	\begin{equation}\nonumber
		A = \left(\frac{\psi}{1- \beta}\right)^{1- \frac{1}{\beta}},
	\end{equation}
	and derive the necessary conditions with respect to $c(t)$, $\varphi(t)$, $l_E(t)$ and $N(t)$, we then have the following equations:
	\begin{equation}
		\frac{\partial \mathcal{H}}{\partial c(t)} = c(t)^{-\gamma} - \lambda(t) = 0,
		\label{ECFOCcS}
	\end{equation}
	\begin{equation}\label{ECFOCphiS}
		\frac{\partial \mathcal{H}}{\partial \varphi(t)} = - \sigma \varphi(t)^{\sigma-1} + \mu(t) \xi \eta N(t)^\zeta \varphi(t)^{\xi-1} l_R(t)^{1-\xi} L(t) = 0,
	\end{equation}
	\begin{equation}\label{ECFOClES}
		\frac{\partial \mathcal{H}}{\partial l_E(t)} = \lambda(t) \beta A N(t)  - \mu(t) \eta (1-\xi) N(t)^\zeta \varphi(t)^\xi l_R(t)^{-\xi} L(t) = 0
	\end{equation}
	and
	\begin{equation}
		\frac{\partial \mathcal{H}}{\partial N(t)} = \lambda(t) A \beta l_E(t) + \mu(t) \eta \zeta N(t)^{\zeta-1} \varphi(t)^\xi l_R(t)^{1-\xi} L(t) = -\dot \mu(t) + (\rho-n)\mu(t).
		\label{ECFOCNS}
	\end{equation}
	This system can be expressed by the seven equations from (\ref{rcwan}) to (\ref{rclabor}) and from (\ref{ECFOCcS}) to (\ref{ECFOCNS}).
	
	Consider (\ref{ECFOClES}), move the second term to the right hand side, change the equation into the form of growth rate, and substitute the growth rate of $\lambda(t)$ and $\mu(t)$ derived from (\ref{ECFOCcS}) and (\ref{ECFOCphiS}), we have\footnote{Here, we expect that $l_E$ and $l_R$ converge to a constant in BGP, as has been shown in Online Appendix \ref{prooflem1}.}
	\begin{align}
		&\frac{\dot \lambda(t)}{\lambda(t)} + \frac{\dot N(t)}{N(t)} = \frac{\dot \mu(t)}{\mu(t)} + \zeta \frac{\dot N(t)}{N(t)} + \xi \frac{\dot \varphi(t)}{\varphi(t)} + n \nonumber\\
		\Rightarrow \quad &-\gamma \frac{\dot c(t)}{c(t)} + (1-\zeta)\frac{\dot N(t)}{N(t)} - \frac{\dot \mu(t)}{\mu(t)} - \xi \frac{\dot \varphi(t)}{\varphi(t)} -n = 0 \nonumber\\
		\Rightarrow \quad &-\gamma \frac{\dot c(t)}{c(t)} + \frac{\dot N(t)}{N(t)} -\sigma \frac{\dot \varphi(t)}{\varphi(t)} =0. \nonumber
	\end{align}
	Notice that the growth rate of $c(t)$ is what we want to derive, which can be defined as $g^*$, then
	\begin{equation}
		\frac{\dot c(t)}{c(t)} = g^* = \frac{1}{\gamma} \left( \frac{\dot N(t)}{N(t)} - \sigma \frac{\dot \varphi(t)}{\varphi(t)}\right).
		\label{ECggrowth1}
	\end{equation}
	Next, we pin down the growth rate of $N(t)$ and $\varphi(t)$. From (\ref{ECFOClES}), we have
	\begin{equation}\nonumber
		\frac{\lambda(t)}{\mu(t)} = \frac{\eta(1-\xi)}{\beta A} N(t)^{\zeta-1} l_R(t)^{-\xi} \varphi(t)^\xi L(t).
	\end{equation}
	Then, (\ref{ECFOCNS}) can be reformed as
	\begin{align}
		&\frac{\lambda(t)}{\mu(t)} A \beta l_E(t) + \eta \zeta N(t)^{\zeta-1} \varphi(t)^\xi l_R(t)^{1-\xi} L(t) = -\frac{\dot \mu(t)}{\mu(t)} + (\rho-n) \nonumber\\
		\Rightarrow \quad &\left[\eta(1-\xi) l_E(t) + \eta \zeta l_R(t) \right] N(t)^{\zeta-1} \varphi(t)^\xi l_R(t)^{-\xi} L(t) = \zeta \frac{\dot N(t)}{N(t)} + (\xi-\sigma) \frac{\dot \varphi(t)}{\varphi(t)} + \rho.
		\label{ECproof1}
	\end{align}
	Since the right hand side is constant in BGP, the left hand side should also be constant. Thus, we have
	\begin{equation}
		(\zeta-1)\frac{\dot N(t)}{N(t)} + \xi \frac{\dot \varphi(t)}{\varphi(t)} + n= 0.
		\label{ECNphi1}
	\end{equation}
	Meanwhile, we can derive the production function (in per capita form) in the social planner's problem as
	\begin{equation}\nonumber
		y_S(t) = A N(t) l_E(t)
	\end{equation}
	Thus, the growth rate of the economy is
	\begin{equation}\label{ECgrowthSP}
		g^* = \frac{\dot N(t)}{N(t)}.
	\end{equation}
	Combine (\ref{ECgrowthSP}) with (\ref{ECggrowth1}) and also (\ref{ECNphi1}), we get
	\begin{equation}\label{ECggrowfin}
		g^*_S = \left[\frac{\sigma}{(1-\zeta)\sigma - \xi (1-\gamma)} \right]n,
	\end{equation}
	which is the same as that in the decentralized model. The growth rates of other related variables can be derived similarly. Q.E.D.
\endproof
\bigskip
We can come back to the following illustrations after reading Appendices \ref{proofSPlsh} and \ref{proofcond}. We should notice that although the result of growth rate here equals to that in the decentralized model, in Online Appendix \ref{proofSPlsh}, the result of labor share in R\&D sector requires the growth rate to be within a certain range. From (\ref{ECthetaSP}) in Online Appendix \ref{proofSPlsh} we know that $g^*_S$ should satisfy the following condition:
\begin{equation}\nonumber
	0< g^*_S < \frac{1}{1-\zeta} \left( n + \frac{\xi}{\sigma-\xi} \rho \right).
\end{equation}
Plug in the expression of $g^*_S$, we have
\begin{equation}\nonumber
	0<\zeta<1-(1-\gamma) \left[ \frac{n(\sigma-\xi)}{\rho \sigma} + \frac{\xi}{\sigma} \right].
\end{equation}
When $\gamma>1$, from (\ref{ECTVC}) in Online Appendix \ref{proofcond} we have $0<\zeta<1-(1-\gamma)\frac{\xi}{\sigma}$, which includes the above condition. However, when $0<\gamma<1$, we have $0<\zeta<1-(1-\gamma)\left( \frac{n}{\rho}+\frac{\xi}{\sigma}\right)$, the condition above is not always satisfied. Since in literature, we usually think that $\gamma>1$, the social planner's problem is always meaningful. 

\subsection{Proof of Proposition 4: Labor Allocations in BGP in Social Planner's Problem}
\label{proofSPlsh}

\proof{Proof of Proposition 4.}
	
	Similar to the decentralized economy, in social planner's problem we also have
	\begin{equation}\nonumber
	    g^* = \eta N(t)^{\zeta-1} \varphi(t)^\xi (l_R^*)^{1-\xi} L(t), \forall t.
	\end{equation}
	Combine (\ref{ECproof1}) and (\ref{ECNphi1}), and plug in the above equation, in BGP we have
	\begin{align}
		&\left[(1-\xi) (1-l_R^*) +  \zeta l_R^* \right] g^* (l_R^*)^{-1}  = \zeta \frac{\dot N(t)}{N(t)} + (\xi-\sigma) \frac{\dot \varphi(t)}{\varphi(t)} + \rho \nonumber\\
		\Rightarrow \quad &\left[(1-\xi) (1-l_R^*) +  \zeta l_R^* \right] g^* (l_R^*)^{-1}  = \left[1-\frac{\sigma}{\xi}(1-\zeta)\right] \frac{\dot N(t)}{N(t)} + \left(\frac{\sigma}{\xi}-1\right) n + \rho \nonumber\\
		\Rightarrow \quad &\left[ (1-\xi) (1-l_R^*) + \left(\frac{\sigma}{\xi}-1\right)(1-\zeta) l_R^*\right] g^* (l_R^*)^{-1} =  \left(\frac{\sigma}{\xi}-1\right)n + \rho.\nonumber
	\end{align}
	As a result, we have
	\begin{equation} \nonumber
		l_R^* = \frac{1}{1+\Theta_{S}},
	\end{equation}
	where,
	\begin{equation}\label{ECthetaSP}
		\Theta_{S} = \frac{(\sigma-\xi)n+\xi \rho }{\xi(1-\xi)g^*_S} - \frac{(\sigma-\xi)(1-\zeta)}{\xi (1-\xi)},
	\end{equation}
	and $g^*_S$ is given in (\ref{ECggrowfin}). Q.E.D.
\endproof

\section{Extended Discussions}
\label{proofaccu}

\subsection{The Existence and Uniqueness of BGP}
\label{proofcond}

\begin{proposition}
	In this data-driven innovation model, there exists a unique BGP which is given in Proposition 1, if the following condition is satisfied
	\begin{equation}
		0<\zeta< \begin{cases} 1 - \frac{\xi}{\sigma}(1-\gamma), & \text{if } \gamma >1; \\
			1 - (1-\gamma) \left( \frac{n}{\rho} + \frac{\xi}{\sigma} \right), & \text{if } 0<\gamma<1. \end{cases} \nonumber
		\label{ECcond}
	\end{equation}
	\label{bgpcond}
\end{proposition}
\proof{Proof of Proposition \ref{bgpcond}.}
	To ensure the existence of BGP, the growth rate of the economy should satisfy two conditions. The first one requires $g^*>0$, which makes
	\begin{align}
		(1-\zeta)\sigma - \xi (1-\gamma) >& 0 \nonumber\\
		\Rightarrow \quad \zeta <& 1- \frac{\xi}{\sigma} (1-\gamma).
		\label{ECcond1}
	\end{align}
	The second one requires that the growth rate should satisfy the transversality condition (TVC), i.e.,
	\begin{equation}\label{ECtvc}
		\lim_{t\to\infty} \left[ \exp{\left(-\int_0^t r(s) \,\mathrm{d}s\right)} \int_0^{N(t)} V(v,t) \,\mathrm{d}v \right] = 0.
	\end{equation}
	In BGP, the TVC is simplified into
	\begin{align}
		\lim_{t\to\infty} e^{-r^*t} N(t) V(t) =& 0 \nonumber\\
		\Rightarrow \quad \lim_{t\to\infty} e^{(g^*+n-r^*)t} N(0)V(0) =& 0, \nonumber
	\end{align}
	which leads to the condition $g^*+n-r^*<0$. Meanwhile, from (\ref{motC}) we can know that $r^* = \gamma g^* +n+\rho$, then
	\begin{align}
		g^* + n - \gamma g^* -n -\rho <& 0 \nonumber\\
		\Rightarrow \quad g^* <& \frac{\rho}{1-\gamma} \nonumber\\
		\Rightarrow \quad \zeta<& 1 - (1-\gamma) \left(\frac{n}{\rho} + \frac{\xi}{\sigma}\right).
		\label{ECcond2}
	\end{align}
	Combine (\ref{ECcond1}) and (\ref{ECcond2}), and $\zeta>0$ (positive technology spillover), we have
	\begin{equation}\label{ECTVC}
		0<\zeta< \begin{cases} 1 - \frac{\xi}{\sigma}(1-\gamma), & \text{if } \gamma >1; \\
			1 - (1-\gamma) \left( \frac{n}{\rho} + \frac{\xi}{\sigma} \right), & \text{if } 0<\gamma<1, \end{cases}
	\end{equation}
	which is what we require the spill-over effect $\zeta$ to be. Q.E.D.
\endproof

\subsection{Data Accumulation}
\label{app:accumulation}

In our baseline model, data are set to be fully depreciated in every period. One might think that in reality, data involved in the innovation possibility frontier can accumulate. This is always the case, since time-series data usually contain useful information. To model this formally, we let the innovation possibility frontier be:
\begin{equation}\nonumber
	\dot N(t) = \eta N(t)^\zeta \Phi(t)^\xi l_R(t)^{1-\xi} L(t),
\end{equation}
where $\Phi(t)$ denotes the total quantity of data accumulated in R\&D activities at time $t$. Its law of motion is then:
\begin{equation}\label{ECmotPhi}
	\dot \Phi(t) = \varphi(t) - \kappa \Phi(t),
\end{equation}
where, $\varphi(t)$ is the same as that in the previous sections, and $\kappa$ here represents the depreciation rate of data, i.e., the rate that data become out of date every period. Other settings are still the same as the baseline model. By introducing this adjustment into the baseline model, we have the following proposition.
\begin{proposition}
	In the case that data involved in the innovation possibility frontier can accumulate, the growth of the economy in BGP remains similar as that in the case without data accumulation.
	\label{accu}
\end{proposition}
\proof{Proof of Proposition \ref{accu}.}
	From (\ref{ECmotPhi}), in BGP, we still need the growth rate of $\Phi(t)$ to be constant, i.e.,
	\begin{equation}\nonumber
		\frac{\dot \Phi(t)}{\Phi(t)} = \frac{\varphi(t)}{\Phi(t)} - \kappa,
	\end{equation}
	which makes the right hand side of the equation be constant as well. Then, the growth rate of information in every period $\varphi(t)$ is the same as that of total data accumulation $\Phi(t)$, i.e.,
	\begin{equation}\nonumber
		\frac{\dot \Phi(t)}{\Phi(t)} = \frac{\dot \varphi(t)}{\varphi(t)}
	\end{equation}
	In the baseline model, all the derivations only involve the growth rate of the endogenous variables, except the innovation possibility frontier, the only place where the level value of $\varphi(t)$ appears. Then, we can still derive the similar results simply by replacing the growth rate of $\varphi(t)$ with $\Phi(t)$ in appropriate places. Q.E.D.
\endproof

\subsection{Derivation of the Differential Equation System in Section 4}
\label{diffsys}

For simplicity, here we assume that $\psi = 1-\beta$. First, combining (\ref{ECFOCphiS}) and (\ref{ECFOClES}), we have
\begin{align}
	&-\frac{\xi(1-\xi)}{\sigma-\xi} \frac{\dot l_R(t)}{l_R(t)} + \xi \frac{\dot l_R(t)}{l_R(t)} \nonumber\\
	= &\left(\zeta-1+\frac{\xi \zeta}{\sigma-\xi}\right) g_N(t) + \left(1+\frac{\xi}{\sigma-\xi}\right)g_\mu(t) + \left(1+\frac{\xi}{\sigma-\xi}\right)n - g_\lambda(t). \nonumber
\end{align}
Consider that
\begin{equation}\nonumber
	\frac{\dot l_R(t)}{l_R(t)} = -\frac{\dot l_E(t)}{1-l_E(t)},
\end{equation}
and from the resource constraint (\ref{rcsim}) and (\ref{ECFOCcS}), we have
\begin{equation}\label{ECclambda}
	g_c(t) = g_N(t) + \frac{\dot l_E(t)}{l_E(t)} = -\frac{1}{\gamma} g_\lambda(t).
\end{equation}
Then, the law of motion of $l_E(t)$ can be derived as
\begin{equation}\label{ECmotlE}
	\dot l_E(t) = \frac{\left(\gamma+\zeta-1+\frac{\xi \zeta}{\sigma-\xi}\right) g_N(t) + \left(1+\frac{\xi}{\sigma-\xi}\right)g_\mu(t) + \left(1+\frac{\xi}{\sigma-\xi}\right)n}{\left(\frac{\xi(1-\xi)}{\sigma-\xi}-\xi\right) \frac{1}{1-l_E(t)} - \frac{\gamma}{l_E(t)}}
\end{equation}
and the growth rate of $\varphi(t)$ can be represented as
\begin{equation}\label{ECgphi}
	g_\varphi(t) = \frac{1}{\sigma-\xi} \left[ g_\mu(t) + \zeta g_N(t) - (1-\xi) \frac{\dot l_E(t)}{1-l_E(t)} + n\right].
\end{equation}
From (\ref{ECFOClES}) and (\ref{ECFOCNS}), we derive the law of motion of $g_\mu(t)$ as
\begin{equation}\label{ECmotgmu}
	\dot g_\mu(t) = (g_\mu(t) - \rho + n) \left[ (\zeta-1)g_N(t) + \xi g_\varphi(t) + \xi \frac{\dot l_E(t)}{1-l_E(t)} + n + \frac{(1-\xi-\zeta)\dot l_E(t)}{\zeta+ (1-\xi-\zeta) l_E(t)} \right].
\end{equation}
Similarly, the law of motion of $g_N(t)$ is
\begin{equation}\label{ECmotgN}
	\dot g_N(t) = g_N(t) \left[ (\zeta-1)g_N(t) + \xi g_\varphi(t) - (1-\xi) \frac{\dot l_E(t)}{1-l_E(t)} + n \right].
\end{equation}

In conclusion, the system of differential equations of the three state-like variables are (\ref{ECmotgN}), (\ref{ECmotgmu}), and (\ref{ECmotlE}). After deriving the transitional dynamics of these three state-like variables, other variables such as $g_c(t)$ and $g_\varphi(t)$ can be derived from (\ref{ECclambda}) and (\ref{ECgphi}), respectively.

\subsection{Additional Results on Transitional Dynamics}
\label{app:addres}

\begin{figure}[htp]
	\begin{minipage}[h]{0.5\linewidth}
		\centering
		\includegraphics[width=\textwidth]{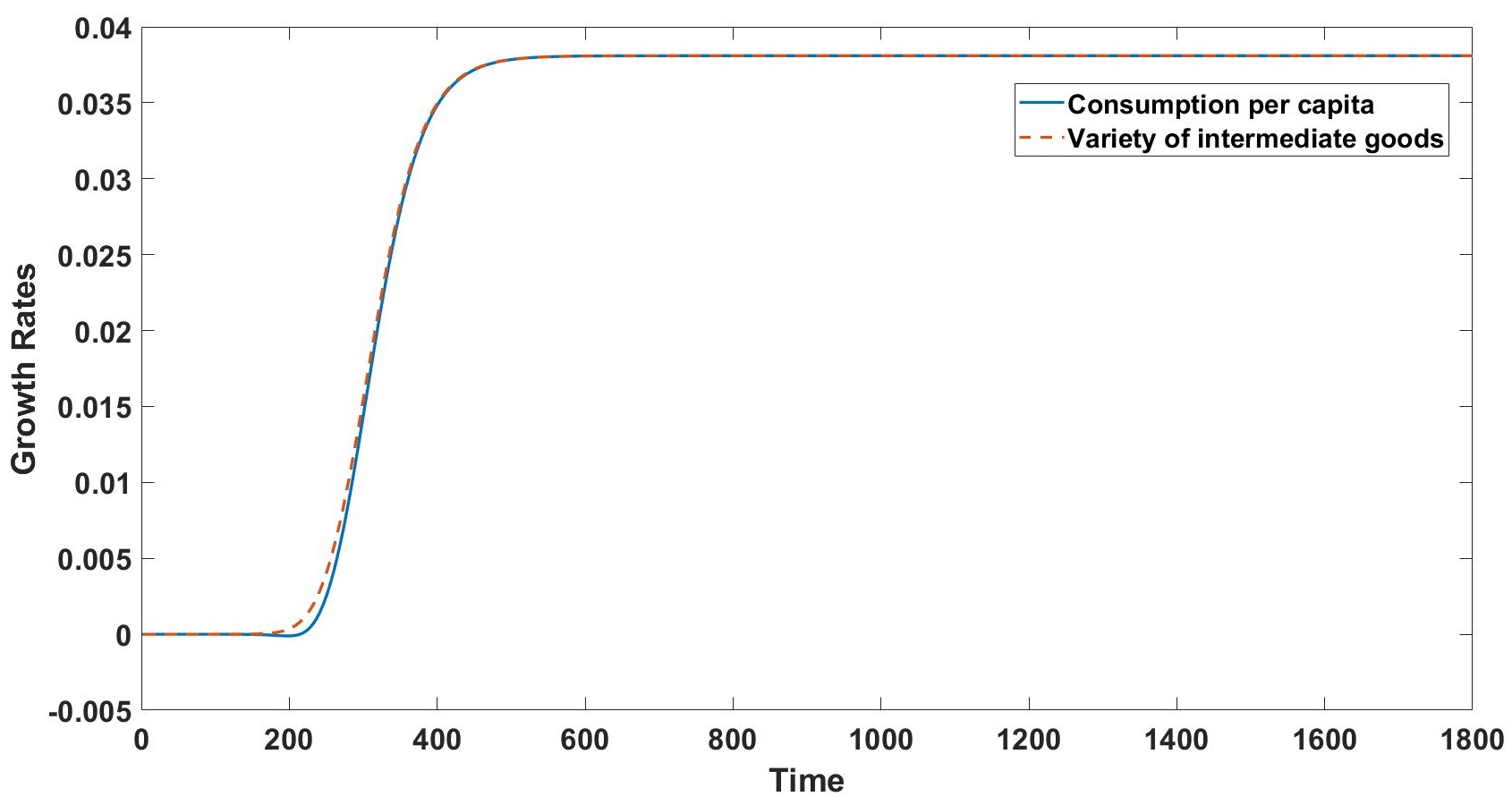}
		\par (a) $\sigma=2.0$ (without constraint)
	\end{minipage}
	\begin{minipage}[h]{0.5\linewidth}
		\centering
		\includegraphics[width=\textwidth]{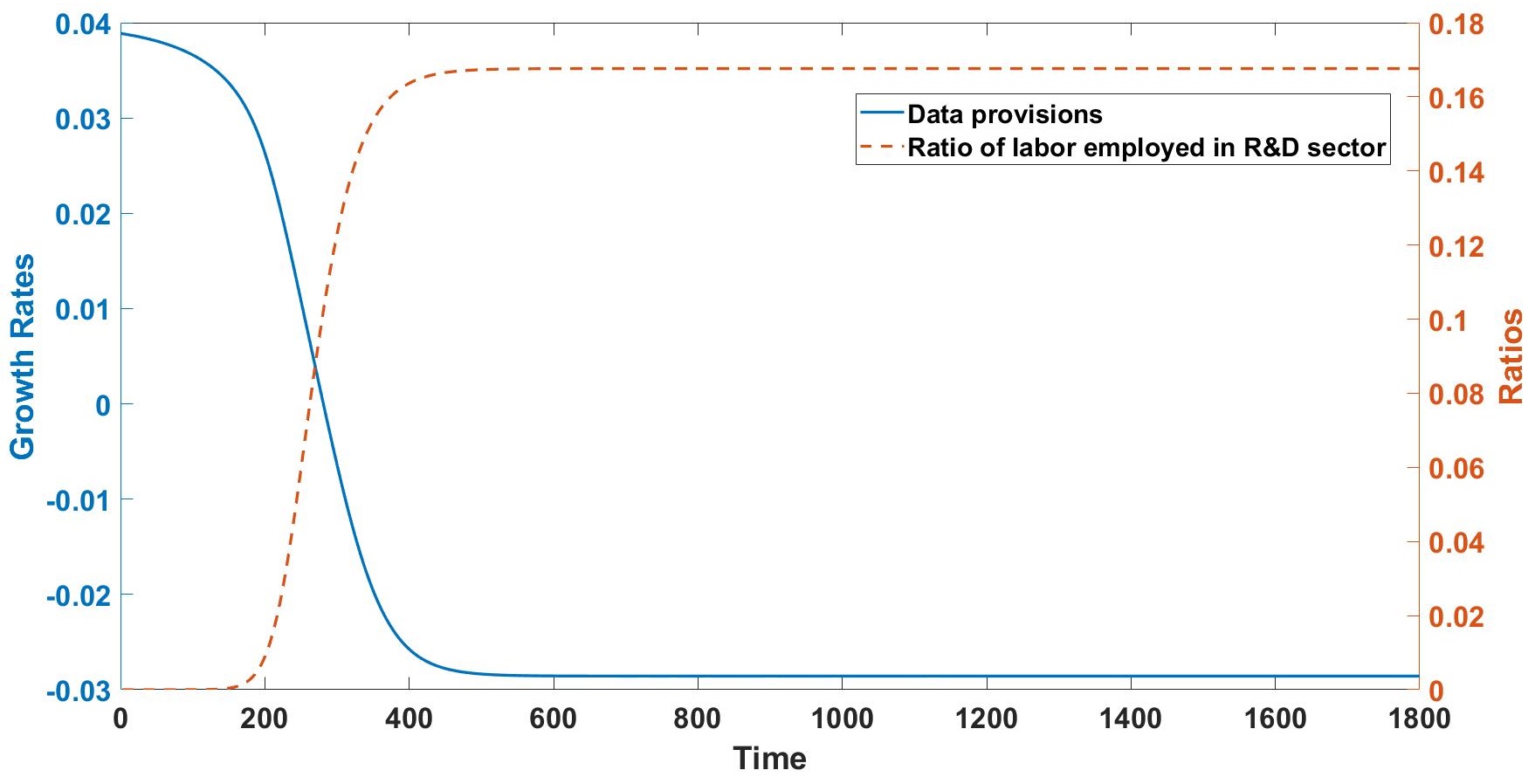}
		\par (b) $\sigma=2.0$ (without constraint)
	\end{minipage}
	\begin{minipage}[h]{0.5\linewidth}
		\centering
		\includegraphics[width=\textwidth]{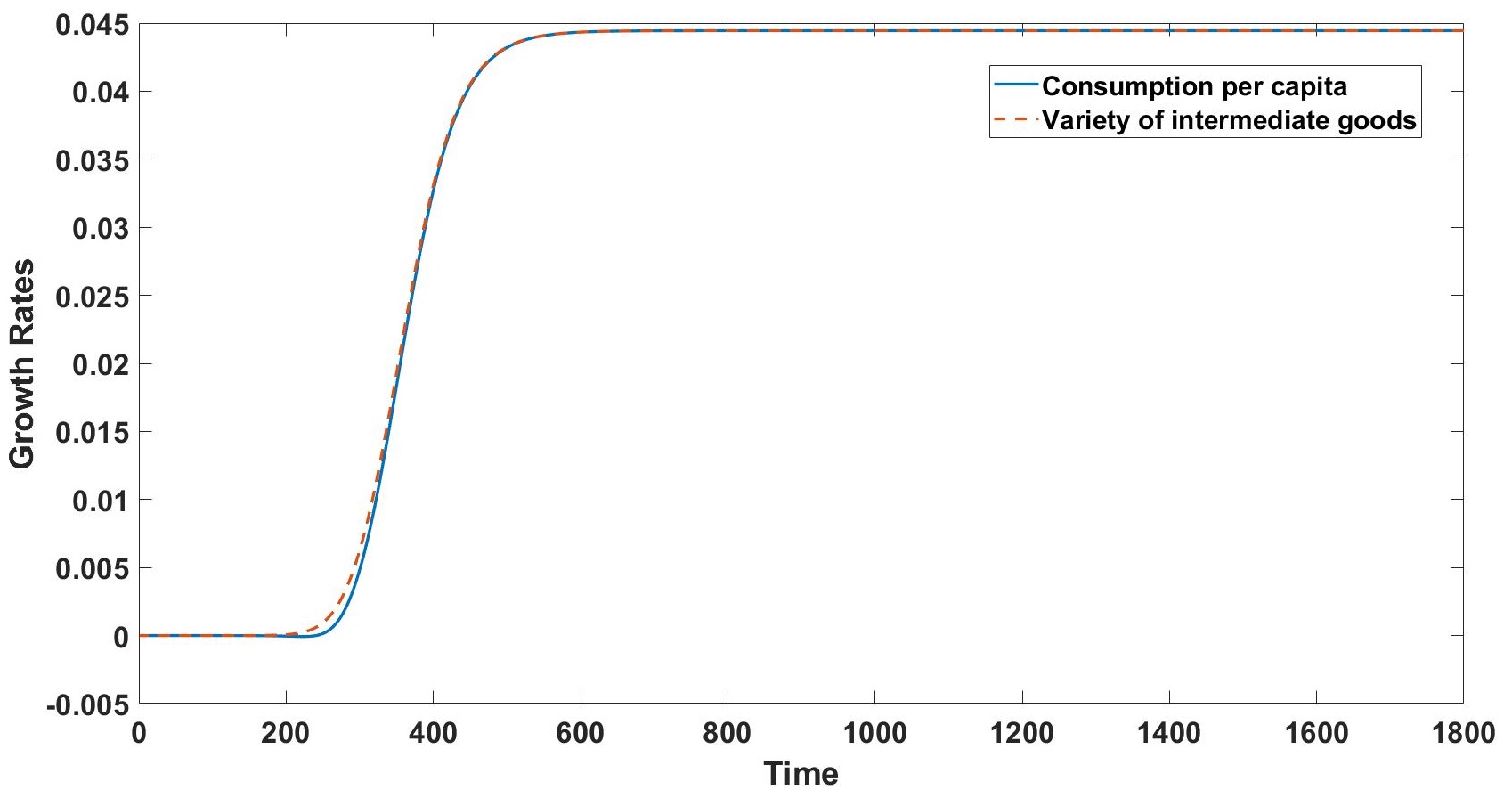}
		\par (c) $\sigma=2.5$ (without constraint)
	\end{minipage}
	\begin{minipage}[h]{0.5\linewidth}
		\centering
		\includegraphics[width=\textwidth]{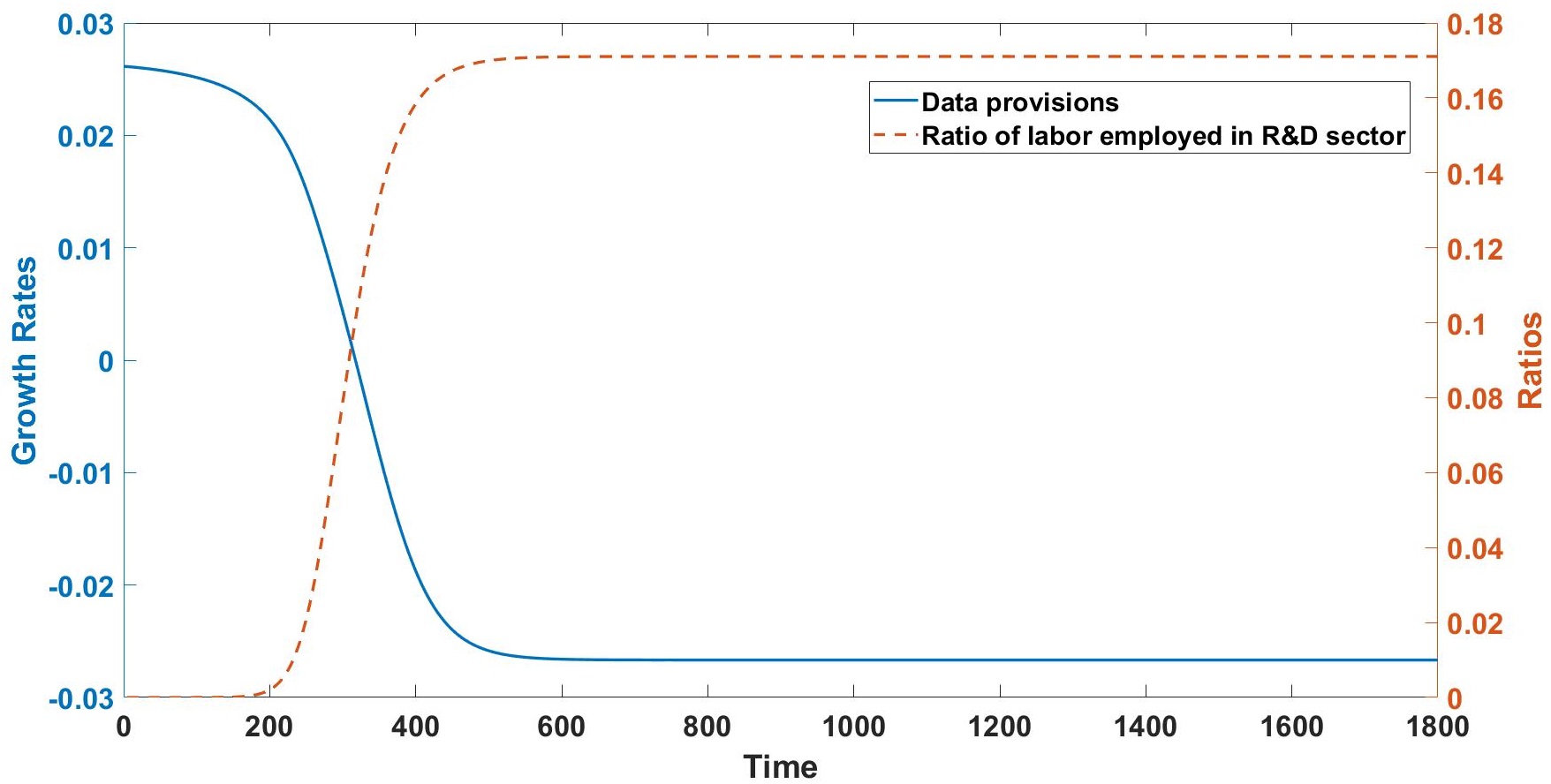}
		\par (d) $\sigma=2.5$ (without constraint)
	\end{minipage}
	\begin{minipage}[h]{0.5\linewidth}
		\centering
		\includegraphics[width=\textwidth]{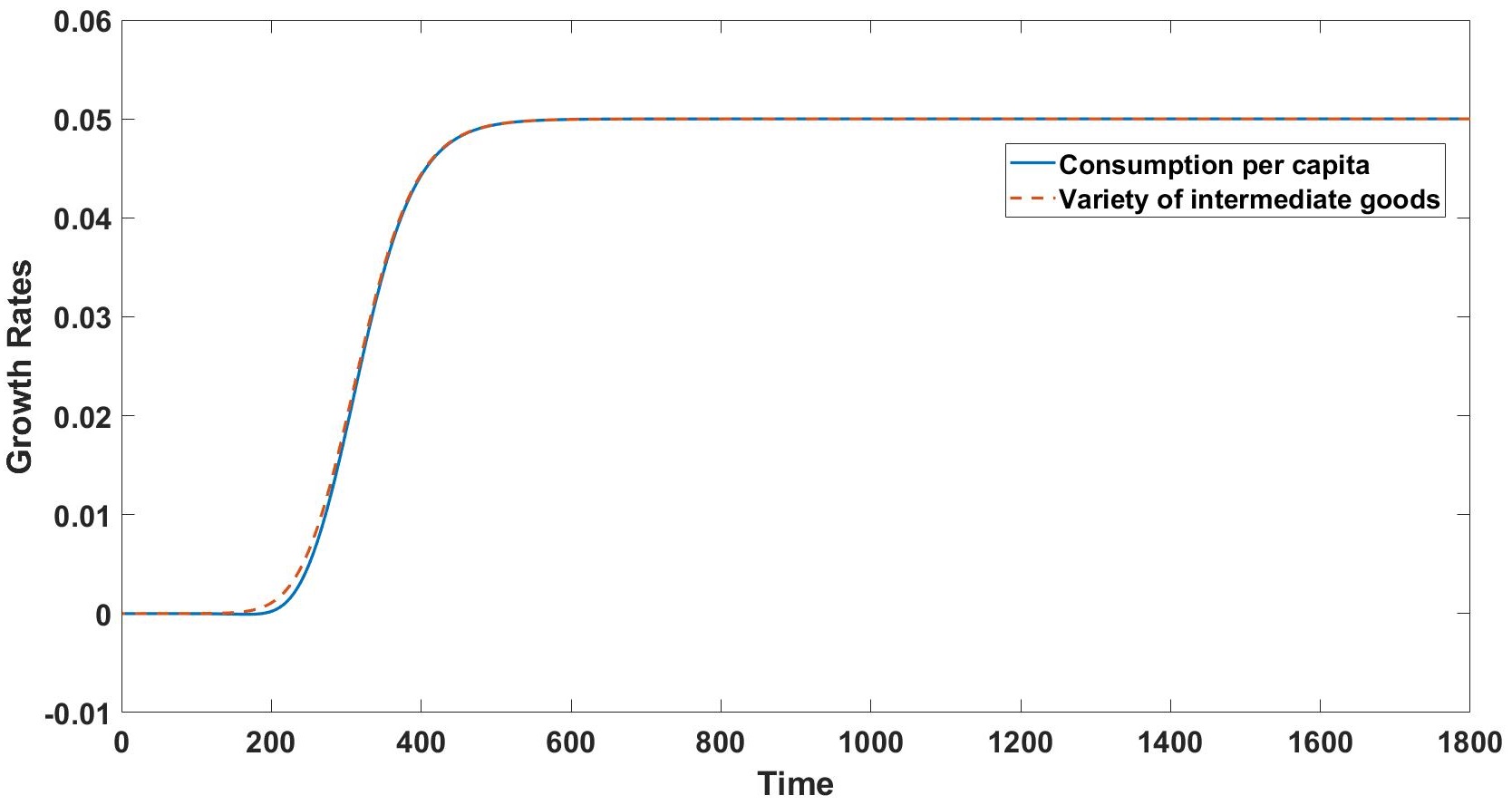}
		\par (e) $\sigma=3.0$ (without constraint)
	\end{minipage}
	\begin{minipage}[h]{0.5\linewidth}
		\centering
		\includegraphics[width=\textwidth]{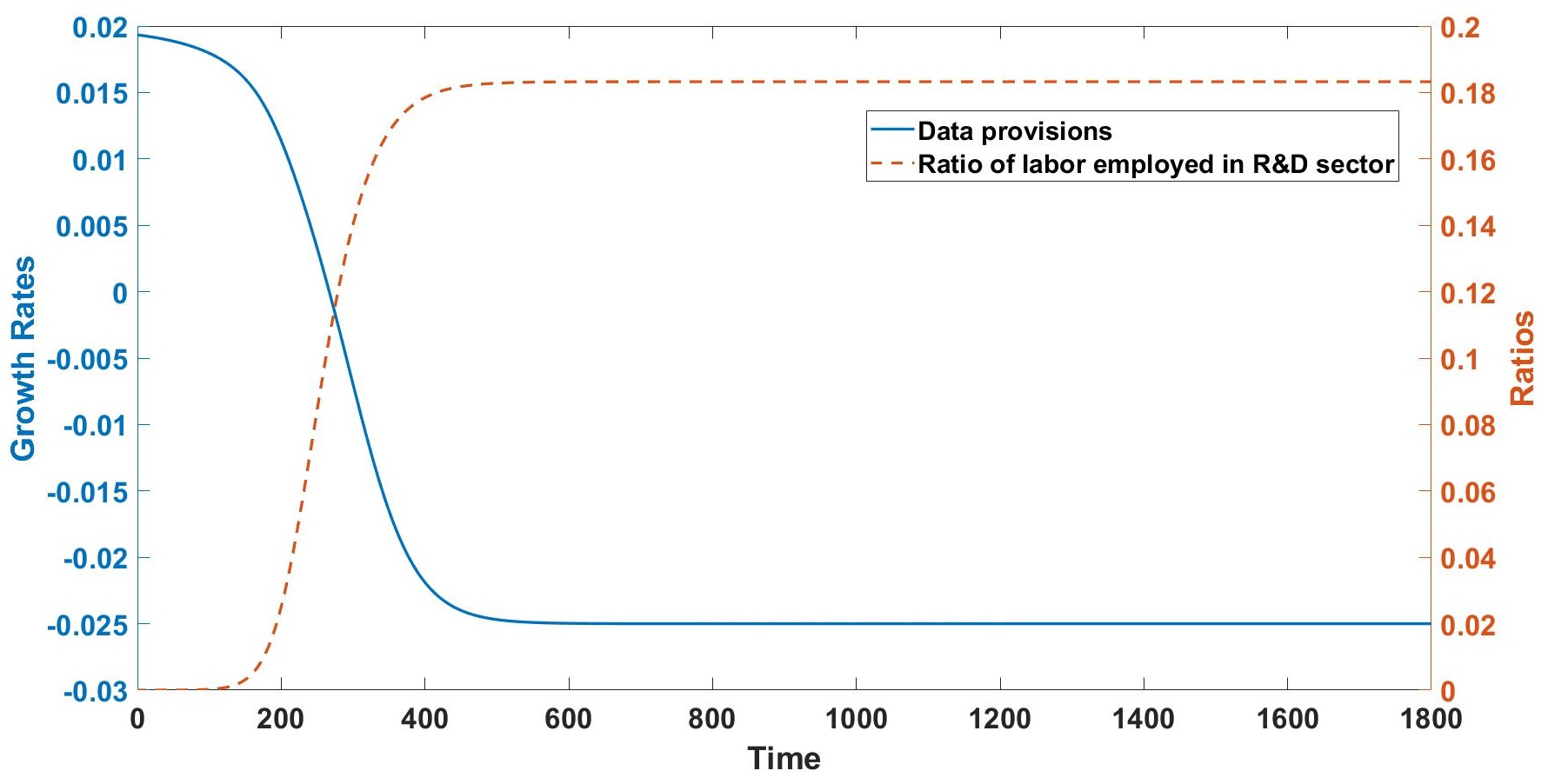}
		\par (f) $\sigma=3.0$ (without constraint)
	\end{minipage}
	\begin{center}
		\caption{Key variables along transitional path with different values of $\sigma$, without data provision constraint}
		\label{ECtrans}
	\end{center}
	\small \qquad \textit{Notes.} These figures show the transitional dynamics of the social planner's problem when $\sigma=2.0, 2.5 \text{ and } 3.0$, without the constraint of data provision. The economies undergo long but relatively steady states before finally reaching BGP at the end points.
\end{figure}

\begin{figure}[htp]
	\begin{minipage}[h]{0.5\linewidth}
		\centering
		\includegraphics[width=\textwidth]{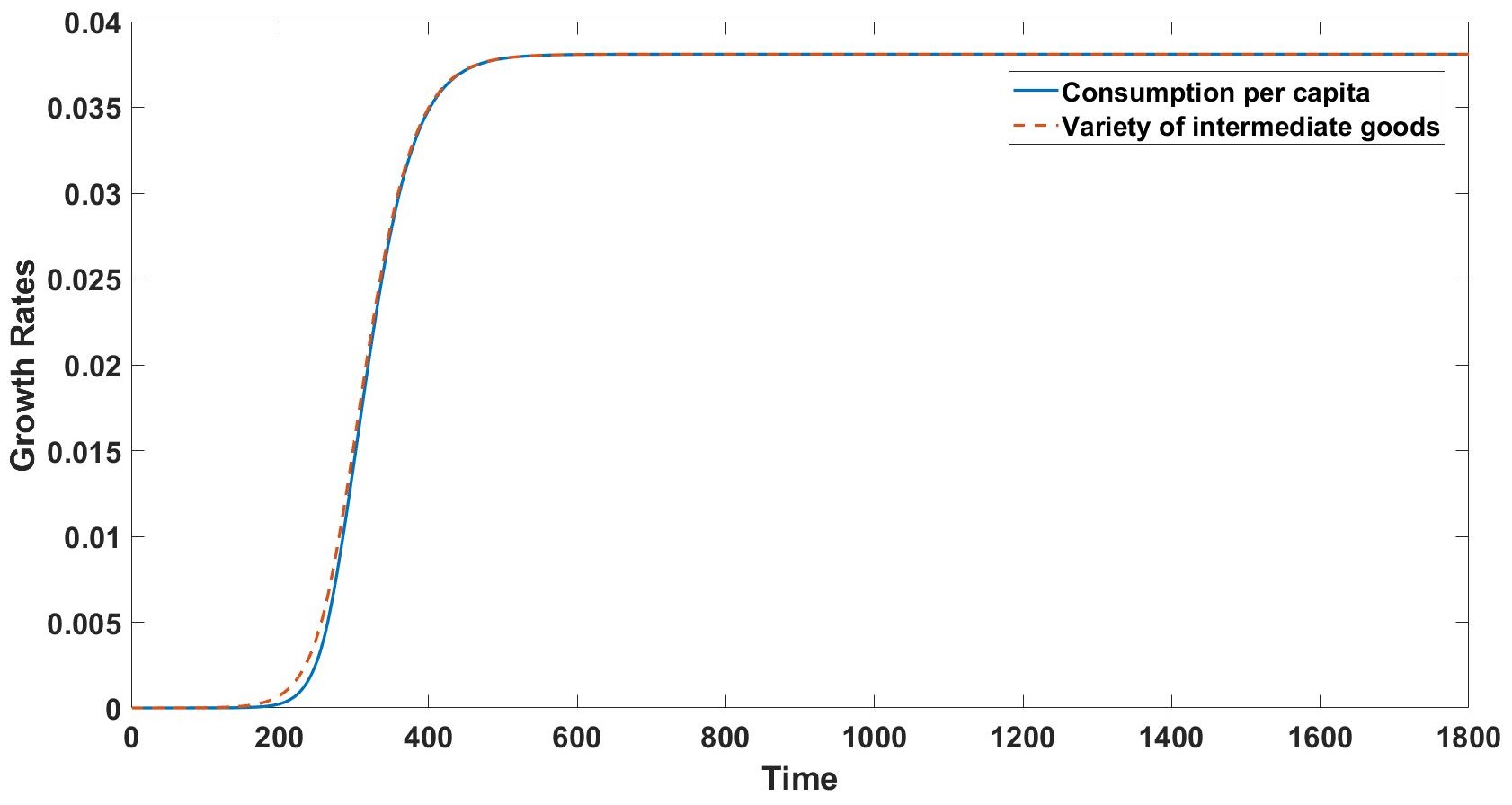}
		\par (a) $\sigma=2.0$ (with constraint)
	\end{minipage}
	\begin{minipage}[h]{0.5\linewidth}
		\centering
		\includegraphics[width=\textwidth]{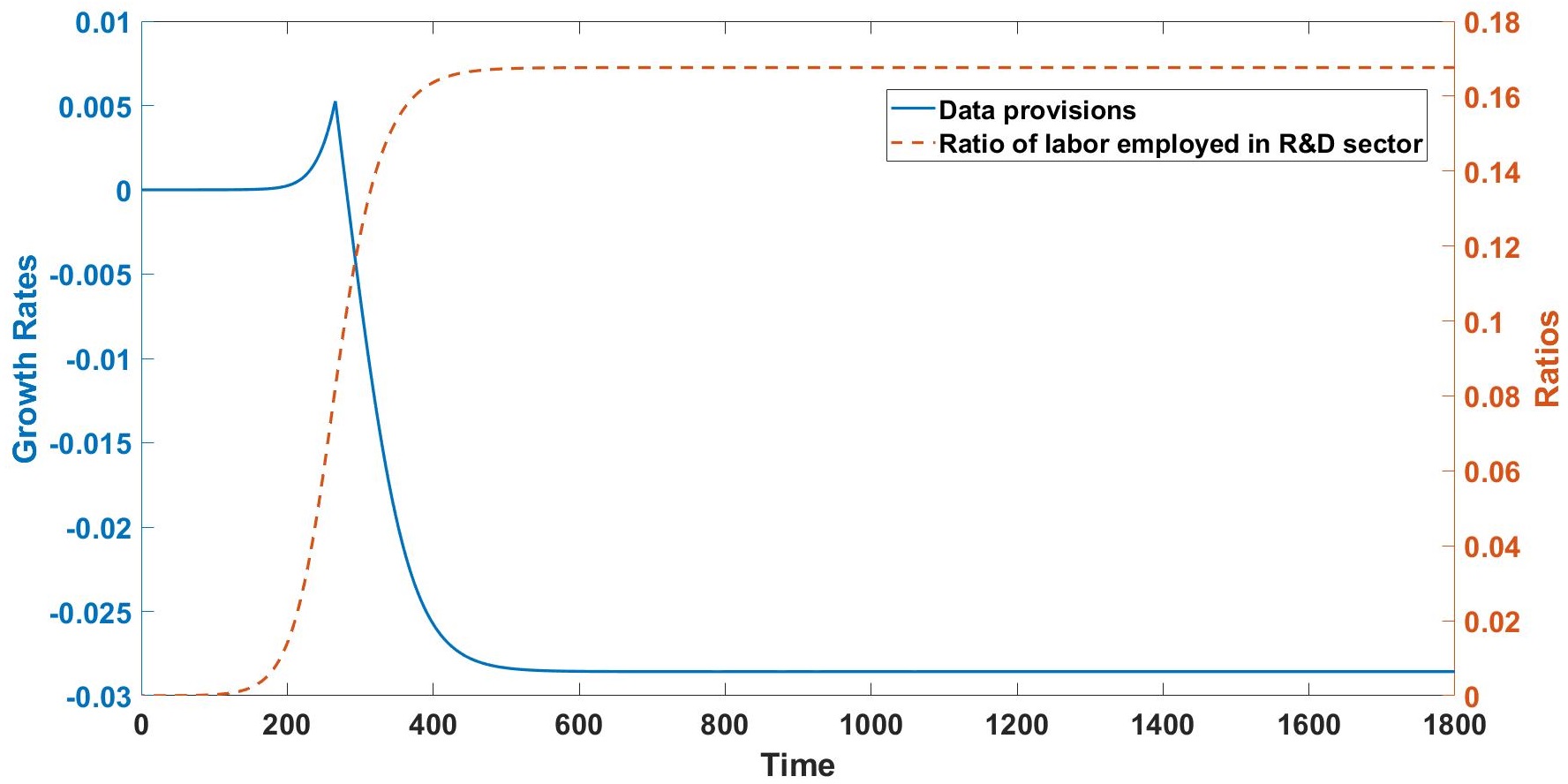}
		\par (b) $\sigma=2.0$ (with constraint)
	\end{minipage} 
	\begin{minipage}[h]{0.5\linewidth}
		\centering
		\includegraphics[width=\textwidth]{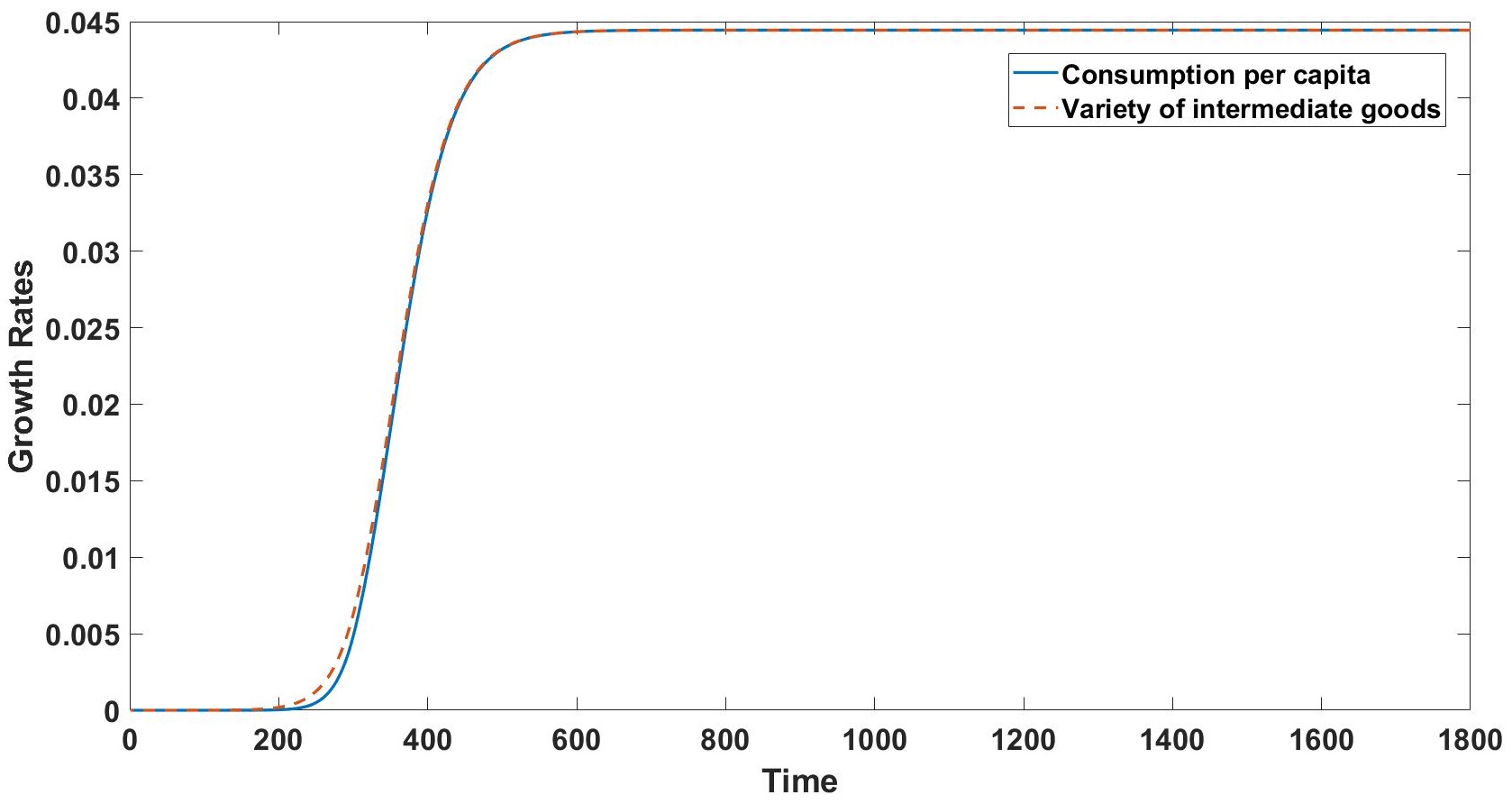}
		\par (c) $\sigma=2.5$ (with constraint)
	\end{minipage}
	\begin{minipage}[h]{0.5\linewidth}
		\centering
		\includegraphics[width=\textwidth]{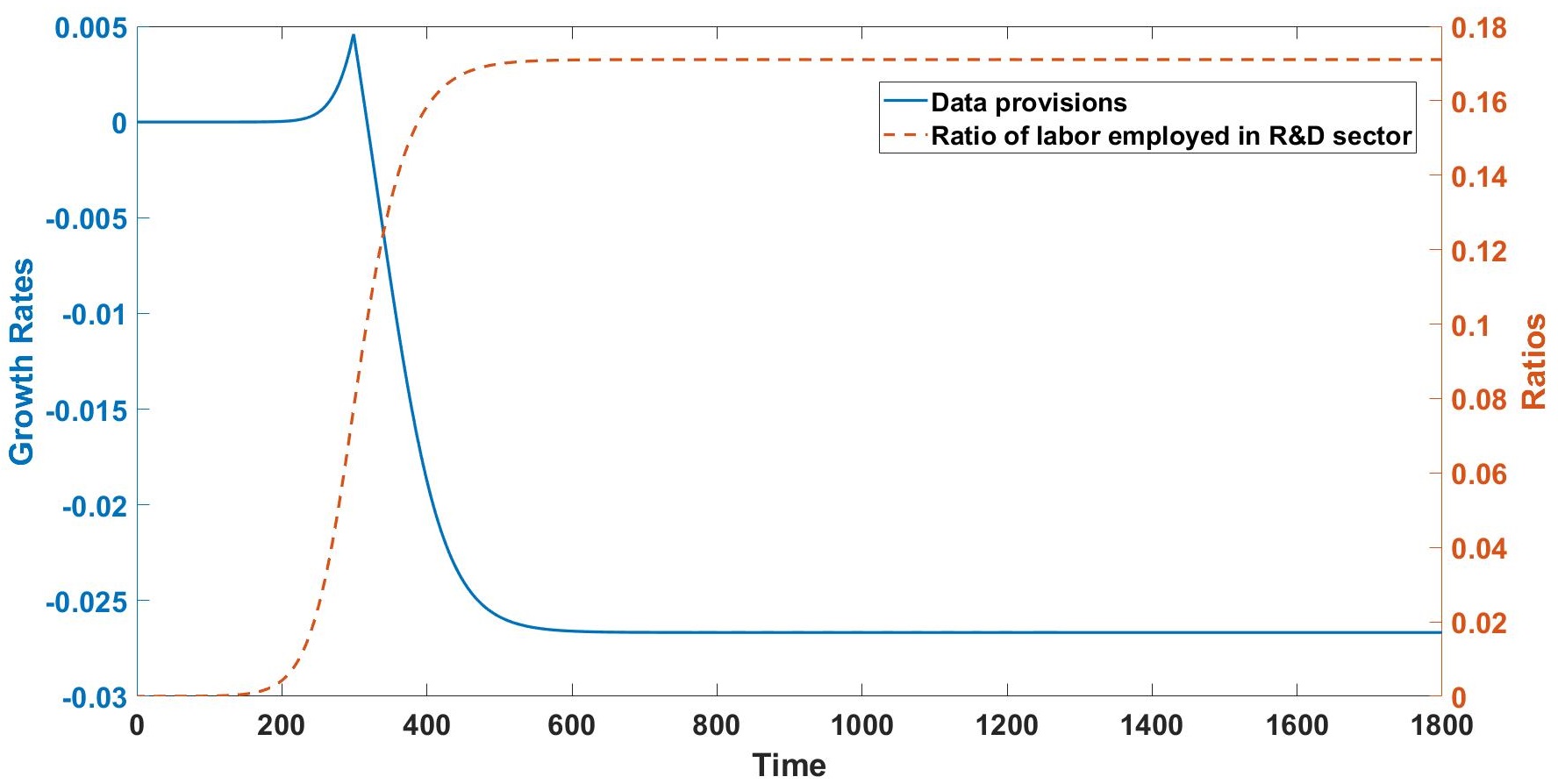}
		\par (d) $\sigma=2.5$ (with constraint)
	\end{minipage}
	\begin{minipage}[h]{0.5\linewidth}
		\centering
		\includegraphics[width=\textwidth]{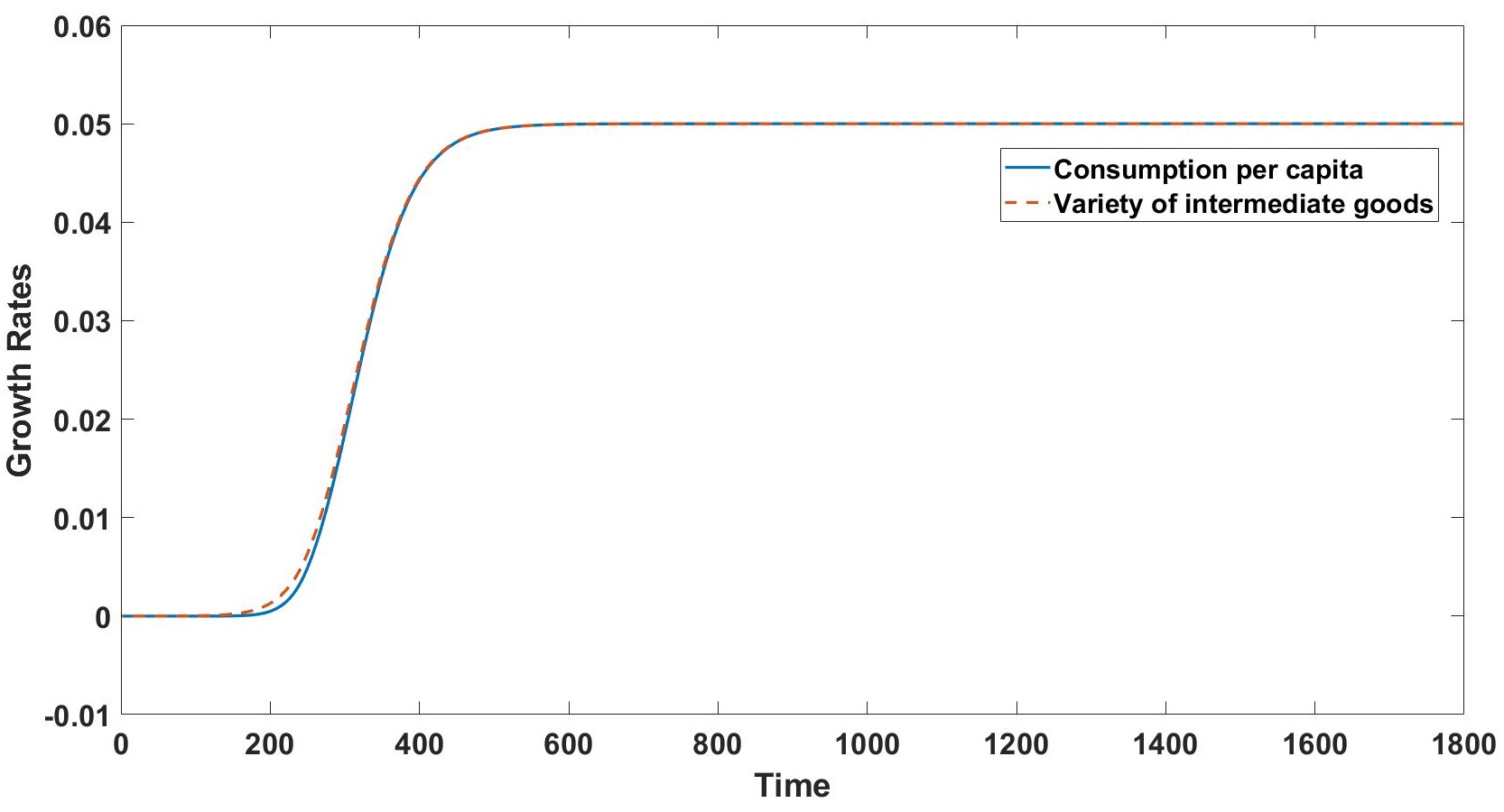}
		\par (e) $\sigma=3.0$ (with constraint)
	\end{minipage}
	\begin{minipage}[h]{0.5\linewidth}
		\centering
		\includegraphics[width=\textwidth]{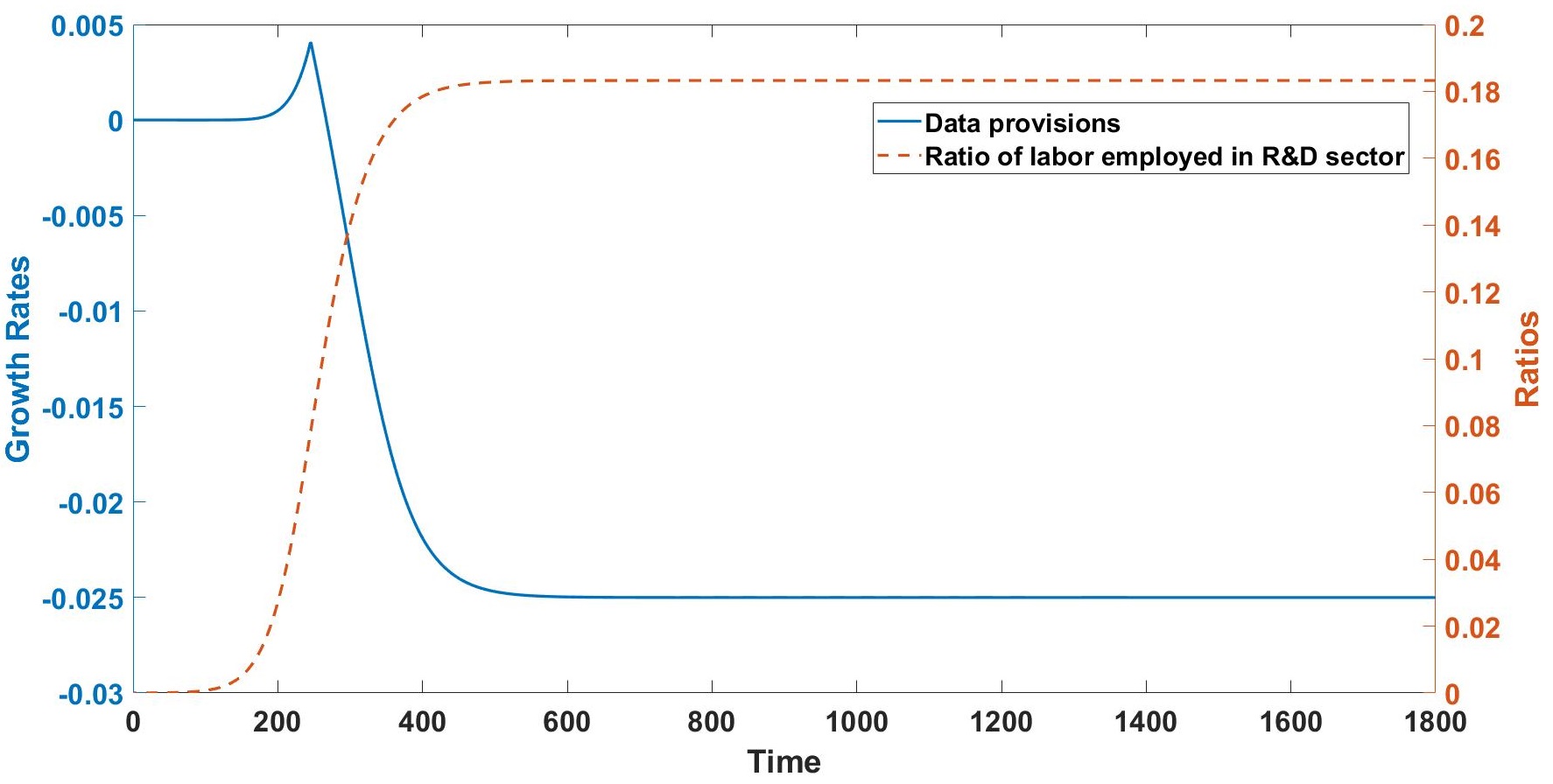}
		\par (f) $\sigma=3.0$ (with constraint)
	\end{minipage}
	\begin{center}
		\caption{Key variables along transitional path with different values of $\sigma$, with data provision constraint}
		\label{ECtranscons}
	\end{center}
	\small \qquad \textit{Notes.} These figures show the transitional dynamics of the social planner's problem when $\sigma=2.0, 2.5 \text{ and } 3.0$, with the constraint of data provision. The economies undergo long but relatively steady states before finally reaching BGP at the end points.
\end{figure}

To demonstrate the robustness of our findings, we show in Figure \ref{ECtrans} and Figure \ref{ECtranscons} the transitional dynamics when $\sigma$, the disutility parameter, takes other values with and without the data provision constraint, respectively.

\begin{figure}[htp]
	\begin{minipage}[h]{0.5\linewidth}
		\centering
		\includegraphics[width=\textwidth]{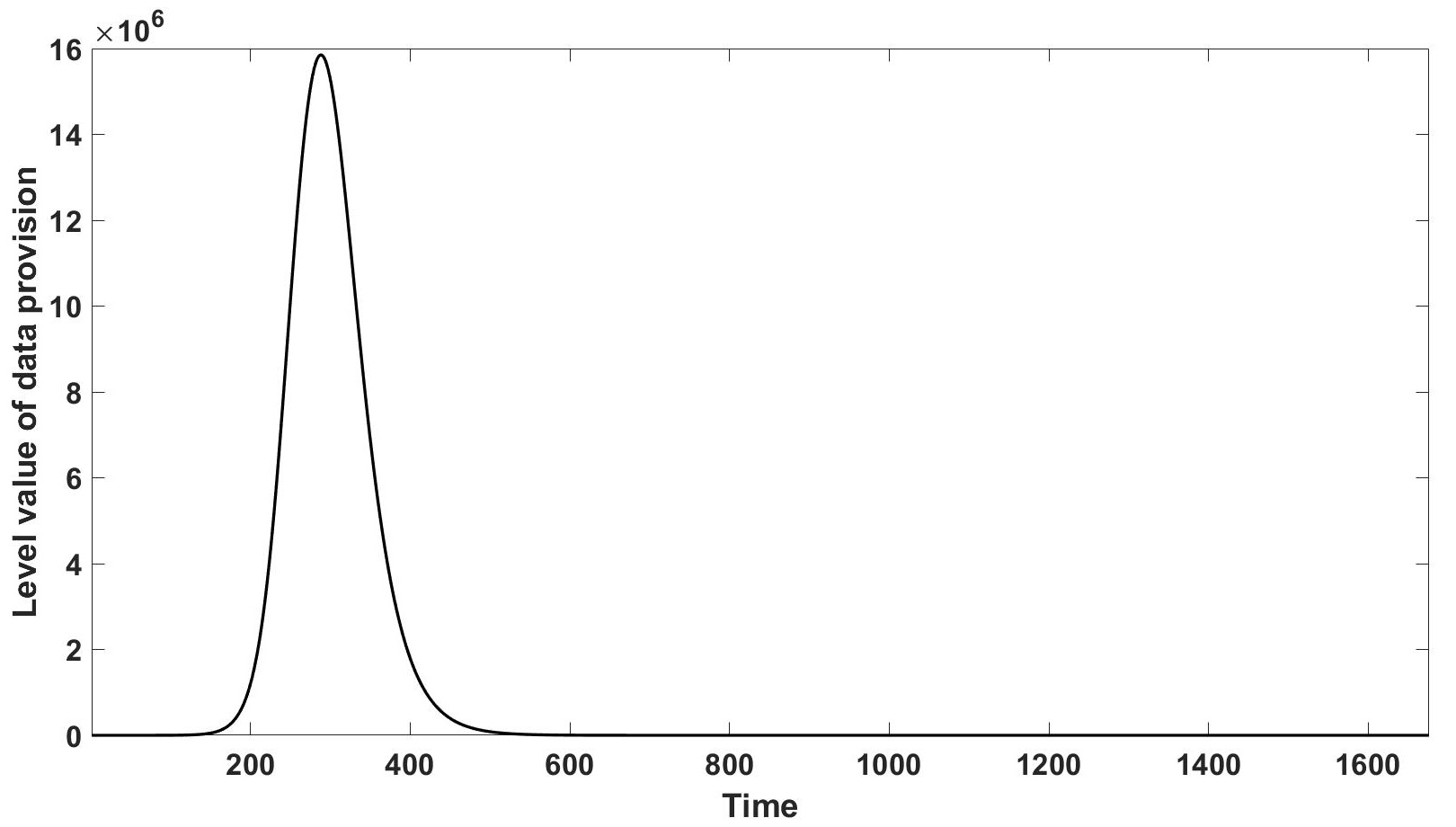}
		\par (a) Provision Level (without constraint)
	\end{minipage}
	\begin{minipage}[h]{0.5\linewidth}
		\centering
		\includegraphics[width=\textwidth]{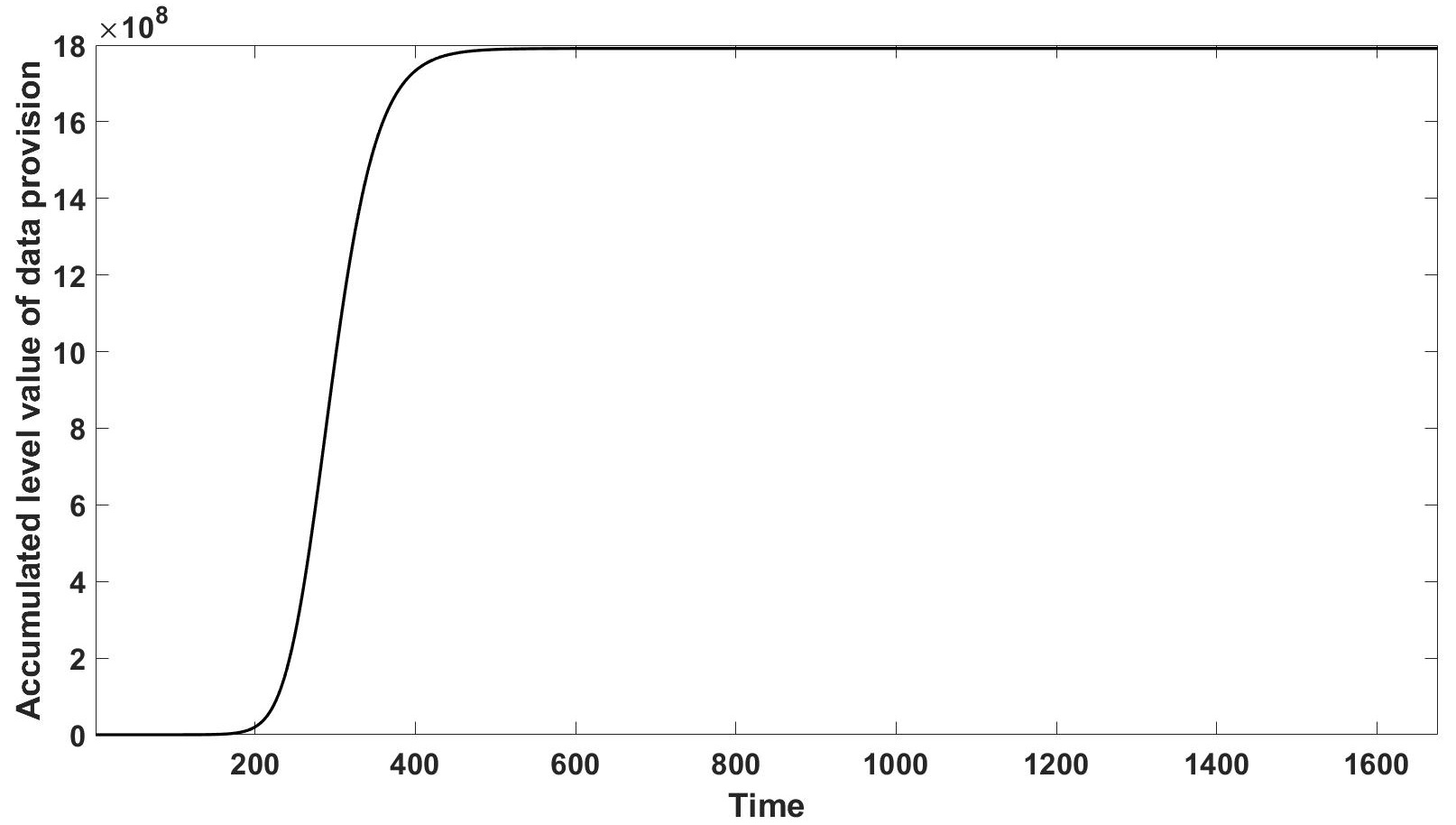}
		\par (b) Cumulative Level (without constraint)
	\end{minipage}
	\begin{minipage}[h]{0.5\linewidth}
		\centering
		\includegraphics[width=\textwidth]{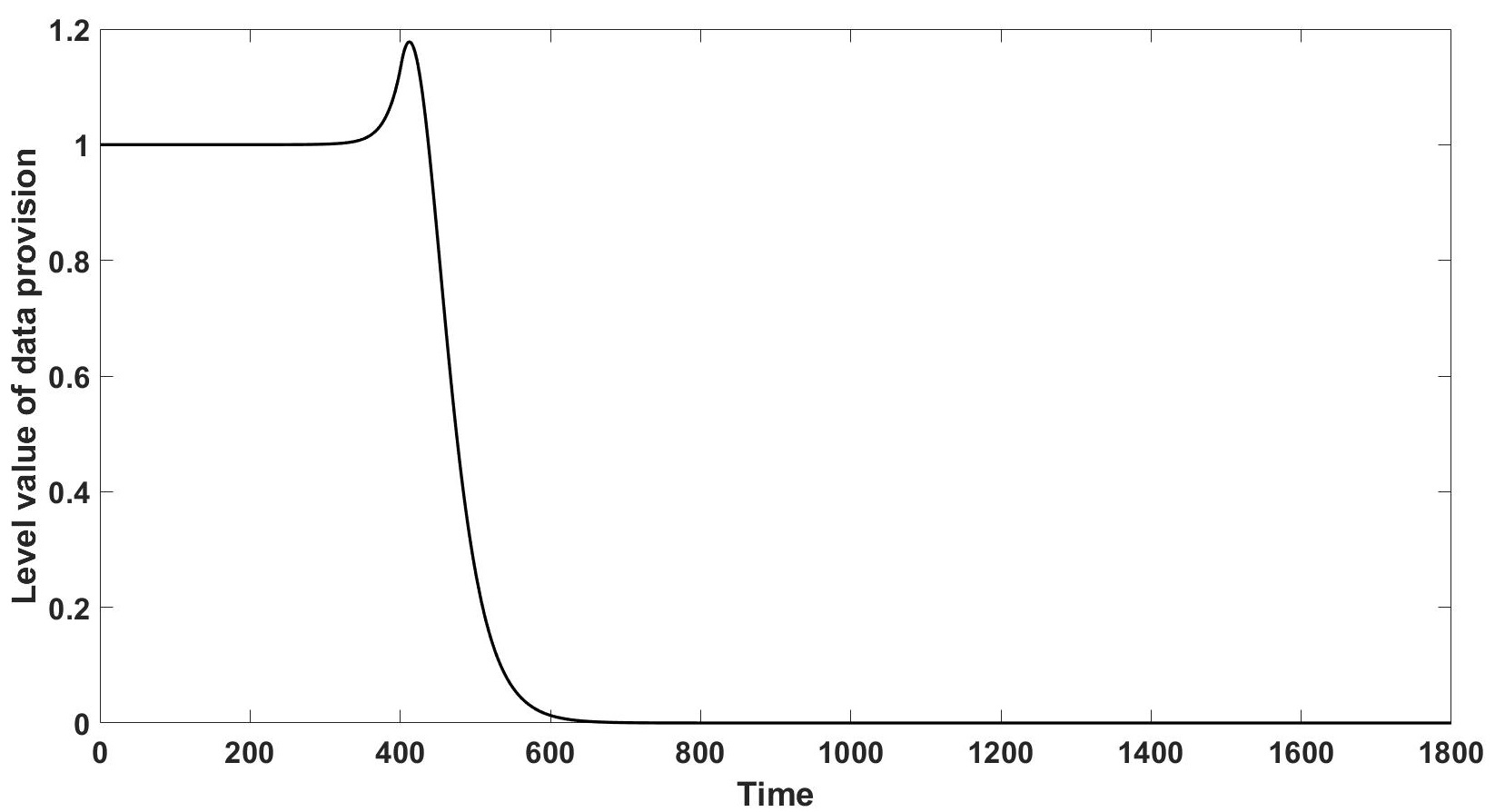}
		\par (c) Provision Level (with constraint)
	\end{minipage}
	\begin{minipage}[h]{0.5\linewidth}
		\centering
		\includegraphics[width=\textwidth]{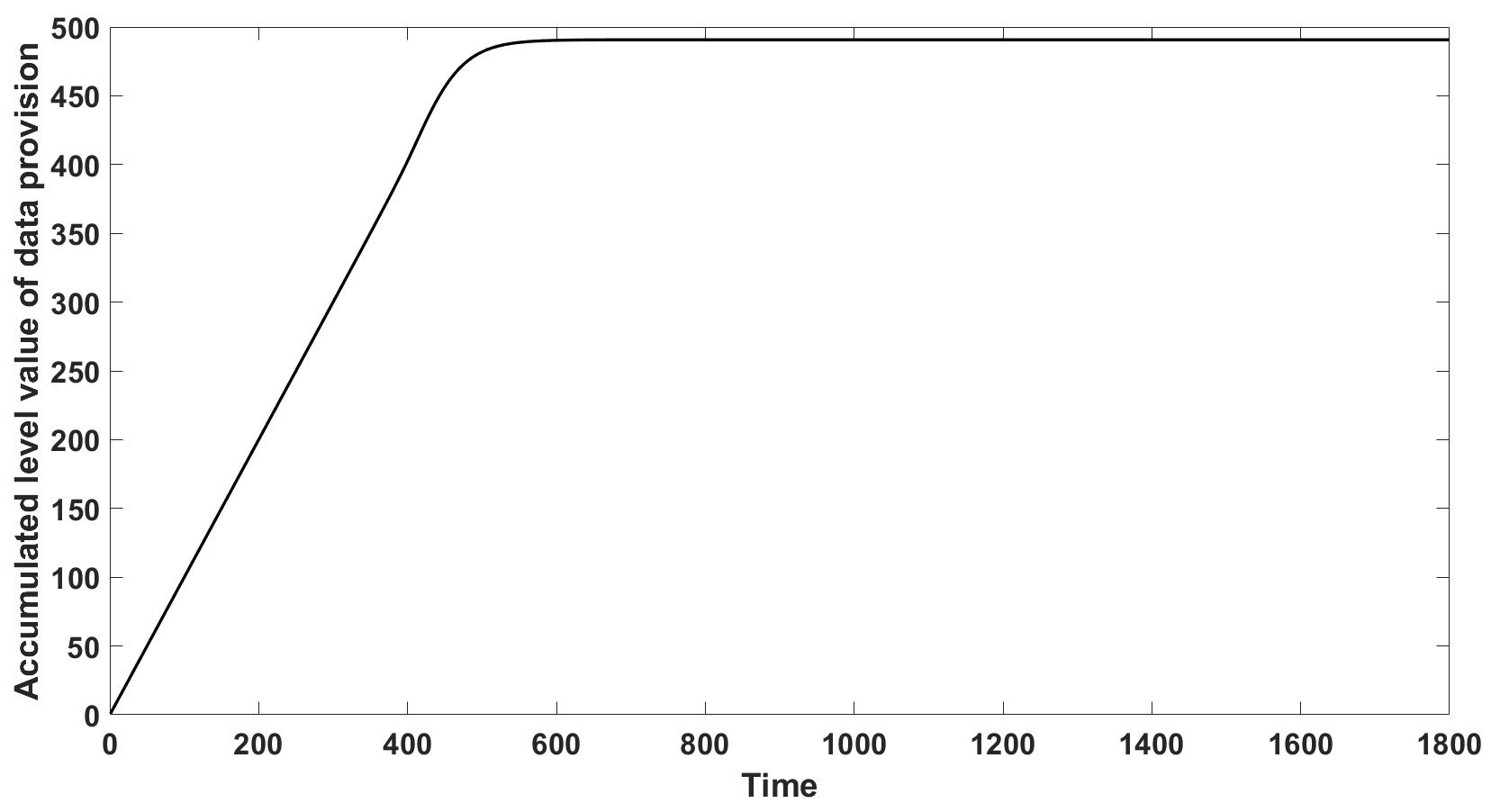}
		\par (d) Cumulative Level (with constraint)
	\end{minipage}
	\begin{center}
		\caption{Flow and cumulative levels of data provision with $\sigma=1.5$, with and without data provision constraint.}
		\label{lv}
	\end{center}
	\small \qquad \textit{Notes.} These figures show the level change and cumulative level change of data provision before reaching BGP when $\sigma=1.5$, with and without the constraint of data provision.
\end{figure}

We further illustrate the evolution of data provision in Figure \ref{lv} by plotting the level of and cumulative level of usage/provision. The initial value of data provision is set to $\varphi(0)=1$ by taking $\sigma=1.5$, we observe a peak of data provision which decreases rapidly to a low level while the cumulative data provision remains on a plateau. As we have discussed previously, the effect of data in our model still exists in the form of new varieties of intermediate goods over time.

\section{Model Extensions: Proofs and Further Discussion}

\subsection{Policy Implications on BGP}

First, as we prove in Online Appendix \ref{consumerOwnership}, reducing $s$ through privacy policy regulation would reduce growth rate, and would not lead to welfare improvement without sacrificing growth, if it improves welfare at all. Solving the case explicitly is difficult, we therefore restrict our discussion here to policy interventions that preserve growth rates.

\subsubsection{Ineffectiveness of Taxing Data Usage}
\label{proofdatatax}

\begin{proposition}\label{propdatatax}
    Levying a tax on the usage of data alters the transitional dynamics but is ineffective in bringing the equilibrium allocations in decentralized economy closer to the social planner's solution because the decentralized economy eventually returns to the original BGP path.
\end{proposition}

\proof{Proof of Proposition \ref{propdatatax}.}

First, we should prove that the optimal tax rate on data usage should be constant to ensure the BGP growth rates not deviate from the level in baseline model. If we add a tax term on the free-entry condition of data provision, i.e.,
\begin{equation}\label{freeentrydatatax}
	\eta \xi N(t)^\zeta \varphi(t)^{\xi-1} l_R(t)^{1-\xi} V(t) = \tau(t) p_\varphi(t)
\end{equation}
where $\tau(t)$ is the tax rate. Rewrite (\ref{freeentrydatatax}) into the form of growth rate, we have
\begin{equation}\nonumber
    g_{p_\varphi}(t) + g_\tau(t) = \zeta g_N(t) + (\xi-1) g_\varphi(t) + (1-\xi) g_{l_R}(t) + g_V(t).
\end{equation}
Then, combine the above equation with (\ref{motphi}), we have (in BGP)
\begin{equation}\label{gtau}
    g_\tau(t) = \zeta g_N(t) + (\xi-\sigma) g_\varphi(t) - r^* +n + \rho.
\end{equation}
Also, recalling (\ref{motC}), (\ref{gtau}) can be further written as
\begin{equation}\nonumber
    g_\tau(t) = \zeta g_N(t) + (\xi-\sigma) g_\varphi(t) - \gamma g_c(t) + n = (\zeta -\gamma) g^* + (\xi-\sigma) g^*_\varphi +n.
\end{equation}
If we want the growth rates $g^*$ and $g^*_\varphi$ to be the same as social optimum, we can derive that
\begin{equation}\nonumber
    g_\tau (t) = (\zeta - \gamma) \left[ \frac{\sigma}{(1-\zeta)\sigma - \xi(1-\gamma)} \right] n + (\xi-\sigma) \left[ \frac{1-\gamma}{(1-\zeta)\sigma - \xi(1-\gamma)} \right] n + n = 0.
\end{equation}
Thus, tax rate $\tau(t)$ should be a constant in BGP. In other words, as long as the tax rate is a constant over time, BGP growth rate remains the same.

Then, we want to show the optimal tax rate that can make the allocations in decentralized economy close to the solution in social planner's problem. Imagine a decentralized economy in BGP, the government suddenly announces a fixed rate of tax on data collection by intermediate producers from consumers, $\tau$, in order to push the economy close to social optimum, i.e., make the labor allocations in both sectors and data provision level the same as social optimum. In this case, we have to derive the labor share in the two sectors in another way, as shown below. Consider that in BGP,
\begin{equation}\nonumber
    g^* = \eta N(t)^{\zeta-1} \varphi(t)^\xi (l_R^*)^{1-\xi} L(t)
\end{equation}
and the free-entry condition of data (in BGP)
\begin{equation}\nonumber
    \eta \xi N(t)^\zeta \varphi(t)^{\xi-1} (l_R^*)^{1-\xi} V(t) = \tau(t) p_\varphi(t).
\end{equation}
Combining the above two equations, and plug in the BGP value of $V(t)$, we have
\begin{align}\label{taxpro}
    g^* \xi N(t) \varphi(t)^{-1} V(t) L(t)^{-1}&= \tau p_\varphi(t) \nonumber \\
    \Rightarrow \quad g^* \xi N(t) \varphi(t)^{-1} \frac{\frac{\psi^{1-\frac{1}{\beta}}\beta}{(1-\beta)^{1-\frac{2}{\beta}}}L_E(t)}{r^*-n} L(t)^{-1}&= \tau p_\varphi(t) \nonumber \\
    \Rightarrow \quad g^* \xi  \frac{\frac{\psi^{1-\frac{1}{\beta}}\beta}{(1-\beta)^{1-\frac{2}{\beta}}}l_E^*}{\gamma g^* + \rho-n} N(t) \varphi(t)^{-1} p_\varphi(t)^{-1}&= \tau
\end{align}
Notice that in BGP, $N(t) \varphi(t)^{-1} p_\varphi(t)^{-1}$ converges to constant since
\begin{equation}\nonumber
    g_N^* - g^*_\varphi - g^*_{p_\varphi} = 0.
\end{equation}
Thus, we can rewrite this term as $N(t_0) \varphi(t_0)^{-1} p_\varphi(t_0)^{-1}$, where $t_0$ is one of the periods in BGP. Go on with (\ref{taxpro}), we finally derive the labor share in R\&D sector as
\begin{equation}\label{lRdata}
    l_R^* = 1- \frac{(\gamma g^* + \rho-n) \tau (1-\beta)^{1-\frac{2}{\beta}}}{\xi g^* \psi^{1-\frac{1}{\beta}} \beta} N(t_0) \varphi(t_0)^{-1} p_\varphi(t_0)^{-1}
\end{equation}

On the other hand, notice that $l^*_R$ should equal to the result coming from the free-entry condition of labor (\ref{freeentrylab}), then, (The derivation process is still the same as we do in Online Appendix \ref{proofDClsh}, with no influences from adding the tax term of data collection.)
\begin{equation}\nonumber
    1- \frac{(\gamma g^* + \rho-n) \tau (1-\beta)^{1-\frac{2}{\beta}}}{\xi g^* \psi^{1-\frac{1}{\beta}} \beta} N(t_0) \varphi(t_0)^{-1} p_\varphi(t_0)^{-1} = \frac{1}{1+\frac{g^*\gamma + \rho-n}{g^*(1-\xi)(1-\beta)}}.
\end{equation}
As a result, the term $\tau N(t_0) \varphi(t_0)^{-1} p_\varphi(t_0)^{-1}$ should always be constant. This means that changes in tax rate $\tau$ are always cancelled by changes in $N(t)$, $\varphi(t)$ and $p_\varphi(t)$ together, while the labor share in the two sectors remain unchanged. Thus, levying tax on data collection will not push the BGP results in decentralized economy close to the social planner's problem. Definitely, levying tax on data collection changes the usage of data and labor employment in the transitional state, but the economy will return to the previous path in BGP. Q.E.D.

\endproof

\subsubsection{Optimal Subsidy on Labor Employment in R\&D Sector}
\label{proofsubsidy}

Although levying a tax directly on data usage comes out to be ineffective, we can still adjust the BGP path of the decentralized economy through indirect ways. First, we consider subsidizing the wage rate of labor employed in R\&D sector. If $\tau(t)$ denotes the gross subsidy rate\footnote{Here, $\tau(t)$ is set together with wage $w(t)$, which is a kind of cost. Then, it is a subsidy rate if $0<\tau(t)<1$, while it is a tax rate if $\tau(t)>1$. In the next section, we set $\tau'(t)$ together with profit $V(t)$, which is a kind of revenue. In this occasion, the meaning becomes opposite: It is a subsidy rate if $\tau'(t)>1$, while it is a tax rate if $0<\tau'(t)<1$.}, the free-entry condition of labor employment (\ref{freeentrylab}) becomes
\begin{equation}\label{subfreeentry}
	\eta(1-\xi)N(t)^\zeta \varphi(t)^\xi l_R(t)^{-\xi} V(t) = \tau(t) w(t).
\end{equation}
Then, we have the following proposition.

\begin{proposition}\label{propsubsidy}
    In order to push labor allocations in decentralized economy close to the solution in the social planner's problem without deviating from the BGP growth rates, the subsidy rate on labor employment in R\&D sector $\tau(t)$ should be constant, which is determined by
    \begin{equation}\nonumber
	\tau(t) = \tau^* = \frac{(1-\beta)[(\sigma-\xi)n+\xi \rho-(\sigma-\xi)(1-\zeta)g^*]}{\gamma g^*+\rho } <1,
	\end{equation}
	where, $g^*$ is BGP growth rate derived in Proposition \ref{dcgrow}.
\end{proposition}

\proof{Proof of Proposition \ref{propsubsidy}.}
First, we prove that the optimal subsidy rate should be constant to ensure that the BGP growth rates do not deviate from previous levels. Rewrite the adjusted free-entry condition (\ref{subfreeentry}) into a form of growth rates, we have
\begin{equation}\nonumber
	\zeta g_N(t) + \xi g_\varphi(t) -\xi g_{l_R}(t) + g_V(t) = g_{\tau}(t) + g_w(t).
\end{equation}
Also, from the determining equation of wage in production sector (\ref{wage}), rewrite it into growth rate form:
\begin{equation}\nonumber
	g_w(t) = g_N(t).
\end{equation}
In BGP, $g_{l_R}(t)=0$, $g_V(t)=n$. Thus, combining the above two equations, we derive the growth rate of subsidy rate as
\begin{equation}\nonumber
	g_{\tau}(t) = (\zeta-1) g_N(t) + \xi g_\varphi(t) + n.
\end{equation}
If we want the growth rates $g^*$ and $g^*_\varphi$ to be the same as the solution in the social planner's problem, we derive that
\begin{equation}\nonumber
	g_{\tau}(t) = (\zeta-1) \left[ \frac{\sigma}{(1-\zeta)\sigma - \xi(1-\gamma)} \right] n + \xi \left[ \frac{1-\gamma}{(1-\zeta)\sigma - \xi(1-\gamma)} \right] n + n = 0.
\end{equation}
Thus, the subsidy rate should also be constant.

Then, we determine the optimal subsidy rate. Go on with the derivation process in Online Appendix \ref{proofDClsh} by replacing the free-entry condition (\ref{freeentrylab}) with (\ref{subfreeentry}), we finally derive the labor share in the R\&D sector with subsidy as
\begin{equation}\nonumber
	(s_{D})' = \frac{1}{1+(\Theta_{D})'},
\end{equation}
where,
\begin{equation}\nonumber
	(\Theta_{D})' = \frac{\tau\left(\gamma + \frac{\rho-n}{g^*}\right)}{(1-\xi)(1-\beta)},
\end{equation}
here, $g^*$ is shown in (\ref{ggrowfinal}). After that, we derive the optimal subsidy rate of wage in R\&D sector by letting the labor share in the decentralized model equal to that in the social planner's problem, i.e.,
\begin{equation}\nonumber
	\tau(t) = \tau^* = \frac{(1-\beta)[(\sigma-\xi)n+\xi \rho-(\sigma-\xi)(1-\zeta)g^*]}{\gamma g^*+\rho-n}.
\end{equation}
Q.E.D.
\endproof

\subsubsection{Optimal Subsidy on Profits from Intermediate Good Productions}
\label{prooftax}

An alternative way of pushing the decentralized model close to the solution of social planner's problem is to subsidize intermediate producers directly. Formally, denote $\tau'(t)$ as the subsidy rate of profit, then in BGP, total value of a patent
\begin{equation}\label{Vsim}
	V(t) = \frac{\pi(t)}{r^* -n}
\end{equation}
becomes
\begin{equation}\label{Vsimadjust}
	V(t) = \frac{\tau'(t)\pi(t)}{r^*-n}.
\end{equation}
Then, we propose the following proposition.

\begin{proposition}\label{proptax}
    In order to push labor allocations in decentralized economy close to the solution in social planner's problem without deviating from the BGP growth rates, the subsidy rate on profit from intermediate good productions $\tau'(t)$ should be constant, which is determined by
    \begin{equation}\nonumber
	\tau'(t) = (\tau')^* = \frac{\gamma g^* + \rho-n}{(1-\beta)[(\sigma-\xi)n + \xi \rho - g^* (\sigma-\xi)(1-\zeta)]} > 1,
	\end{equation}
	where, $g^*$ is BGP growth rate derived in Proposition \ref{dcgrow}.
\end{proposition}

\proof{Proof of Proposition \ref{proptax}.}
Similarly, we should first prove the constancy of subsidy rate $\tau'(t)$. Combining the free-entry condition of labor (\ref{freeentrylab}) and the wage determining equation in production side (\ref{wage}), and then plugging it into the BGP form of $V(t)$ shown in (\ref{Vsimadjust}), we have
\begin{align}\label{tax}
    \eta (1-\xi) N(t)^\zeta \varphi(t)^\xi l_R(t)^{-\xi} V(t) &= w(t) = \beta \left[ \frac{(1-\beta)^2}{\psi} \right]^{\frac{1}{\beta}-1} N(t) \nonumber \\
	\Rightarrow \quad \eta (1-\xi) N(t)^{\zeta-1} \varphi(t)^\xi l_R(t)^{-\xi} \frac{\tau'(t)\frac{\psi^{1-\frac{1}{\beta}}\beta}{(1-\beta)^{1-\frac{2}{\beta}}}L_E(t)}{r^*-n} &= \beta \left[ \frac{(1-\beta)^2}{\psi} \right]^{\frac{1}{\beta}-1}
\end{align}
Rewrite this equation into the form of growth rate, we have
\begin{equation}\nonumber
    (\zeta-1) g_N(t) + \xi g_\varphi(t) + g_{\tau'}(t) + n = 0.
\end{equation}
If we want the growth rate $g^*$ and $g^*_\varphi$ to be the same as the solution in the social planner's problem, we derive
\begin{equation}\nonumber
    g_{\varphi'}(t) = (1-\zeta) g^* - \xi g^*_\varphi - n = 0.
\end{equation}
Thus, the subsidy rate should be constant.

Then we determine the optimal subsidy rate. Go on with (\ref{tax}), we have
\begin{align}
	g^* (1-\xi) (l_R^*)^{-1} \frac{\tau'\frac{\psi^{1-\frac{1}{\beta}}\beta}{(1-\beta)^{1-\frac{2}{\beta}}}l_E(t)}{r^*-n} &= \beta \left[ \frac{(1-\beta)^2}{\psi} \right]^{\frac{1}{\beta}-1} \nonumber \\
	\Rightarrow \quad g^* (1-\xi) (l_R^*)^{-1} \frac{\tau'\frac{\psi^{1-\frac{1}{\beta}}\beta}{(1-\beta)^{1-\frac{2}{\beta}}}l_E^*}{\gamma g^* + \rho-n} &= \beta \left[ \frac{(1-\beta)^2}{\psi} \right]^{\frac{1}{\beta}-1} \nonumber \\
	\Rightarrow \quad (l_R^*)^{-1} -1  &= \frac{\gamma g^* + \rho-n}{g^* (1-\xi)\tau'(1-\beta)}. \nonumber
	\end{align}
	To make the labor employment the same as that in the social planner's problem, we need
	\begin{align}
	(l_R^*)^{-1} -1  &= \frac{\gamma g^* + \rho-n}{g^* (1-\xi)\tau'(1-\beta)}  = \frac{(\sigma-\xi)n+\xi \rho}{\xi(1-\xi)g^*} - \frac{(\sigma-\xi)(1-\zeta)}{\xi(1-\xi)} \nonumber \\
	\Rightarrow \quad \tau' &= \frac{\gamma g^* + \rho-n}{(1-\beta)[(\sigma-\xi)n + \xi \rho - g^* (\sigma-\xi)(1-\zeta)]}. \nonumber
	\end{align}
	Q.E.D.
\endproof

\subsection{Data Nonrivalry and Creative Destruction}
\label{app:nonrivalry}

Formally, the innovation possibility frontier becomes
\begin{equation}\label{extendIPF}
	\dot N(t) = \eta N(t)^\zeta D(t)^\xi L_R(t)^{1-\xi},
\end{equation}
where, $D(t)$ is the composite data used by potential intermediate producers, which consist of new data collected from consumers, $\varphi(t)$, and historical data bundles bought from existing intermediate producers in the past $M$ periods, $B(t)$. The composite data are represented as
\begin{equation}\nonumber
	D(t) = \alpha\varphi(t) L(t) + (1-\alpha) B(t),
\end{equation}
where $\alpha$ represents the relative importance between historical and new data in the data composite. If we rearrange the existing intermediate producers by the order of entering time, and set $\delta$ as the depreciation rate, we can define the historical data bundle as:
\begin{equation}\nonumber
	B(t) = \left[M^{-\frac{1}{\epsilon}} \int_{t-M}^{t-\mathrm{d}t} \left(\delta^{t-s} \underline{d}(t,s) \varphi(s) L(s)\right)^{\frac{\epsilon-1}{\epsilon}} \mathrm{d}s \right]^{\frac{\epsilon}{\epsilon-1}},
\end{equation}
where, $\underline{d}(t,s)$ stands for the proportion of time $t$ usage of data that are generated at time $s$ ($s<t$), and $\epsilon$ is the elasticity coefficient of data collected from different periods. We underline those variables related to data generated in the past and traded in current period, from the perspective of potential intermediate producers in the current period. Similarly, we overline those variables related to data generated in current period and traded in future periods.
	
For consistency with the baseline model, let us consider the case when consumers own data. A potential intermediate producer can not only collect new data from consumers, but also buy historical data from existing intermediate producers with a depreciation rate of $\delta$. Thus, given the prices of data generated from time $(t-M)$ to $(t+M)$, a potential intermediate producer decides the quantity of data to collect from consumers, the quantity of data to buy from existing intermediate producers entering in the past $M$ periods, and the quantity of data to sell to potential intermediate producers who will enter in the following $M$ periods.
	
The consumer's utility maximization problem now becomes
\begin{equation}\label{preferextend}
	\max_{c(t),\varphi(t)} \int_0^\infty e^{-(\rho-n)t} \left[ \frac{c(t)^{1-\gamma}}{1-\gamma} - \varphi(t)^\sigma \int_t^{t+M} \delta^{s-t} \bar d(s,t) \mathrm{d}s \right] \mathrm{d}t,
\end{equation}
where $\bar d(s,t)$ means the proportion of data generated at time $t$ and sold to intermediate producers at time $s$ ($s>t$). It is taken as given by consumers but is endogenously determined by the corresponding intermediate producers. The budget constraint (\ref{budget}) and the data provision constraint (\ref{dataBound}) remain the same. Meanwhile, it is straightforward that firms use all the new data they collect from consumers, so we need $d(t,t)=1$. We can expect $\bar d(s,t)$ to be constant in BGP since $\bar d(s,t)\in[0,1]$.\footnote{Refer to the conclusion that labor employment share in R\&D and production sectors is constant in BGP, which is shown in Online Appendix \ref{prooflem1}.} Then, the term $\int_t^{t+M}\delta^{s-t} \bar d(s,t) \mathrm{d}s$ should also be a constant in BGP. Thus, the necessary conditions with respect to $c(t)$ and $\varphi(t)$ are still the same as our baseline model in BGP, which are shown in (\ref{motC}) and (\ref{motphi}).

\subsubsection{The Decentralized Economy}
	
The production process is still the same as that in our baseline model. Now, consider the R\&D process of potential intermediate producers. Their profit maximization problem becomes
\begin{align}\label{extendmaxi}
	\max_{L_R(t),\left\{d(s,t)\right\}_{s=t+\mathrm{d}t}^{t+M},\atop \left\{d(t,s)\right\}_{s=t-M}^{t-\mathrm{d}t},\varphi(t)}& \eta N(t)^\zeta D(t)^\xi L_R(t)^{1-\xi} \left[ V(t) - \int_{t+\mathrm{d}t}^{t+M} e^{-\lambda(s,t)(s-t)} V(s) \mathrm{d}s \right] - w(t) L_R(t)  \nonumber \\
	&  - p_\varphi(t) \varphi(t) L(t) + L(t) \int_{t+\mathrm{d}t}^{t+M} \bar p_\varphi(s,t) \bar d(s,t) \varphi(t) \mathrm{d}s  \nonumber \\
	& - \int_{t-M}^{t-\mathrm{d}t} \underline{p_\varphi}(t,s) \underline{d}(t,s) \varphi(s) L(s) \mathrm{d}s,
\end{align}
where $\underline{p_\varphi}(t,s)$ is the price of data collected at time $s$ and sold to potential intermediate producers at time $t$, ($t>s$), $\bar p_\varphi(s,t)$ is the price of data collected at time $t$ and sold to potential intermediate producers at time $s$, ($t<s$), and $V(t)$ is the total discounted production profit of the intermediate good invented at time $t$. $\varphi(t)$ and $p_\varphi(t)$ are new data collected from consumers at time $t$ and the corresponding price, respectively. Selling data to future intermediate producers makes the seller face the risk of creative destruction: Sellers' own patents become valueless according to a Poisson process with an arrival rate $\lambda(s,t)$, which is assumed to have the following form:
\begin{equation}\nonumber
	\lambda(s,t) = c_0 \bar d(s,t)^2.
\end{equation}
Notice that intermediate producers will never collect new data again if they succeed in entering the market, thus the data they can sell to future intermediate producers are always within the range they collect from consumers at current period, i.e.,
\begin{equation}\nonumber
	\bar d(s,t) \in [0,1], \forall s, t.
\end{equation}
Also, we assume that only incumbent intermediate producers can sell historical data to entrant ones, that is, consumers cannot resell their historical data.

The profit maximization problem of a potential intermediate producer (\ref{extendmaxi}) consists of five terms. The first term is the total benefit that it can get if it succeeds in R\&D and enters the market, minus the expected total loss of creative destruction coming from selling data to future intermediate producers. The second and third terms are the same as those in our baseline model: the wage cost and the cost of new data collected from consumers. The fourth term is the revenue from selling data to future intermediate producers. The last term is the cost of buying historical data from the intermediate producers who entered in earlier periods.

From the potential intermediate producers' profit maximization in (\ref{extendmaxi}), we know that the first order condition with respect to $L_R(t)$ is similar to the baseline model, i.e.,
\begin{equation}\label{FOCw}
	\eta (1-\xi) N(t)^\zeta D(t)^\xi L_R(t)^{-\xi} V(t) = w(t).
\end{equation}

The first order condition with respect to $\varphi(t)$ is
\begin{align}\label{FOCphitt}
	&\eta \xi N(t)^\zeta D(t)^{\xi-1} L_R(t)^{1-\xi} V(t) \alpha L(t) = p_\varphi(t) L(t) - L(t) \int_{t+\mathrm{d}t}^{t+M} \bar p_\varphi(s,t) \bar d(s,t)\mathrm{d}s \nonumber \\
	\Rightarrow \quad & \alpha \eta \xi N(t)^\zeta D(t)^{\xi-1} L_R(t)^{1-\xi} V(t) = p_\varphi(t) - \int_{t+\mathrm{d}t}^{t+M} \bar p_\varphi(s,t) \bar d(s,t)\mathrm{d}s .
\end{align}
	
For any $s\in(t,t+M]$, the first order condition with respect to $\bar d(s,t)$ is
\begin{align}\label{FOCt1d}
	L(t) \bar p_\varphi(s,t) \varphi(t) = 2 \eta N(t)^\zeta D(t)^\xi L_R(t)^{1-\xi} e^{-c_0 \bar d(s,t)^2 (s-t)} c_0 \bar d(s,t) (s-t) V(s).
\end{align}
	
For any $s\in[t-M,t)$, the first order condition with respect to $\underline{d}(t,s)$ is
\begin{align}\label{FOCt2d}
	&(1-\alpha) \eta \xi N(t)^\zeta D(t)^{\xi-1} L_R(t)^{1-\xi} B(t)^{\frac{1}{\epsilon}} M^{-\frac{1}{\epsilon}} \left[ \delta^{t-s} \underline{d}(t,s) \varphi(s) L(s) \right]^{-\frac{1}{\epsilon}}... \nonumber\\
	&\delta^{t-s} \varphi(s) L(s) \left[ V(t) - \int_{t+\mathrm{d}t}^{t+M} e^{-\lambda(s,t)(s-t)}V(s)\mathrm{d}s \right]= \underline{p_\varphi}(t,s) \varphi(s) L(s)
\end{align}
	
In equilibrium, the demand and supply of historical data are equal, i.e.,
\begin{equation}\nonumber
	\underline{d}(t,s) = \bar d(t,s), \forall t>s.
\end{equation}
Also, the prices of historical data follow:
\begin{equation}\nonumber
	\underline{p_\varphi}(t,s) = \bar p_\varphi(t,s), \forall t>s.
\end{equation}
Combining (\ref{FOCt1d}) with (\ref{FOCt2d}), we solve for $d(t,s)$ and their corresponding prices given other variables, i.e., ($s<t, t-s\leq M$)
\begin{align}\label{dDC}
	&(1-\alpha) \eta \xi N(t)^\zeta D(t)^{\xi-1} L_R(t)^{1-\xi}  B(t)^{\frac{1}{\epsilon}} M^{-\frac{1}{\epsilon}} \left[ \delta^{t-s} \underline{d}(t,s) \varphi(s) L(s) \right]^{-\frac{1}{\epsilon}} \delta^{t-s} \varphi(s) L(s)... \nonumber\\
	& \left[ V(t) - \int_{t+\mathrm{d}t}^{t+M} e^{-\lambda(s,t)(s-t)}V(s)\mathrm{d}s \right] = 2 \eta N(s)^\zeta D(s)^\xi L_R(s)^{1-\xi} e^{-c_0 \bar d(t,s)^2 (t-s)} c_0 \bar d(s,t) (t-s) V(t) \nonumber \\
	\Rightarrow \quad &d(t,s)^{1+\frac{1}{\epsilon}} e^{-c_0 d(t,s)^2 (t-s)} = \frac{(1-\alpha) \xi}{2 c_0(t-s)} \delta^{\left(1-\frac{1}{\epsilon}\right) (t-s)} e^{\bar g_D (s-t)}  D(t)^{-1} B(t)^{\frac{1}{\epsilon}} M^{-\frac{1}{\epsilon}} [\varphi(t)L(t)]^{1-\frac{1}{\epsilon}}... \nonumber \\
	&\left[ 1 - \int_{t+\mathrm{d}t}^{t+M} e^{-\lambda(s,t)(s-t)} \frac{V(s)}{V(t)}\mathrm{d}s \right] \nonumber \\
	\Rightarrow \quad &d(t,s)^{1+ \frac{1}{\epsilon}} e^{-c_0 \bar d(t,s)^2 (t-s)} = \frac{(1-\alpha) \xi}{2 c_0(t-s)} \delta^{\left(1-\frac{1}{\epsilon}\right) (t-s)} e^{\bar g_D (s-t)}  D(t)^{-1} B(t)^{\frac{1}{\epsilon}} M^{-\frac{1}{\epsilon}} [\varphi(t)L(t)]^{1-\frac{1}{\epsilon}}... \nonumber \\
	&\left[ 1 - \int_{t+\mathrm{d}t}^{t+M} e^{-(c_0 d(s,t)^2 -n)(s-t)} \mathrm{d}s \right],
\end{align}
where $\bar g_D = \zeta g_N + \xi g_D + (1-\xi)n.$

Note that in BGP, $g_D=\max{\{g_\varphi +  n, g_B\}}$, $g_V=n$, and
\begin{align}
	g_B &= \frac{\mathrm{d}}{\mathrm{d}t} \left[ M^{\frac{1}{1-\epsilon}} + \frac{\epsilon}{\epsilon-1} \ln{\left( \int_{t-M}^{t-\mathrm{d}t} (\delta^{t-s} d(t,s) \varphi(s) L(s) )^{\frac{\epsilon-1}{\epsilon}} \mathrm{d}s \right)} \right] \nonumber \\
	&= \frac{\epsilon}{\epsilon-1} \frac{\mathrm{d}}{\mathrm{d}t} \left[ \ln{\left( \int_{t-M}^{t-\mathrm{d}t} (\delta^{t-s} d(t,s) \varphi(t-M) L(t-M) e^{(g^*_\varphi+n)(s-t+M)} )^{\frac{\epsilon-1}{\epsilon}} \mathrm{d}s \right)} \right] \nonumber \\
	&= \frac{\epsilon}{\epsilon-1} \frac{\mathrm{d}}{\mathrm{d}t} \left[ \frac{\epsilon-1}{\epsilon} (\ln{\varphi(t-M)} + \ln{L(t-M)}) + \ln{\left( \int_{t-M}^{t-\mathrm{d}t} (\delta^{t-s} d(t,s)  e^{(g^*_\varphi+n)(s-t+M)} )^{\frac{\epsilon-1}{\epsilon}} \mathrm{d}s \right)} \right] \nonumber \\
	&= g_\varphi + n + \frac{\epsilon}{\epsilon-1} \frac{\mathrm{d}}{\mathrm{d}t} \left[ \ln{\left( \int_{t-M}^{t-\mathrm{d}t} (\delta^{t-s} d(t,s)  e^{(g^*_\varphi+n)(s-t+M)} )^{\frac{\epsilon-1}{\epsilon}} \mathrm{d}s \right)} \right] \nonumber \\
	&= g_\varphi + n. \nonumber
\end{align}
The last equation follows from the fact that the term $\delta^{t-s} d(t,s)  e^{(g^*_\varphi+n)(s-t+M)}$, $s\in[t-M,t)$ is not time varying. Thus, we have $g_D=g_\varphi+n$, and $\bar g_D = \zeta g_N +\xi g_\varphi +n$.
	
Now, let us consider (\ref{FOCphitt}). First, it is obvious that in BGP,
\begin{equation}\nonumber
	\frac{\dot p_\varphi(s,t)}{p_\varphi(s,t)} \leq \frac{\dot p_\varphi(t)}{p_\varphi(t)}.
\end{equation}
Otherwise, after a certain period, when the price of historical data exceeds the price at which they are newly sold, intermediate producers can get arbitrage by buying and selling historical data. Thus, the growth rate of the term
\begin{equation}\nonumber
	p_\varphi(t) - \int_{t+\mathrm{d}t}^{t+M} \bar p_\varphi(s,t) \bar d(s,t) \mathrm{d}s
\end{equation}
is just the same as that of $p_\varphi(t)$, which is denoted as $g_{p_\varphi}$.\footnote{We can further discuss the growth rates of historical data and new data prices. If $\dot p_\varphi(s,t) / p_\varphi(s,t) < \dot p_\varphi(t) / p_\varphi(t)$, then we can expect that in a certain period, the price of new data becomes much larger than that of historical data, or we can say that the price of historical data becomes trivial compared with that of the new data. In this case, it is obvious that intermediate producers demand all of the historical data since they are very cheap, i.e., $d(s,t)\to 1$. Then, we can derive the optimal $d(s,t)$ directly and the results of other variables remain unchanged. As a result, in the following analysis, we just consider the case that $\dot p_\varphi(s,t) / p_\varphi(s,t) = \dot p_\varphi(t) / p_\varphi(t)$.} Now, transforming (\ref{FOCphitt}) into the form of growth rate and plugging in the relation of growth rates derived in BGP, we get
\begin{align}\label{gpphi1}
	\zeta g_N + \xi g_D + (1-\xi)n + n &= g_\varphi + n + g_{p_\varphi} \nonumber \\
	\Rightarrow \quad \zeta g_N + \xi \left(g_\varphi + n \right) + (1-\xi)n &= g_\varphi + g_{p_\varphi} \nonumber \\
	\Rightarrow \quad \zeta g_N + \left( \xi -1 \right) g_\varphi + n &= g_{p_\varphi}.
\end{align}
Combining (\ref{gpphi1}) with the Euler equation derived from consumers' problem (\ref{motphi}), we get (note that in BGP, $g_c=g_N=g^*$)
\begin{align}\label{equ1}
	\zeta g_N + \left( \xi -1 \right) g_\varphi + n &= (\sigma-1) g_\varphi + \gamma g_c \nonumber \\
	\Rightarrow \quad (\zeta - \gamma) g^* + ( \xi - \sigma ) g^*_\varphi + n &= 0.
\end{align}
	
Meanwhile, from (\ref{FOCw}), as well as the wage determining equation from the final good producer side (\ref{wage}), we have
\begin{align}\label{equ2}
	\zeta g_N + \xi g_D - \xi n + g_V &= g_N \nonumber \\
	\Rightarrow \quad (\zeta-1)g_N + \xi (g_\varphi + n) - \xi n + n &=0 \nonumber \\
	\Rightarrow \quad ( \zeta-1 ) g^* +  \xi g^*_\varphi + n &=0.
\end{align}
Combining (\ref{equ1}) and (\ref{equ2}), we derive the BGP growth rates of variety and data provision, which are the same as those in our baseline model.
\begin{equation}\nonumber
	g^* = \left[\frac{\sigma}{(1-\zeta)\sigma - \xi(1-\gamma)} \right]n
\end{equation}
and
\begin{equation}\nonumber
g^*_\varphi = \left[\frac{1-\gamma}{(1-\zeta)\sigma - \xi(1-\gamma)} \right]n.
\end{equation}
Since there is no change in the evolution of labor allocations, the expression of labor employment in R\&D sector is also the same.

Now, we derive the proportion of data sharing. From (\ref{dDC}) we have the growth rate of the right hand side as
\begin{align}\nonumber
	-g_D + \frac{1}{\epsilon} g_B +\left(1-\frac{1}{\epsilon}\right) (g_\varphi+n) = 0.
\end{align}
Thus, the left hand side is a constant determined by
\begin{align}
	d_{D}^{1+\frac{1}{\epsilon}} e^{-c_0 d_{D}^2 (t-s)} =& \frac{(1-\alpha) \xi}{2 c_0(t-s)} \delta^{\left(1-\frac{1}{\epsilon}\right) (t-s)} e^{\bar g_D (s-t)} M^{-\frac{1}{\epsilon}} \left[ 1 - \int_{t+\mathrm{d}t}^{t+M} e^{-(c_0 d_{D}(r,t)^2 -n)(r-t)} \mathrm{d}r \right]... \nonumber \\
	&D(t)^{-1} B(t)^{\frac{1}{\epsilon}} [\varphi(t)L(t)]^{1-\frac{1}{\epsilon}}. \nonumber
\end{align}
The terms $D(t)^{-1} B(t)^{\frac{1}{\epsilon}} [\varphi(t)L(t)]^{1-\frac{1}{\epsilon}}$ can be further represented by
\begin{align}\label{extendexp}
	&D(t)^{-1} B(t)^{\frac{1}{\epsilon}} [\varphi(t)L(t)]^{1-\frac{1}{\epsilon}} \nonumber \\
	=& \left\{ \alpha B(t)^{-\frac{1}{\epsilon}} [\varphi(t) L(t)]^{\frac{1}{\epsilon}} + (1-\alpha) B(t)^{1-\frac{1}{\epsilon}} [\varphi(t) L(t)]^{\frac{1}{\epsilon}-1} \right\}^{-1} \nonumber \\
	\equiv& \left[ \alpha \Delta_D^{-\frac{1}{\epsilon}} + (1-\alpha) \Delta_D^{1-\frac{1}{\epsilon}} \right]^{-1},
\end{align}
where,
\begin{align}\label{Delta}
    \Delta_D =& B(t) [\varphi(t) L(t)]^{-1} \nonumber \\
    =& [\varphi(t) L(t)]^{-1} \left[ M^{-\frac{1}{\epsilon}} \int_{t-M}^{t-\mathrm{d}t} (\delta^{t-r} d_D(t,r) \varphi(r) L(r))^{\frac{\epsilon-1}{\epsilon}} \mathrm{d}r \right]^{\frac{\epsilon}{\epsilon-1}} \nonumber \\
    =& M^{\frac{1}{1-\epsilon}} \left[ \int_{t-M}^{t-\mathrm{d}t} (\delta^{t-r} d_D(t,r) e^{(g^*_\varphi +n)(r-t)})^{\frac{\epsilon-1}{\epsilon}} \mathrm{d}r \right]^{\frac{\epsilon}{\epsilon-1}}.
\end{align}

In order to derive the proportion of historical data trading, suppose that data generated at time $t$ can only be sold to intermediate producer entering at time $t+M$. Also, it is reasonable to guess that $d(t+M,t)=d(t+2M,t+M)=...=d_D$, since they are all proportions within the range of $[0,1]$ and have the same time interval. Then, (\ref{dDC}) is simplified as
\begin{equation}\nonumber
	d_D^{1+\frac{1}{\epsilon}} e^{-c_0 d_D^2 M} = \frac{(1-\alpha)\xi}{2c_0 M} \delta^{\left(1-\frac{1}{\epsilon}\right)M} e^{-\bar g_D M} M^{-\frac{1}{\epsilon}} \left[ 1- e^{-(c_0 d_D^2 -n)M} \right] \left[ \alpha (\Delta_D')^{-\frac{1}{\epsilon}} + (1-\alpha) (\Delta_D')^{1-\frac{1}{\epsilon}} \right]^{-1},
\end{equation}
where,
\begin{equation}
    \Delta_D' = M^{\frac{1}{1-\epsilon}} \delta^M d_D e^{-(g^*_\varphi +n)M}.
\end{equation}
Thus, $d_D$ can be derived numerically given the values of parameters. We refer to \citet{Jones2020} to set $\alpha=0.5$, $\epsilon=50$, and $c_0=0.2$ and we take $M=1$ and $\delta=0.9$, $\xi=0.5$, $\zeta=0.85$, $\gamma=2.5$, $n=0.02$, and $\sigma=1.5$. We then get $d_D=0$ or $d_D=3.70>1$. We omit the trivial result of zeros and focus on the non-zero result. It seems that the negative effect of creative destruction is too trivial to make the optimal choice of incumbent intermediate producers be within an appropriate range, if we take $c_0=0.2$ as \citet{Jones2020} do in their paper. However, if we increase $c_0$ to 4, the optimal choice of $d_D$ becomes 0.92, which is now a reasonable value. We discuss this further after deriving the corresponding result for the social planner's problem.

\subsubsection{The Social Planner's Problem}

The social planner's problem is similar to the baseline model shown from (\ref{prefer2}) to (\ref{rclabor}), except that the consumers' utility function (\ref{prefer2}) is now (\ref{preferextend}), and the innovation possibility frontier (\ref{rcwan}) is now (\ref{extendIPF}). 

The maximization problem is shown as
\begin{equation}\nonumber
	\max_{c(t), \left\{d(s,t)\right\}_{s=t+\mathrm{d}t}^{t+M},\atop \left\{d(t,s)\right\}_{s=t-M}^{t-\mathrm{d}t},\varphi(t)} \int_0^\infty e^{-(\rho-n)t} \left[ \frac{c(t)^{1-\gamma}-1}{1-\gamma} - \varphi(t)^\sigma \int_t^{t+M} \delta^{s-t} \bar d(s,t) \mathrm{d}s \right] \mathrm{d}t,
\end{equation}
s.t.
\begin{equation}\nonumber
	\dot N(t) = \eta N(t)^\zeta D(t)^\xi L_R(t)^{1-\xi},
\end{equation}
\begin{equation}\nonumber
	c(t) = \left( \frac{\psi}{1-\beta} \right)^{1-\frac{1}{\beta}} \beta N(t) l_E(t),
\end{equation}
and $l_E(t)+l_R(t)=1$. Let
\begin{equation} \nonumber
	A=\left( \frac{\psi}{1-\beta} \right)^{1-\frac{1}{\beta}},
\end{equation}
we set up the current-value Hamiltonian equation as
\begin{equation}\nonumber
	\mathcal{H} = \frac{c(t)^{1-\gamma}-1}{1-\gamma} - \varphi(t)^\sigma \int_t^{t+M} \delta^{s-t} \bar d(s,t) \mathrm{d}s + \lambda(t)\left[ A\beta N(t) l_E(t) - c(t) \right] + \mu(t) \eta N(t)^\zeta D(t)^\xi L_R(t)^{1-\xi}.
\end{equation}
The necessary conditions with respect to $c(t)$, $\varphi(t)$, $l_E(t)$, and $N(t)$ are
\begin{equation}\label{SPFOCc}
	\frac{\partial \mathcal{H}}{\partial c(t)} = c(t)^{-\gamma} - \lambda(t) = 0,
	\end{equation}
\begin{equation}\label{SPFOCphi}
\frac{\partial \mathcal{H}}{\partial \varphi(t)} = -\sigma \varphi(t)^{\sigma-1}\int_t^{t+M} \delta^{s-t} \bar d(s,t) \mathrm{d}s + \mu(t) \alpha \eta \xi N(t)^\zeta D(t)^\xi L_R(t)^{1-\xi} \frac{1}{\varphi(t)} = 0,
\end{equation}
\begin{equation}\label{SPFOClE}
	\frac{\partial \mathcal{H}}{\partial l_E(t)} = \lambda(t)\beta A N(t) - \mu(t) \eta (1-\xi) N(t)^\zeta D(t)^\xi l_R(t)^{-\xi} L(t)^{1-\xi} = 0
\end{equation}
and
\begin{equation}\label{SPFOCN}
	\frac{\partial \mathcal{H}}{\partial N(t)} = \lambda(t)A\beta l_E(t) + \mu(t) \eta \zeta N(t)^{\zeta-1} D(t)^\xi l_R(t)^{1-\xi} L(t)^{1-\xi} = -\dot \mu(t) + (\rho-n) \mu(t).
\end{equation}
	
For the necessary conditions of $\bar d(s,t)$ and $\underline{d}(t,s)$, we should derive them in other ways. For certain periods $t$ and $s$ ($s<t$, $t-s \leq M$), the Hamiltonian equation is rewritten into the following form
\begin{align}
	\widetilde{\mathcal{H}} =& \frac{c(s)^{1-\gamma}-1}{1-\gamma} - \varphi(s)^\sigma \int_s^{s+M} \delta^{r-s} \bar d(r,s) \mathrm{d}r + e^{-(\rho-n)(t-s)} \left[ \frac{c(t)^{1-\gamma}-1}{1-\gamma} - \varphi(t)^\sigma \int_t^{t+M} \delta^{r-t} \bar d(r,t) \mathrm{d}r \right]... \nonumber \\
	&+ \lambda(s)\left[ A\beta N(s) l_E(s) - c(s) \right] + \mu(s) \eta N(s)^\zeta D(s)^\xi L_R(s)^{1-\xi} + \lambda(t)\left[ A\beta N(t) l_E(t) - c(t) \right]... \nonumber \\
	&+ \mu(t) \eta N(t)^\zeta D(t)^\xi L_R(t)^{1-\xi}. \nonumber
\end{align}
In equilibrium, we have $\bar d(t,s)= \underline{d}(t,s) = d(t,s)$. Then, the first order condition with respect to $d(t,s)$ is
\begin{align}\label{SPFOCd}
	\frac{\partial \widetilde{ \mathcal{H}}}{\partial d(t,s)} =& - \varphi(s)^\sigma \delta^{t-s} + \mu(t) (1-\alpha) \eta \xi N(t)^\zeta D(t)^{\xi-1} L_R(t)^{1-\xi} B(t)^{\frac{1}{\epsilon}} M^{-\frac{1}{\epsilon}} [\delta^{t-s} d(t,s) \varphi(s) L(s)]^{-\frac{1}{\epsilon}}... \nonumber \\
	&\delta^{t-s} \varphi(s) L(s) = 0.
\end{align}
Consider that in BGP, $d(t,s), \forall t,s$ should be constants, thus
\begin{equation}\nonumber
	\int_t^{t+M} \delta^{t-s} \bar d(s,t) \mathrm{d}s
\end{equation}
should also be a constant, which we denote by $\bar M$. Then, from (\ref{SPFOCphi}), we derive the shadow price $\mu(t)$ as
\begin{equation}\nonumber
	\mu(t) = \frac{\sigma \bar M}{\alpha \eta \xi} \varphi(t)^{\sigma} N(t)^{-\zeta} D(t)^{-\xi} l_R(t)^{\xi-1} L(t)^{\xi-1}.
\end{equation}
Substitute this into (\ref{SPFOCd}), we have
\begin{align}\label{SPd}
	\varphi(s)^\sigma \delta^{t-s} &= \frac{\bar M\sigma(1-\alpha)}{\alpha} \varphi(t)^\sigma D(t)^{-1}  B(t)^{\frac{1}{\epsilon}} M^{-\frac{1}{\epsilon}} [\delta^{t-s} d(t,s) \varphi(s) L(s)]^{-\frac{1}{\epsilon}} \delta^{t-s} \varphi(s) L(s) \nonumber \\
	\Rightarrow \quad d(t,s) &= \left[ \frac{\bar M \sigma (1-\alpha)}{\alpha} \varphi(t)^\sigma \varphi(s)^{-\sigma} D(t)^{-1} B(t)^{\frac{1}{\epsilon}} M^{-\frac{1}{\epsilon}} \delta^{-\frac{1}{\epsilon}(t-s)} [\varphi(s) L(s)]^{1-\frac{1}{\epsilon}} \right]^{\epsilon} \nonumber \\
	\Rightarrow \quad d(t,s) &= \left[ \frac{\bar M \sigma (1-\alpha)}{\alpha} \delta^{-\frac{1}{\epsilon}(t-s)} e^{\bar g_S (s-t)} M^{-\frac{1}{\epsilon}} \varphi(t)^{1-\frac{1}{\epsilon}} D(t)^{-1} B(t)^{\frac{1}{\epsilon}} L(t)^{1-\frac{1}{\epsilon}} \right]^{\epsilon},
\end{align}
where,
\begin{equation}\nonumber
	\bar g_S =  \left(-\sigma + 1-\frac{1}{\epsilon} \right) g^*_\varphi + \left(1-\frac{1}{\epsilon} \right) n.
\end{equation}
	
From (\ref{SPFOClE}), by moving the second term to the right hand side, changing the equation into the form of growth rate, and substituting the growth rate of $\lambda_t$ and $\mu_t$ derived from (\ref{SPFOCc}) and (\ref{SPFOCphi}), we have
\begin{align}
	&\frac{\dot \lambda(t)}{\lambda(t)} + \frac{\dot N(t)}{N(t)} = \frac{\dot \mu(t)}{\mu(t)} + \zeta \frac{\dot N(t)}{N(t)} + \xi \frac{\dot D(t)}{D(t)} - \xi n \nonumber\\
	\Rightarrow \quad &-\gamma \frac{\dot c(t)}{c(t)} + (1-\zeta)\frac{\dot N(t)}{N(t)} - \frac{\dot \mu(t)}{\mu(t)} - \xi \frac{\dot D(t)}{D(t)} + \xi n = 0 \nonumber\\
	\Rightarrow \quad &-\gamma \frac{\dot c(t)}{c(t)} + \frac{\dot N(t)}{N(t)} -\sigma \frac{\dot \varphi(t)}{\varphi(t)} =0. 
\end{align}
Notice that the growth of $c(t)$ is what we want to derive, which can be defined as $g^*$, then
\begin{equation}
	\frac{\dot c(t)}{c(t)} = g^* = \frac{1}{\gamma} \left( \frac{\dot N(t)}{N(t)} - \sigma \frac{\dot \varphi(t)}{\varphi(t)} \right).
	\label{extendggrowth1}
\end{equation}
Next, we work to pin down the growth rate of $N(t)$ and $\varphi(t)$. From (\ref{SPFOClE}), we have
\begin{equation} \nonumber
	\frac{\lambda(t)}{\mu(t)} = \frac{\eta(1-\xi)}{\beta A} N(t)^{\zeta-1} l_R(t)^{-\xi} D(t)^\xi L(t)^{1-\xi}.
\end{equation}
Then, (\ref{SPFOCN}) can be reformed as
\begin{align}
	\frac{\lambda(t)}{\mu(t)} A \beta l_E(t) + \eta \zeta N(t)^{\zeta-1} D(t)^\xi l_R(t)^{1-\xi} L(t)^{1-\xi} &= -\frac{\dot \mu(t)}{\mu(t)} + (\rho-n) \nonumber\\
	\eta \left[(1-\xi) l_E(t) + \zeta l_R(t) \right] N(t)^{\zeta-1} D(t)^\xi l_R(t)^{-\xi} L(t)^{1-\xi} &= \zeta \frac{\dot N(t)}{N(t)} + \xi \frac{\dot D(t)}{D(t)} - \sigma \frac{\dot \varphi(t)}{\varphi(t)} + \rho - (1+\xi)n \nonumber \\
	\eta \left[(1-\xi) l_E(t) + \zeta l_R(t) \right] N(t)^{\zeta-1} D(t)^\xi l_R(t)^{-\xi} L(t)^{1-\xi} &= \zeta g_N + (\xi-\sigma) g_\varphi + \rho - n.
	\label{extendproof1}
\end{align}
Since the right hand side is constant in BGP, the left hand side should also be constant. Thus, we have
\begin{equation}
	(\zeta-1)\frac{\dot N(t)}{N(t)} + \xi \frac{\dot D(t)}{D(t)} (1-\xi) n = (\zeta-1 ) g_N + \xi g_\varphi + n= 0.
	\label{extendNphi1}
\end{equation}
We can easily derive that the BGP growth rate of variety $g^*_N$ is that of the economy $g^*$. Combining (\ref{extendggrowth1}) with (\ref{extendNphi1}), we still get the same BGP growth rates as in our baseline model
\begin{equation}\nonumber
	g^* = \left[\frac{\sigma}{(1-\zeta)\sigma - \xi(1-\gamma)} \right]n,
\end{equation}
	and
\begin{equation}\nonumber
	g^*_\varphi = \left[\frac{1-\gamma}{(1-\zeta)\sigma - \xi(1-\gamma)} \right]n,
\end{equation}
	
We then calculate the labor employment in R\&D sector in BGP. Eq. (\ref{extendproof1}) can be rewritten as:
\begin{align} \nonumber
	\eta \left[(1-\xi) l_E(t) + \zeta l_R(t) \right] N(t)^{\zeta-1} D(t)^\xi l_R(t)^{-\xi} L(t)^{1-\xi} &=\zeta g_N + (\xi-\sigma) g_\varphi + \rho - n.
\end{align}
Then, let $g^*=\eta N(t)^{\zeta-1} D(t)^\xi (l_R^*)^{1-\xi} L(t)^{1-\xi}$, (\ref{ECproof1}) can be further rewritten as:
\begin{align}\label{laborpro}
	[(1-\xi)(1-l_R^*) + \zeta l^*_R] g^* (l_R^*)^{-1} &= \zeta g^* + \xi (g^*_\varphi +n) - \sigma g^*_\varphi + \rho -(1+\xi)n \nonumber\\
	\Rightarrow \quad [(1-\xi)(1-l_R^*) + \zeta l^*_R] g^* (l_R^*)^{-1} &= \left[ \zeta + \frac{(\xi-\sigma)(1-\gamma)}{\sigma} \right] g^* + \rho - n \nonumber \\
	\Rightarrow \quad [(1-\xi)(1-l_R^*) + \zeta l^*_R] g^* (l_R^*)^{-1} &= \left[ 1+ \frac{\sigma(\zeta-1)}{\xi} \right] g^* + \left( \frac{\sigma}{\xi} -1 \right) n +\rho.
\end{align}
We have the same labor allocations as those in our baseline model.
	
Finally, we derive the historical data trading proposition. From (\ref{SPd}) we have the growth rate of the right hand side as
\begin{align}\nonumber
	\left(1-\frac{1}{\epsilon} \right) g_\varphi + \left(\frac{1}{\epsilon} \right)g_B - g_D + \left(1-\frac{1}{\epsilon} \right) n =0.
\end{align}
Thus, $d(t,s)$ is a constant determined by (\ref{SPd}). The terms $\varphi(t)^{1-\frac{1}{\epsilon}} B(t)^{\frac{1}{\epsilon}-1} L(t)^{1-\frac{1}{\epsilon}}$ can be further represented by expression similar to (\ref{extendexp}), except that $d(t,s)$ may be different, i.e.,
\begin{equation}\nonumber
	\varphi(t)^{1-\frac{1}{\epsilon}} D(t)^{-1} B(t)^{\frac{1}{\epsilon}} L(t)^{1-\frac{1}{\epsilon}} = \left[ \alpha \Delta_S^{-\frac{1}{\epsilon}} + (1-\alpha) \Delta_S^{1-\frac{1}{\epsilon}} \right]^{-1},
\end{equation}
where,
\begin{equation}
    \Delta_S = M^{\frac{1}{1-\epsilon}} \left[ \int_{t-M}^{t-\mathrm{d}t} (\delta^{t-r} d_S(t,r) e^{(g^*_\varphi +n)(r-t)})^{\frac{\epsilon-1}{\epsilon}} \mathrm{d}r \right]^{\frac{\epsilon}{\epsilon-1}}.
\end{equation}

\bigskip

Similar to the last section, a simplified version of this model can help us understand this problem further. We also assume that data generated at time $t$ can only be sold to intermediate producer entering at time $t+M$. Also, it is reasonable to guess that $d(t+M,t)=d(t+2M,t+M)=...=d_S$ since they are all proportions within the range of $[0,1]$ and have the same time interval. Then, (\ref{SPd}) is simplified as
\begin{equation}\nonumber
	d_S^{\frac{1}{\epsilon}} = \frac{\delta^{-M} \sigma (1-\alpha)}{\alpha} \delta^{-\frac{1}{\epsilon} M} e^{-\bar g_S M} M^{-\frac{1}{\epsilon}} \left[ \alpha (\Delta_S')^{-\frac{1}{\epsilon}} + (1-\alpha) (\Delta_S')^{1-\frac{1}{\epsilon}} \right]^{-1},
\end{equation}
where,
\begin{equation}
    \Delta'_S = M^{\frac{1}{1-\epsilon}} \delta^M d_S e^{-(g^*_\varphi +n)M}.
\end{equation}
We use the same parameters as those in the decentralized economy, which leads to $d_S=0$ or $d_S=0.53\in (0,1)$. We also focus on the non-zero result. Notice that creative destruction does not affect the social planner's solution, so this result still holds when we increase the value of $c_0$. Comparing the optimal selling proportion derived in these two different situations ($d_D$ and $d_S$), we conclude that incumbent intermediate producers are always inclined to sell more historical data to entrants than the social optimal level, unless the negative effect of creative destruction becomes too large.\footnote{Given the parameters we use, we find that $d_D$ becomes smaller than $d_S$ when $c_0$ increases to 14.}

\subsection{Data Ownership: Firms versus Consumers}
\label{appendixOwnership}

We extend the model to allow different specifications of data ownership. As described earlier, the firm ownership problem becomes trivial with zero data processing cost ($\theta=0$). Also, it is not tractable to analyze the case that consumers pay the firms for decreasing the usage of data in order to lower their disutilities from privacy violation. For both reasons, we focus on $\theta>0$ in (\ref{processingCost})---a friction representing the cost of handling data by firms. 

\subsubsection{Analysis with Consumer Ownership of Data}
\label{consumerOwnership}

Let us first consider what happens to our baseline model once we have non-trivial ``data processing cost'' by firms. When consumers own data, the budget constraint of consumers is still the same as (\ref{budget}), and so are other conditions that we do not mention in this section. What makes the model different here is the profit maximization problem of potential intermediate producers who now have to pay the ``data processing cost'' in addition to the fees given to consumers. Then, we have the following proposition.
\begin{proposition}\label{propfirm2}
    With $\theta>0$ and consumer ownership of data, the BGP growth rates and labor allocations are the same as those in the baseline model, provided $\sigma(2-\zeta) + (\xi+\phi) (\gamma-1) > 0$.
\end{proposition}

\proof{Proof of Proposition \ref{propfirm2}.}
The condition $\sigma(2-\zeta) + (\xi+\phi) (\gamma-1) > 0$ holds easily when the knowledge spillover effect is not so large and consumers' EIS are rather small.

The maximization problem of intermediate producers now becomes
\begin{equation}\nonumber
\max_{L_R(t),\varphi(t)} \eta N(t)^\zeta \varphi(t)^\xi l_R(t)^{1-\xi} L(t) V(t) - w(t) l_R(t) L(t) - \theta \varphi(t)^\phi - p_\varphi(t) \varphi(t) L(t).
\end{equation}
Free-entry conditions with respect to labor is still the same as (\ref{freeentrylab}), while the other free-entry condition becomes
\begin{equation}\label{freeentrycon}
\eta \xi N(t)^\zeta \varphi(t)^{\xi-\phi} l_R(t)^{1-\xi} V(t) L(t) = \frac{p_\varphi(t) L(t)}{\varphi(t)^{\phi-1}} + \theta \phi.
\end{equation}
Since $p_\varphi(t)$, $\varphi(t)$ and $L(t)$ all grow in constant rates in BGP, we go on with deriving the BGP growth rates by discussing different situations separately.

\paragraph{Case 1.} In BGP, if $g^*_{p_\varphi} + n > (\phi-1) g^*_\varphi$, it can be derived that the right hand side of (\ref{freeentrycon}) converges to the first term, i.e.,
\begin{equation}\nonumber
\eta \xi N(t)^\zeta \varphi(t)^{\xi-\phi} l_R(t)^{1-\xi} V(t) L(t) \approx \frac{p_\varphi(t) L(t)}{\varphi(t)^{\phi-1}}.
\end{equation}
Then, the free-entry condition becomes the same as that in (\ref{freeentryinfo}). Consequently, the BGP growth rates become the same as those shown in Proposition \ref{dcgrow}. Furthermore, from
\begin{equation}\nonumber
\frac{\dot p_\varphi(t)}{p_\varphi(t)} = (\sigma-1)\frac{\dot \varphi(t)}{\varphi(t)} + \gamma \frac{\dot c(t)}{c(t)} = \left[ \frac{\sigma+\gamma-1}{(1-\zeta)\sigma-\xi(1-\gamma)} \right]n,
\end{equation}
we can derive the condition as
\begin{align}
g^*_{p_\varphi} + n >& (\phi-1) g^*_\varphi \nonumber \\
\Rightarrow \quad \left[ \frac{\sigma+\gamma-1}{(1-\zeta)\sigma-\xi(1-\gamma)} \right]n + n >& (\phi-1)  \left[ \frac{1-\gamma}{(1-\zeta)\sigma-\xi(1-\gamma)} \right]n \nonumber \\
\Rightarrow \quad \sigma(2-\zeta) + (\xi+\phi) (\gamma-1) > &0, \nonumber
\end{align}
which is always the case given the values of the above parameters.

\paragraph{Case 2.} If $g^*_{p_\varphi} + n \leq (\phi-1) g^*_\varphi$, it can be derived that the right hand side of (\ref{freeentrycon}) converges to a constant. Then, this problem becomes the same as the case when firms own data. Then, the BGP growth rate of $p_\varphi(t)$ becomes
\begin{equation}\nonumber
\frac{\dot p_\varphi(t)}{p_\varphi(t)} = (\sigma-1)\frac{\dot \varphi(t)}{\varphi(t)} + \gamma \frac{\dot c(t)}{c(t)} = \left[ \frac{(\sigma-1)(2-\zeta)+\gamma(\xi+\phi)}{\phi(1-\zeta) - \xi} \right] n.
\end{equation}
And the condition becomes
\begin{align}
g^*_{p_\varphi} + n \leq& (\phi-1) g^*_\varphi \nonumber \\
\Rightarrow \quad \left[ \frac{(\sigma-1)(2-\zeta)+\gamma(\xi+\phi)}{\phi(1-\zeta)-\xi} \right]n + n \leq& (\phi-1)  \left[ \frac{2-\zeta}{\phi(1-\zeta)-\xi} \right]n \nonumber \\
\Rightarrow \quad \sigma(2-\zeta) + (\xi+\phi) (\gamma-1) \leq &0, \nonumber
\end{align}
which does not exist given the standard values of the parameters.

Thus, the BGP growth rates when consumers own data are
\begin{equation}\label{BGPgrowth}
g^* = \left[\frac{\sigma }{(1-\zeta)\sigma - \xi (1-\gamma)}\right] n
\end{equation}
and
\begin{equation}\nonumber
\frac{\dot \varphi(t)}{\varphi(t)} = -\frac{1}{\xi} n + \frac{1-\zeta}{\xi}\frac{\dot N(t)}{N(t)} = \left[\frac{ 1-\gamma }{(1-\zeta)\sigma-\xi (1-\gamma)}\right] n,
\end{equation}
which are the same as those shown in Proposition \ref{dcgrow}. Obviously, the labor allocations in BGP is still the same as those shown in Proposition \ref{DClsh}. Q.E.D.
\endproof

\bigskip

In general, we assume $\sigma>0$, $\zeta\in [0,1]$, $\xi\in[0,1]$, $\phi>1$, $\gamma>1$. Thus, the proposition exists when we take standard values of the parameters. The BGP growth rate of consumption is now larger than that of the data provision. However, if we make the data provision constraint (\ref{dataBound}) tighter, i.e., let $s$ be smaller so that the constraint binds in BGP, we have lower growth rates. Formally, if $s$ satisfies
\begin{equation}\nonumber
    s < g^*_\varphi - g^* =  \left[\frac{ 1-\gamma -\sigma }{(1-\zeta)\sigma-\xi (1-\gamma)}\right] n,
\end{equation}
we always have
\begin{equation}\label{consumercons}
    (g^*_\varphi)' = (g^*)'+s.
\end{equation}
Then, Eq. (\ref{ECgrphi2}) derived from the free-entry condition of data (\ref{freeentryinfo}) fails. Thus, we combine (\ref{ECgrphi}) with (\ref{consumercons}) to derive the BGP growth rates in this case as
\begin{equation}
    (g^*)' = \frac{\xi s + n}{1-\zeta-\xi} < \left[\frac{\sigma }{(1-\zeta)\sigma - \xi (1-\gamma)}\right] n= g^*,
\end{equation}
which means the BGP growth rate in this case is always lower than that in our baseline model.

In addition, we consider the social planner's problem, which leads to the following proposition.
\begin{proposition}\label{propfirm3}
    With $\theta>0$ in the social planner's problem, the BGP growth rate as well as labor allocations still do not change compared with the baseline model, provided $\gamma \sigma + (\phi-1) ( \gamma-1) +1> 0$.
\end{proposition}
\proof{Proof of Proposition \ref{propfirm3}.}

The condition $\gamma \sigma + (\phi-1) ( \gamma-1) +1> 0$ holds easily when the ``data processing cost'' is convex and consumers' consumption growth is rather insensitive to real interest rates.

In the social planner's problem, the resource constraint now becomes
\begin{equation}\nonumber
C(t) = \widetilde Y(t) \equiv L_E(t)^\beta \int_0^{N(t)} x(v,t)^{1-\beta}\,\mathrm{d}v - \int_0^{N(t)} \psi x(v,t)\,\mathrm{d}v - \theta \varphi(t)^\phi.
\end{equation}
Then, the optimal net output becomes
\begin{equation}\label{rcnew}
c(t) = \left(\frac{\psi}{1-\beta}\right)^{1-\frac{1}{\beta}} \beta N(t) l_E(t)- \frac{\theta \varphi(t)^\phi}{L(t)}.
\end{equation}
Thus, (\ref{ECFOCphiS}) becomes
\begin{equation}\label{FOCphinew}
\frac{\partial \mathcal{H}}{\partial \varphi(t)} = - \sigma \varphi(t)^{\sigma-1} - \theta \phi \frac{\lambda(t) \varphi(t)^{\phi-1}}{L(t)} + \mu(t) \xi \eta N(t)^\zeta \varphi(t)^{\xi-1} l_R(t)^{1-\xi} L(t) = 0.
\end{equation}
Since $c(t)^{-\gamma}=\lambda(t)$, we should first discuss the value of the additional term in BGP. This term converges to zero when
\begin{equation}\nonumber
-\gamma \frac{\dot c(t)}{c(t)} + (\phi-1)\frac{\dot \varphi(t)}{\varphi(t)} < n.
\end{equation}
If the above inequality is satisfied, the BGP growth rates will be the same as those we derived in Proposition \ref{growsp}. Then, we further derive the inequality as
\begin{align}
-\gamma \left[\frac{\sigma }{(1-\zeta)\sigma - \xi (1-\gamma)}\right] n + (\phi-1) \left[\frac{1-\gamma }{(1-\zeta)\sigma - \xi (1-\gamma)}\right] n <& n \nonumber \\
\Rightarrow \quad \gamma \sigma + (\phi-1) ( \gamma-1) >& -1,
\end{align}
which is always the case given standard values of the above parameters. Thus, we can verify that Proposition \ref{growsp} is still satisfied in this new setting. Obviously, labor allocations in BGP are still the same as those shown in Proposition \ref{SPlsh}. Q.E.D.
\endproof

\bigskip

Similarly, this proposition also holds when we take standard values of the parameters. Propositions \ref{propfirm2} and \ref{propfirm3} both show that our baseline model is robust under general data processing cost $\theta$.

\subsubsection{Analysis with Firm Ownership of Data}

When firms own data, consumers' budget constraint (\ref{budget}) now becomes
\begin{equation}\nonumber
\dot a(t) = (r(t)-n) a(t) + w(t) - c(t), \forall t \in [0,\infty).
\end{equation}
Then, we have the following proposition about the BGP growth rates and labor allocations in this model.
\begin{proposition}\label{propfirm1}
    If firms own data, the BGP growth rate becomes
    \begin{equation}\label{firmg}
        g^*_c = g^*_N = g^*_y = g^* = \left[\frac{\xi+\phi}{\phi(1-\zeta)-\xi}\right] n,
    \end{equation}
    \begin{equation}\label{firmgphi}
        g^*_\varphi = \left[ \frac{2-\zeta}{\phi(1-\zeta)-\xi} \right]n > 0.
    \end{equation}
    Labor allocations in this case have the same form as those shown in Proposition \ref{DClsh}, except that the corresponding BGP growth rates take (\ref{firmg}) and (\ref{firmgphi}) here. 
\end{proposition}

\proof{Proof of Proposition \ref{propfirm1}.}
When firms own data, the profit maximization problem of a potential intermediate producer becomes
\begin{equation}\nonumber
\max_{L_R(t),\varphi(t)} \eta N(t)^\zeta \varphi(t)^\xi l_R(t)^{1-\xi} L(t) V(t) - w(t) l_R(t) L(t) - \theta \varphi(t)^\phi.
\end{equation}
Thus, the free-entry conditions in this case are:
\begin{equation}\label{firmfreeentryl}
(1-\xi) \eta N(t)^\zeta \varphi(t)^\xi l_R(t)^{-\xi} V(t) = w(t)
\end{equation}
and
\begin{equation}\label{firmfreeentryfirm}
\xi \eta N(t)^\zeta \varphi(t)^{\xi-1} l_R(t)^{1-\xi} L(t) V(t) = \theta \phi \varphi(t)^{\phi-1}.
\end{equation}

Next, we derive the BGP growth rate in this case. Similarly, we still have (\ref{ECVgrow}) and (\ref{ECgrphi}). Now, from (\ref{firmfreeentryfirm}), we rewrite it as
\begin{equation}\nonumber
\zeta \frac{\dot N(t)}{N(t)} + (\xi-1) \frac{\dot \varphi(t)}{\varphi(t)} + (1-\xi) \frac{\dot l_R(t)}{l_R(t)} + n + \frac{\dot V(t)}{V(t)} = (\phi-1) \frac{\dot \varphi(t)}{\varphi(t)}
\end{equation}
In BGP, this equation can be further derived as
\begin{equation}\label{grphifirm}
\zeta \frac{\dot N(t)}{N(t)} + (\xi-\phi) \frac{\dot \varphi(t)}{\varphi(t)} + 2n = 0.
\end{equation}
Combining (\ref{grphifirm}) with (\ref{ECgrphi}), we derive the BGP growth rate as
\begin{equation}\nonumber
\frac{\dot N(t)}{N(t)} = \left[ \frac{\xi+\phi}{\phi(1-\zeta) - \xi} \right] n.
\end{equation}
Meanwhile, the BGP growth rate of data provision is
\begin{equation}\nonumber
\frac{\dot \varphi(t)}{\varphi(t)} = - \frac{n}{\xi} + \frac{1-\zeta}{\xi} \frac{\dot N(t)}{N(t)}= \left[ \frac{2-\zeta }{\phi(1-\zeta) - \xi} \right] n>0.
\end{equation}

The labor allocations in this case have the same form as those shown in Proposition \ref{DClsh}, except that the BGP growth rate is different here. Q.E.D.
\endproof

\bigskip

In (\ref{dataBound}), we require that the growth rate of data provision should not exceed the growth rate of consumption. As a result, when $2-\zeta > \xi + \phi$, the constraint binds. In this case, free-entry condition of data fails since firms always require excessive data. We can then calculate the growth rates from the free-entry condition of labor (\ref{freeentrylab}) as
\begin{equation}\nonumber
    g^* = g^*_\varphi = \frac{n}{1-\zeta-\xi},
\end{equation}
which means the constraint (\ref{dataBound}) affects the BGP growth rates in this case. Also, following \citet{Jones2020}, the innovation possibility frontier in this paper is set as
\begin{equation}\nonumber
    \dot N_t = \frac{1}{\chi} L_E(t),
\end{equation}
which can be viewed as setting the knowledge spillover effect $\zeta$ to be 0, while data are set as different characters. Then, the condition $\xi + \phi < 2-\zeta$ is likely to be satisfied when we take proper values of $\xi$ and $\phi$. As a result, we can derive similar conclusion under this condition: When firms own data, they are inclined to use up all the data they have.

\end{document}